\documentclass[aps,prd,preprintnumbers,showpacs,nofootinbib]{revtex4}

\usepackage{overpic}
\usepackage{subfigure}
\usepackage[english]{babel}
\usepackage{amsmath}
\usepackage{amssymb}
\usepackage{epsfig}
\usepackage{graphics,psfrag,rotating}
\usepackage{graphicx}
\usepackage{dcolumn}
\usepackage{bm}
\usepackage{hyperref}

\begin{document}
\newcommand{\Od}{{\cal O}}
\newcommand{\lsim}   {\mathrel{\mathop{\kern 0pt \rlap
  {\raise.2ex\hbox{$<$}}}
  \lower.9ex\hbox{\kern-.190em $\sim$}}}
\newcommand{\gsim}   {\mathrel{\mathop{\kern 0pt \rlap
  {\raise.2ex\hbox{$>$}}}
  \lower.9ex\hbox{\kern-.190em $\sim$}}}
\newcommand{\be}{\begin{equation}}
\newcommand{\ee}{\end{equation}}
\newcommand{\bea}{\begin{eqnarray}}
\newcommand{\eea}{\end{eqnarray}}
\newcommand{\beaa}{\begin{eqnarray*}}
\newcommand{\eeaa}{\end{eqnarray*}}
\newcommand{\Lhat}{\widehat{\mathcal{L}}}
\newcommand{\nn}{\nonumber \\}
\newcommand{\e}{{\rm e}}

\title{Black Holes, Cosmological Solutions, Future Singularities, \\and Their Thermodynamical Properties in Modified Gravity Theories}
\author{A. de la Cruz-Dombriz\footnote{E-mail: alvaro.delacruzdombriz@uct.ac.za}$^{(1, 2)}$ and D. S\'aez-G\'omez\footnote{E-mail: diego.saez@ehu.es}$^{(3)}$}

\address{%
$^{1}$ Astrophysics, Cosmology and Gravity Centre (ACGC), University of Cape Town, Rondebosch 7701, Cape Town, South Africa\\
$^{2}$ Department of Mathematics and Applied Mathematics, University of Cape Town,  Rondebosch 7701, Cape Town, South Africa\\
$^{3}$ Fisika Teorikoaren eta Zientziaren Historia Saila, Zientzia eta Teknologia Fakultatea, Euskal Herriko Unibertsitatea, 644 Posta Kutxatila, 48080 Bilbao, Spain, EU}

\begin{abstract}Along this review, we focus on the study of several properties of modified gravity theories, in particular on black-hole solutions and its comparison with those solutions in General Relativity, and on Friedmann--Lema\^itre--Robertson--Walker metrics. The thermodynamical properties of fourth order gravity theories are also a subject of this investigation with special attention on  local and global stability of paradigmatic $f(R)$ models. In addition, we revise some attempts to extend the Cardy--Verlinde formula, including modified gravity, where a relation between entropy bounds is obtained. Moreover, a deep study on cosmological singularities, which appear as a real possibility for some kind of modified gravity theories, is performed, and the validity of the entropy bounds is studied.
\end{abstract}
\pacs{98.80.-k; 04.50.+h}

\maketitle 


\section{Introduction}

General Relativity (GR) has been the most successful gravitational theory of the last century, fully accepted as a theory that describes the macroscopic geometrical properties of space-time. For an isotropic and homogeneous geometry, GR leads to Friedmann equations which describe in an appropriate way the cosmological evolution with radiation and then matter dominated epochs. Nevertheless, the development of observational cosmology in the last decades with experiments of increasing precision like supernovae observations \cite{SIa,SIa1,SIa2} has revealed that the Universe is in a stage of accelerated expansion. GR provided with usual matter sources is not able to explain this phenomenon. Moreover, GR does not account either for the cosmological era known as inflation \cite{inflac}, believed to have taken place before the radiation stage and that could alleviate
some problems of standard cosmology like the horizon and the flatness problem \cite{Peebles}. In addition, GR with usual baryonic matter cannot explain the observed matter density 
determined by fitting the standard $\Lambda\text{CDM}$ model to the WMAP7 data \cite{Komatsu}, the latest measurements from the BAO (Baryon Acoustic Oscillations) in the distribution of galaxies \cite{Percival} and the Hubble constant ($H_0$) measurement \cite{Riess}. Thus, GR requires the introduction of an extra component called dark matter that accounts for about $20\%$ of the energy content of our Universe.

A more puzzling problem is associated to the present accelerated expansion of the Universe. There are also a large amount of different explanations. One of them, assuming the validity of GR, postulates the existence of an extra cosmic fluid, the dark energy (DE), whose state equation $p=\omega_{\text{DE}} \rho$ (where $p$ and $\rho$ are the pressure and the energy density of the fluid) demand $\omega_{\text{DE}}<-1/3$ in order to provide an accelerated cosmic expansion \cite{DE,DE1,DE2}. The cosmological constant is the simplest model of DE, corresponding to an equation of state $\omega_{\text{DE}}=-1$. However, if we assume that the cosmological constant represents the quantum vacuum energy, its value seems to be many orders of magnitude bigger than the observed one \cite{cosmoproblema}.

Thus, alternative explanations to the cosmological constant have been largely studied in the last years. These theories essentially
modify GR by considering actions different from the Einstein--Hilbert \linebreak one \cite{varios,varios1,varios3,varios4,varios5,varios6,varios7,varios8,varios9,varios10,varios11,varios12}. Examples are Lovelock theories, free of ghosts and whose field equations contain second derivatives of the metric at most; string theory inspired models, which include a Gauss--Bonnet term in the Lagrangian (see \cite{Dombriz-Saez} and references therein); scalar-tensor theories like Brans--Dicke one, in which gravitational interaction is mediated by both a scalar field and GR tensor field; or the so-called $f(R)$ theories (see \cite{Reviews,Reviews1,Reviews2,Reviews3,Reviews4,Reviews5,Reviews6,Reviews7} for a recent and extensive review). 
In this investigation we shall restrict ourselves to $f(R)$ theories in the metric formalism (where the connection depends on the metric, so the present fields in the gravitational sector of the action come only from the metric tensor) in the Jordan frame. In this frame, the gravitational Lagrangian is given by $R+f(R)$, where $f(R)$ is an arbitrary function of the scalar curvature $R$, and Einstein's equations usually become fourth order on the metric derivatives.
The $f(R)$ theories were proved (see \cite{Capozziello:2002rd,Capozziello:2002rd1,Reviews,Reviews1,Reviews2,Reviews3,Reviews4,Reviews5,Reviews6,Reviews7,Odintsov,reconstruction3,Dunsby1} among others) to be able to mimic the whole cosmological history, from inflation to the actual accelerated expansion era.
Diverse applications of these theories on gravitation and cosmology have been also widely studied  \cite{Tsujikawa,Tsujikawa1,Tsujikawa2}, as well as multiple ways to observationally and experimentally distinguish them from GR.  Concerning local tests of gravity and other cosmological constraints, see \cite{varia,varia1,varia2,varia3,varia4}.


In fact, most of the modified gravity theories present the so-called {\it degeneracy problem}: from large scale observations (Ia type supernova, BAO, or the cosmic microwave background) which depend uniquely on the evolution history of the Universe, the nature and the origin of DE cannot be determined due to the fact that identical evolutions can be explained by a diverse number of theories. However, it has been proved in \cite{Dombriz_perturbaciones_PRD,Dombriz_perturbaciones_PRD1} and more recently in \cite{Amare} that when scalar cosmological perturbations are studied, $f(R)$ theories, even mimicking the standard cosmological expansion, provide a different matter power spectrum from that predicted by the $\Lambda$CDM model \cite{Comment}.

The study of alternative gravitational theories to GR requires to confirm or discard their validity by obtaining solutions that can describe correctly, e.g., the
cosmological evolution, the  growth factor of  cosmological perturbations and the existence of GR-predicted astrophysical objects such as black holes (BH).
Therefore, it is interesting to study the properties of BH in this kind of theories, since some of their features might be either exclusive of Einstein's gravity or intrinsic features of any covariant gravitational theory. On the other hand, obtained results could provide a method to discard models  that disagree with expected physical results. In this sense, research of BH thermodynamics may shed some light about the viability of alternative gravity theories since local and global stability regions, and consequently the existence itself of BH, depend on the values of the parameters of the model under consideration.


BH properties have been widely studied in other modified gravity theories: for instance \cite{cvetic,Cai_GaussBonet_AdS} studied BH in Einstein's theory with a Gauss--Bonnet term and a cosmological constant. Gauss--Bonnet and/or quadratic Riemann interaction terms are studied in \cite{Cho}, where it is found that for a negative curvature of the horizon phase transitions might occur. BH in Lovelock gravitational theories were studied in \cite{Matyjasek,Matyjasek1}, where the corresponding entropy was calculated. Other recent works have \linebreak studied \cite{Horava,Horava1,Horava2} BH in the context of Ho\u{r}ava--Lifshitz gravity and static solutions in the context of $f(T)$ theories \cite{Wang:2011xf} as well.

Previous works concerning BH in $f(R)$ theories proved \cite{Whitt} that for a Lagrangian $R+aR^2$ the only spherically symmetric solution is Schwarzschild's one provided that one works in the Einstein's frame. Again in Einstein's frame, \cite{Mignemi} proposed uniqueness theorems for spherically symmetric solutions with an arbitrary number of dimensions (see \cite{Multamaki} for additional results).

Spherical solution with sources were also studied in \cite{olmo} whereas \cite{Nzioki:2009av} developed a new covariant \linebreak formalism to treat spherically symmetric space-times claiming that Schwarzschild solution is not a unique static spherically symmetric solution. Spherically symmetric $f(R)$-Maxwell and $f(R)$-Yang--Mills BH were studied in \cite{Taeyoon1}, confirming the existence of numerical asymptotic solution for the second ones. Concerning axially symmetric solutions, authors in \cite{Capozziello:2009jg} showed that these solutions can be derived by generalizing Newman and Janis method to $f(R)$ theories. An scalar-tensor approach is used in \cite{Myung} to show that Kerr BH are unstable in a subset of $f(R)$ models because of the superradiant instability. In \cite{Palatini_Noether} the entropy of BH is calculated in the Palatini formalism by using the Noether charge approach.
Anti-de Sitter (AdS) BH have been studied \cite{Cognola} in $f(R)$ models using the Euclidean action method \linebreak(see, e.g., \cite{Hawking&Page, Witten}) to determine different thermodynamic quantities.
In \cite{Briscese}, the entropy of Schwarzschild--de Sitter ($SdS$) BH is calculated in vacuum for certain cosmologically viable models, and their stability is discussed.

%
%
Different aspects of BH in $f(R)$ have been studied recently, including in particular stability issues \cite{Taeyoon1,Myung:2011ih}, showing an increasing interest in these topics,  static and spherically symmetric  solutions \cite{PerezBergliaffa:2011gj, Nelson:2010ig, Moon:2011sz},
rotating configurations   \cite{Larranaga:2011fv, Myung:2011we} and the existence of anomalies \cite{Hendi:2012nj}.

In fact, since the establishment of thermodynamical properties for black holes along the seventies, a special interest has been focused on the relation of the Einstein gravity and laws of thermodynamics, whose efforts have been increasing along recent years. Particularly, Jacobson showed in 1995~\cite{Jacobson:1995ab} that field equations in General Relativity can be recovered from thermodynamics identities by assuming the relation, given by mechanical laws of black holes, between the entropy and the area of the horizon, and assuming that the relation $\delta Q=TdS$ holds for all local Rindler causal horizons through each spacetime point. Such an important result points to the hypothesis that gravitational equations may be just equations of state of the spacetime rather than fundamental laws of gravitation. Furthermore, this suggestion has been reinforced since the relationship between gravitational equations and thermodynamical ones has been extended to more general gravitational theories, as for instance, $f(R)$ gravity (see~\cite{Elizalde:2008pv}).

Moreover, Friedmann--Lema\^itre--Robertson--Walker (FLRW) equations have been successfully derived from the thermodynamical properties of the apparent horizon (see~\cite{Cai:2005ra,Cai:2005ra1,Cai:2005ra2}). In this sense, modified FLRW equations of more general gravitational theories as Lovelock gravity or $F(R)$ gravity have been also related with the thermodynamical properties of the apparent horizon~\cite{Akbar:2006er,Wu:2007se,Wu:2007se1,Wu:2007se2,Wu:2007se3,Wu:2007se4,Wu:2007se5,Wu:2007se6,Wu:2007se7,Wu:2007se8,Wu:2007se9,Wu:2007se10}. In addition, for a radiation dominated universe, the FLRW equations can be rewritten as an entropy relation, which gives an equivalence with the Cardy formula from a two dimensional conformal field theory (CFT) (see~\cite{Cardy}). This relation, proposed initially by Verlinde in \cite{Verlinde}, assumes an entropy bound for the universe related with the entropy proportional to the area of the size of the universe, a kind of holographic principle. Finally, the formula may be rewritten as a dynamical entropy bound from which a number of  entropy bounds, proposed earlier, follow. However, such a relation
 with the 2d CFT has been attempted to be extended for general perfect fluids (see \cite{Youm}), and more general scenarios including modified gravity~\cite{Brevik:2010jv}, but the equivalence could not be extended. Furthermore, it was found that the bound on the entropy may be violated in some particular scenarios, mainly when the universe moves close to a future singularity. Moreover, in spite of that in general the bound is violated sufficiently close to the singularity, even when some quantum effects are included, the violation can be also produced much before the singularity occurs~\cite{Brevik:2010jv}.

Hence, all of these suggest a deep connection between gravitational physics and thermodynamics, which is still an open issue nowadays, and could help to improve our knowledge on the way to reconstruct consistent gravitational theories that may solve some of the most important problems on theoretical physics. Along this work, some of these results are reviewed, and we may suggest new scenarios where gravitational physics may be tested

The present review is organized as follows: In Section \ref{Generalities_fR} we revised the rudiments of the $f(R)$ modified gravity theories that will be used throughout the paper. Then, Section \ref{Static_BH} is devoted to study static and spherically symmetric configurations in these theories, with special focus on electromagnetic solutions and perturbative approach. The following Section \ref{Section_KN_BH} deals with the Kerr--Newman black holes in $f(R)$ theories. Section \ref{Thermodynamics} encompasses a general revision of thermodynamics of $f(R)$ black holes for both anti-de Sitter and Kerr--Newman scenarios. The thermodynamical stability is then studied for two paradigmatic examples of $f(R)$ models in Section \ref{AdS_Thermodynamics}.  Section \ref{CosmoSol} deals with the reconstruction of $F(R)$ actions for some particular cosmological solutions and how this kind of theories are capable to reproduce the whole cosmological evolution. In Section \ref{ThermFLRW}, we review some of the thermodynamical properties of FLRW equations, and its derivation from the first law of thermodynamics, which can be easily extended to $F(R)$ gravity. Section \ref{CVformula} is devoted to some extensions of the Cardy--Verlinde formula to more complex configurations of FLRW universes, including modified gravity. In Section  \ref{Singularities}, future singularities are studied in the context of entropy bounds, and the possible violation of the proposed bounds on the energy and entropy of the universe. Finally,
we conclude the paper by giving our conclusions in Section~\ref{Conclusions}.


\section{$F(R)$ Theories of Gravity}
\label{Generalities_fR}

In this section, $F(R)$ gravity is introduced, which basically consists in an extension of the \mbox{Einstein--Hilbert} action by considering more complex gravitational actions consisting on functions of the Ricci scalar, and which can be rewritten in terms of a Brans--Dicke-like theory as shown below. The action for $F(R)$-gravity is given by:
\be
S=\int \text{d}^4 x
\sqrt{-g}\left(F(R)+2\kappa^2\mathcal{L}_m\right)\  \label{In.1}
\ee
Here $\mathcal{L}_m$ denotes the Lagrangian of some kind of matter, and $\kappa^2=8\pi G$ is the gravitational coupling constant. In the present work we employ the natural units system in which $\hbar=c=1$. Note also that our definition for the Riemann tensor is:
    \be
    R_{\mu\nu\kappa}^{\sigma}=\partial_{\kappa}\Gamma_{\mu\nu}^{\sigma}-\partial_{\nu}\Gamma_{\mu\kappa}^{\sigma}+
    \Gamma_{\kappa\lambda}^{\sigma}\Gamma_{\mu\nu}^{\lambda}-\Gamma_{\nu\lambda}^{\sigma}\Gamma_{\mu\kappa}^{\lambda}
    \ee

Field equations are obtained by varying the action Equation~\eqref{In.1} with respect to $g_{\mu\nu}$,
\be
R_{\mu\nu}F'(R)-\frac{1}{2}g_{\mu\nu}F(R)+g_{\mu\nu}\Box F'(R)-\nabla_{\mu}\nabla_{\nu}F'(R)=\kappa^2T^{(m)}_{\mu\nu}\
\label{In.2}
\ee
where $T^{(m)}_{\mu\nu}=-\frac{2}{\sqrt{-g}}\frac{\delta\mathcal{L}_m}{\delta g^{\mu\nu}}$ is the energy-momentum tensor and prime denotes here derivatives with respect to $R$.  Nevertheless, in general it is very difficult to get exact solutions directly from field Equation~(\ref{In.2}), although some particular solutions can be obtained, as shown in the next sections. However, the action Equation~\eqref{In.1} is equivalent to a kind of scalar-tensor theory with a null kinetic term, which can be used to reconstruct the appropriate action for some particular solutions, specially for cosmological solutions, as shown in Section~\ref{CosmoSol}. Then, the action Equation~(\ref{In.1}) is rewritten as follows \cite{ScalarFR}:
\be
S=\int \text{d}^4 x
\sqrt{-g}\left(P(\phi)R+Q(\phi)+2\kappa^2\mathcal{L}_m\right)\
\label{In.3}
\ee
By the variation of the action with respect to the metric tensor $g_{\mu\nu}$, the field equation is obtained:
\be
-\frac{1}{2}g_{\mu\nu}\left( P(\phi)R+Q(\phi)\right)+P(\phi)R_{\mu\nu}+g_{\mu\nu}\Box P(\phi)-\nabla_{\mu}\nabla_{\nu}P(\phi)= \kappa^2T^{(m)}_{\mu\nu}\
\label{In.4}
\ee
In addition, the action Equation~(\ref{In.3}) gives an extra equation for the scalar field $\phi$, obtained directly from the action by varying it with respect to $\phi$:
\be
P'(\phi)R+Q'(\phi)=0\
\label{In.5}
\ee
where the primes denote derivatives with respect to $\phi$. This equation can be solved such that  the scalar field is a function of the Ricci scalar $R$, $\phi=\phi(R)$, and then, by replacing this result in the action Equation~(\ref{In.3}), the action Equation~(\ref{In.1}) is recovered:
\be
F(R)=P\left(\phi(R)\right)R+Q\left(\phi(R)\right)\
\label{In.6}
\ee
With the aim of proposing a realistic alternative to GR, a possible modification consists of adding a function of the scalar curvature, $f(R)$
to the Einstein--Hilbert (EH) Lagrangian. Therefore the gravitational sector of the action Equation~\eqref{In.1} becomes:
\begin{equation}
S_G=\int \text{d}^4x \sqrt{\vert g \vert}\left(R+f(R)\right)
\label{S_g}
\end{equation}
By performing variations with respect to the metric, the modified Einstein Equation~\eqref{In.2} turns out to be:
\begin{equation}
(1+f_R)R_{\mu\nu}-\frac{1}{2}(R+f(R))g_{\mu\nu}+{\cal D}_{\mu\nu}f_R\,=\,8\pi G \,T_{\mu\nu}
\label{fieldtensorialequation}
\end{equation}
where $f_R\,\equiv\,\text{d}f(R)/\text{d}R$ and
${\cal D}_{\mu\nu}\equiv  g_{\mu\nu}\square- \nabla_{\mu}\nabla_{\nu}$ with $\square\,\equiv\,\nabla_{\alpha}\nabla^{\alpha}$ and $\nabla$ is the usual
covariant derivative.

These equations may be written \textit{\`{a} la Einstein} by isolating on the l.h.s. the Einstein tensor and the $f(R)$ contribution on the r.h.s. as follows
\cite{Dombriz_thesis}:
\begin{eqnarray}
R_{\mu\nu}-\frac{1}{2}R g_{\mu\nu}
\,=\,\frac{1}{(1+f_R)}\left[\vphantom{\frac{1}{2}}\,8\pi G T_{\mu\nu}
-
{\cal D}_{\mu\nu}f_R+\frac{1}{2}\left(f(R)-Rf_R\right)g_{\mu\nu}\right]
\end{eqnarray}
We can also find the expression for the scalar curvature by contracting Equation~\eqref{fieldtensorialequation} with $g^{\mu\nu}$, which~gives:
\begin{eqnarray}
(f_R-1)R-2f(R)+3\square f_R\,=8\pi G \,T
\label{}
\end{eqnarray}
Note that, unlike GR where $R$ and $T$ are related algebraically, for a general $f(R)$ those two quantities are dynamically related. Thus, in principle, non-null curvature solutions are possible in vacuum for certain $f(R)$ models.

In addition, one may study the possible corrections at local scales of the Newtonian law and the violations of the equivalence principle by using again an auxiliary scalar field $A$ and applying a conformal transformation. Then, we may rewrite the action Equation~\eqref{S_g} as:
\be
S_G=\int \text{d}^4 x
\sqrt{-g}\left[(1+f'(A))(R-A)+A+f(A)\right]\
\label{In.7}
\ee
Then, by varying this action with respect to A, it gets $R=A$, and the action Equation~\eqref{S_g} is recovered. By the conformal transformation $\tilde{g}_{\mu\nu}=\e^{\sigma}g_{\mu\nu}$, where $\sigma=-\ln(1+f'(A))$, the action Equation~\eqref{In.7} is rewritten in the so-called Einstein frame:
\be
S_{G}^E=\int \text{d}^4 x
\sqrt{-\tilde{g}}\left[\tilde{R}-\frac{3}{2}\partial^{\mu}\sigma\partial_{\mu}\sigma -V(\sigma)\right]
\label{In.8}
\ee
where the tilde represents that the scalar curvature is evaluated with respect to the transformed metric $\tilde{g}_{\mu\nu}$, and the potential $V(\sigma)$ is given by:
\be
V(\sigma)=\frac{A(\sigma)}{1+f'(A(\sigma))}-\frac{A+f(A(\sigma))}{(1+f'(A(\sigma)))^2}\
\label{In.9}
\ee
Here, one has to solve the equation $\sigma=-\ln(1+f'(A))$ in order to get $A=A(\sigma)$. This conformal transformation affects the matter sector in the gravitational action by a term proportional to $\e^{2\sigma}$, which can imply large corrections to the Newtonian law unless the potential Equation~\eqref{In.9} is constructed in order to avoid them at local scales by the so-called chameleon mechanism \cite{khoury,Khoury:2003rn}, which implies that the mass of the scalar field $\sigma$ should be enough large at local scales in order to avoid large corrections to the Newtonian law \cite{Nojiri:2007as}:
\be
m_{\sigma}^2=\frac{1}{2}\frac{\text{d}^2 V
(\sigma)}{\text{d}\sigma^2}=\frac{1}{2}\left\{\frac{A(\sigma)}{1+f'(A(\sigma))}-4\frac{A+f(A(\sigma))}{(1+f'(A(\sigma)))^2}+\frac{1}{f''(A(\sigma))}\right\}\
\label{In.10}
\ee
Hence, if the mass is much larger at local scales than the curvature, the possible corrections to the Newtonian law will be negligible. In this sense, several models that accomplish this rule have been proposed (see~\cite{Hu&Sawicki2007,Viable,Viable1,Viable2,Viable3}), which  are also capable to reproduce the cosmological history.

\section{Static and Spherically Symmetric Black Holes in $f(R)$ Gravities}
\label{Static_BH}

Let us study in this section the $f(R)$ solutions under staticity and spherical symmetry assumptions.
The most general $D\geq 4$ dimensional metric external to static and spherically symmetric---therefore non-rotating---configurations can be written as (see \cite{ortin}):
\begin{eqnarray}
\text{d}s^2\,=\,e^{-2\Phi(r)} A(r)\text{d}t^2-A^{-1}(r)\text{d}
r^2-r^2\text{d}\Omega_{D-2}^2
\label{metric_D_v1}
\end{eqnarray}
or alternatively:
\begin{eqnarray}
\text{d}s^2\,=\,\lambda(r)\text{d}t^2-\mu^{-1}(r)\text{d}r^{2}-r^2\text{d}\Omega_{D-2}^2
\label{metric_D_v2}
\end{eqnarray}
where $\text{d}\Omega_{D-2}^2$ is the metric on the $S^{D-2}$ sphere and identification
$\lambda(r)=e^{-2\Phi(r)}A(r)$ and $\mu(r)=A(r)$ can be straightforwardly established if required.
%
%
%
Since the metric is static, the scalar curvature $R$ in $D$ dimensions depends only on  $r$ and
it is given, for the metric parametrization Equation~\eqref{metric_D_v1}, by:
\begin{eqnarray}
R(r)\,  &=& \,\frac{1}{r^2}[D^2-5 D+6+r A'(r) \left(-2 D+3 r
\Phi '(r)+4\right)\nonumber\\&-&r^2 A''(r)-A(r) \left(D^2-5 D+2
r^2 \Phi '(r)^2-2 (D-2) r \Phi '(r)-2 r^2 \Phi ''(r)+6\right)]
\label{Dcurv}
\end{eqnarray}
%
where the prime denotes derivative with respect to $r$ coordinate. 

\subsection{Spherically Symmetric and Static Constant Curvature Solutions: Generalities} 

At this stage it is interesting to ask about which are the most
general static and spherically symmetric metrics with constant
scalar curvature $R_{0}$. This curvature can be found solving the
Equation \eqref{Dcurv} \linebreak$R=R_0$. Thus, it is straightforward to check that, for a constant $\Phi(r)=\Phi_{0}$,  the
general solution is~\cite{Dombriz_PRD_BH,Dombriz_PRD_BH1}:
\begin{eqnarray}
A(r)\,=\,1+\frac{a_{1}}{r^{D-3}}+\frac{a_{2}}{r^{D-2}}-\frac{R_0}{D(D-1)}r^2
\label{A_solution_R_constant_Dobado_procedure}
\end{eqnarray}
with $a_{1}$ and $a_{2}$ being arbitrary integration constants. In fact,
for the particular case $D=4$, $R_{0}=0$ and $\Phi_{0}=0$, the
metric can be written exclusively in terms of the function:
\begin{eqnarray}
A(r)\,=\,1+\frac{a_{1}}{r}+\frac{a_{2}}{r^{2}}
\label{RN_solution_R_constant_Dobado_procedure}
\end{eqnarray}
By establishing the identifications $a_{1}\,=\,-2G_{N}M$ and
$a_{2}\,=\,Q^2$, this solution corresponds to a \linebreak Reissner--Nordstr\"{o}m solution, \emph{i.e}., a charged massive BH solution with mass $M$
and charge $Q$. Further comments about this result will be made
in Section \ref{fR_constant_solutions}.

\subsection{Spherically Symmetric and Static Constant Curvature Solutions 
}
\label{fR_constant_solutions}

Let us study here which are the vacuum constant curvature solutions for $f(R)$ theories provided that the metric under
consideration is of the form Equation~\eqref{metric_D_v1}---or alternatively Equation~\eqref{metric_D_v2}.
By inserting the metric Equation~\eqref{metric_D_v1} into the general
$f(R)$ gravitational action $S_g$ in Equation~(\ref{S_g}), and making variations with
respect to the metric functions, $A(r)$ and $\Phi(r)$, the equations of motion become:
\begin{eqnarray}
(2-D ) (1+f'(R)) \Phi'(r)-r\left[f'''(R)
R'(r)^2+f''(R)(\Phi'(r)R'(r)+ R''(r))\right]\,=\,0
\label{eqn_A}
\end{eqnarray}
and
\begin{eqnarray}
&& 2 r A(r) f'''(R) R'(r)^2+ f''(R)[2 D A(r)R'(r)
- 4 A(r) R'(r) + 2 r A(r) R''(r)+A'(r) r R'(r)] \nonumber\\
&&- r(R+f(R))
+g'(R)[-2 r A(r) \Phi'(r)^{2} + 2 D A(r) \Phi'(r) -4 A(r)
\Phi'(r) - r A''(r)  + 2 r A(r) \Phi''(r)\nonumber\\
&& + A'(r)(2 - D  + 3 r \Phi'(r)) ] \,=\,0
\label{eqn_phi}
\end{eqnarray}
or alternatively in by using the metric parametrization given by
Equation~\eqref{metric_D_v2} the equations of motion now are expressed as:

\begin{eqnarray}
&&\lambda (r) (1+f'(R)) \left\{2 \mu (r) \left[(D -2) \lambda '(r)
+r \lambda ''(r)\right]+r \lambda '(r) \mu '(r)\right\}\nonumber \\
&&-2 \lambda (r)^2 \left\{2 \mu(r)[(D -2) R'(r) f''(R)
+r f^{(3)}(R) R'(r)^2+r R''(r) f''(R)]+r R'(r) \mu '(r) f''(R)\right\}\nonumber\\
&&-r \mu (r) \lambda '(r)^2 (1+f'(R))+2 r \lambda (r)^2(R+f(R))=\,0
\label{eqn_lambda}
\end{eqnarray}
and
\begin{eqnarray}
&-&\lambda (r) \mu '(r) \left[2 (D -2) \lambda (r)
+r \lambda '(r)\right] (1+f'(R))   -2 r \lambda (r)^2 (R+f(R))\nonumber \\
&+&\mu (r) \Big\{2 \lambda (r) R'(r)
\left[2 (D -2) \lambda(r)+r\lambda '(r)\right] f''(R)
+r(1+f'(R))(\lambda '(r)^2-2 \lambda (r) \lambda ''(r))\Big\}\,=\,0
\nonumber\\
&&
\label{eqn_mu}
\end{eqnarray}
where $f'$, $f''$ and $f'''$ denote derivatives of $f(R)$ with
respect to the curvature $R$. In the following calculations in this section, we use the set of Equations~\eqref{eqn_A} and \eqref{eqn_phi}.
In order to simplify the difficulty of the above equations, let us first consider here
the case of constant scalar curvature
$R=R_0$ solutions. Then the equations of motion Equations~\eqref{eqn_A} and \eqref{eqn_phi} reduce to:
\begin{equation}
(2-D)\,(1+f'(R))\Phi'(r)=0
\label{eq_motion_phi}
\end{equation}
and
\begin{eqnarray}
&&R+f(R)
+(1+f'(R))\left[A''(r)+(D-2)\frac{A'(r)}{r}
-(2D-4)\frac{A(r)\Phi'(r)}{r}\right. \nonumber\\
&&\left. -3A'(r)\Phi'(r)+2A(r)\Phi'^2(r)-2A(r)\Phi''(r)\right]\,=\,0
\label{eq_motion_A}
\end{eqnarray}
As commented in Section \ref{Generalities_fR}, the constant curvature solutions of
$f(R)$ theories in vacuum satisfy:
\begin{eqnarray}
R_0=\frac{D\,f(R_0)}{2(1+f'(R_0))-D}
\label{const_R0}
\end{eqnarray}
provided that $2(1+f'(R_0))\neq D$. Thus from Equation~\eqref{eq_motion_phi}  one concludes (excluding pathologic cases) that $\Phi'(r)=0$ and then Equation \eqref{eq_motion_A} becomes:
%
\begin{equation}
A''(r)+(D-2)\frac{A'(r)}{r}=-\frac{2}{D}R_0
\label{A_eq_R_constant}
\end{equation}
This is a $f(R)$-independent, linear second order inhomogeneous differential
equation. This statement must be understood in the sense that Equation \eqref{A_eq_R_constant} does not involve explicitly any $f(R)$ parameter. Nonetheless, it is obvious from Equation~\eqref{const_R0} that the value of $R_0$ does depend upon the $f(R)$ model under study. Therefore the metric solution $A(r)$ will also inherit information about the model under consideration. Then, Equation (\ref{A_eq_R_constant}) can be easily integrated to give the general solution:
\begin{equation}
A(r)\,=\,C_1\,+\,\frac{C_2}{r^{D-3}}-\frac{R_0}{D(D-1)}r^2\,
\label{A_solution_R_constant}
\end{equation}
which depends on two arbitrary constants $C_1$ and $C_2$. However, two important remarks need to be done at this stage:
\begin{itemize}

\item Firstly, by comparison with Equation~\eqref{A_solution_R_constant_Dobado_procedure}, one can see that the term with the power $r^{2-D}$ is absent. This fact will be studied in Section \ref{fR_EM}.

\item Secondly, this solution has no constant curvature in
the general case since, as we found above, the constant curvature requirement
demands $C_{1}=1$. This issue  just requires a constant fixing (or equivalently a time reparametrization) and does not affect the solution.
\end{itemize}
Then, for negative $R_0$, the found solution in Equation~\eqref{A_solution_R_constant} with $C_1=1$
is basically the $D$-dimensional generalization obtained by Witten \cite{Witten}
of the BH in AdS space-time solution considered by Hawking and Page \cite{Hawking&Page}. With
the natural choice $\Phi_0=0$ the solution can be written as:
\begin{equation}
A(r)=1-\frac{R_{S}^{D-3}}{r^{D-3}}+\frac{r^2}{l^2}
\label{A_Witten}
\end{equation}
where
\begin{equation}
R_S^{D-3}=\frac{16\pi G_D M}{(D-2)\mu_{D-2}} \;\;\;,\;\;\;
\mu_{D-2}=\frac{2\pi^{\frac{D-1}{2}}}{\Gamma\left(\frac{D-1}{2}\right)}
\label{BHmass}
\end{equation}
%
where $\mu_{D-2}$ is the area of the $D-2$ sphere, $l^2\equiv-D(D-1)/R_0 $ is the asymptotic
AdS space scale squared and $M$ is the mass parameter usually found in the literature.

Thus, the main conclusion of this analysis \cite{Dombriz_PRD_BH,Dombriz_PRD_BH1} is that the only static and
spherically symmetric vacuum solutions with constant (negative) curvature of any
$f(R)$ gravity is just the Hawking--Page BH in AdS space.
However this kind of solution is not the most general static and
spherically symmetric metric with constant curvature, as can be seen by
comparison with the solutions found in Equation~\eqref{A_solution_R_constant_Dobado_procedure}.
Therefore there exist constant curvature
BH solutions that cannot be obtained as vacuum solutions of any $f(R)$
model. In particular, the term with $r^{2-D}$ power cannot be mimicked by any $f(R)$ model.
As we shall show immediately, 
the most general solution as given by Equation~\eqref{A_solution_R_constant_Dobado_procedure}
can be described as a charged BH solution in a $D=4$ $f(R)$-Maxwell theory.

\subsection{$f(R)$ Solutions Combined with Electromagnetism}
\label{fR_EM}

Let us briefly mention here for completeness with the previous results the case of charged BH in $f(R)$ theories that we shall
study in detail in Section \ref{Section_KN_BH}. We shall limit ourselves to  the $D=4$ case, 
%
since otherwise ($D\neq4$) 
the $r^{2-D}$ term appearing in Equation~\eqref{A_solution_R_constant_Dobado_procedure} cannot be originated by the Maxwell action to be presented below. 
Let us consider here a $f(R)$ gravitational Lagrangian supplemented with the usual Maxwell action for standard Electromagnetism. Therefore, the considered action represents a 
generalization of the Einstein--Maxwell action:
\begin{equation}
S_g=\frac{1}{16 \pi G_N}\int \text{d}^{4}x\sqrt{\mid g\mid}\,(R+f(R)-F_{\mu\nu}F^{\mu\nu})
\label{Action_RN}
\end{equation}�
where as usual $F_{\mu\nu}=\partial_\mu A_\nu-\partial_\nu A_\mu$.
Considering an electromagnetic potential of the form \linebreak$A_\mu=(V(r),\vec 0)$ and the  static
spherically symmetric metric Equation~(\ref{metric_D_v2}), one finds that the solution with
constant curvature $R_0$ reads \cite{Dombriz_PRD_BH,Dombriz_PRD_BH1}:
\begin{eqnarray}
V(r)=\frac{Q}{r}\;\;\;;\;\;\;
\lambda(r)=\mu(r)=1-\frac{2G_{N}M}{r}+\frac{1}{1+f'(R_0)}\frac{Q^2}{r^2}-\frac{R_0}{12}r^2
\label{RN_simple_solution}
\end{eqnarray}
Notice that unlike the EH case, the contribution of the BH charge
to the metric tensor is corrected by a $(1+f'(R_0))^{-1}$ factor. This fact is of particular importance in order
to constrain $f(R)$ models as astrophysically viable: the sign of the factor $1+f'(R_0)$ needs
to be positive in order to guarantee that the solution Equation~\eqref{RN_simple_solution} keeps the same sign as in the
Reissner--Nordstr\"{o}m  solution. Otherwise, for a charge $Q$ the third term in the solution Equation~\eqref{RN_simple_solution} will be negative and therefore this fact becomes experimentally observable (affecting for instance to the geodesic movements around a charged BH). In fact, the requirement $1+f'(R_0)>0$ has been usually accepted \cite{silvestri} as one of the gravitational viability conditions for $f(R)$ models and is also related, as will be shown here, with thermodynamical viability.

\subsection{Perturbations Around Schwarzschild--(anti)-de Sitter Solutions}
\label{Perturbations}		

In the previous section we have revised the static spherically symmetric
solutions with constant curvature. In EH theory, the constant curvature ansatz  Equation~\eqref{const_R0} provides the most general
static and spherical symmetric solution straightforwardly. Without entering into details, this is the well-known Birkhoff--Jebsen theorem result \cite{Birkhoff,Birkhoff2}.
However, it is not transparent for this to also be the case in $f(R)$ theories, \emph{i.e}., that the most general static and spherically symmetric
solution for $f(R)$ theories possesses constant curvature. Furthermore, it seems that a generic solution would behave differently in the Jordan and Einstein frames, both related by a conformal transformation, at least in the perturbative analysis (see~\cite{Capozziello:2011wg,Capozziello:2011wg1}).

In this section we revise the basic rudiments of the perturbative analysis of the problem originally studied in \cite{Dombriz_PRD_BH,Dombriz_PRD_BH1}, only in the Jordan frame. In that analysis, it was assumed that the modified gravitational Lagrangian is a small perturbation around the EH Lagrangian.
Therefore, the arbitrary $f(R)$ function is of the form:
\begin{eqnarray}
f(R)\,=-(D-2)\Lambda_{D}+\alpha g(R)
\label{expansion_en_alpha_fR}
\end{eqnarray}
where $\alpha\ll 1$ is a dimensionless parameter and $g(R)$ is assumed
to be analytic in $\alpha$. This last assumption is based on the fact that
the scalar curvature $R$ is also $\alpha$-dependent as will be seen in expressions Equation~\eqref{expansion_en_alpha_lambda&mu}. Therefore $g(R)$ may also be expanded in powers of $\alpha$. By assumption and for the sake of simplicity,  $g(R)$ is assumed analytic in the expansion in powers of $\alpha$.

%
%

Now, assuming that the $\lambda(r)$ and
$\mu(r)$ functions appearing in the metric Equation~\eqref{metric_D_v2} are also
analytical in $\alpha$, they can
be written as follows:
\begin{eqnarray}
\lambda(r)\,=\,\lambda_{0}(r)+\sum_{i=1}^{\infty}\alpha^{i}\lambda_{i}(r)\;\;\;,\;\;\;
\mu(r)\,=\,\mu_{0}(r)+\sum_{i=1}^{\infty}\alpha^{i}\mu_{i}(r)
\label{expansion_en_alpha_lambda&mu}
\end{eqnarray}
where $\{\lambda_{0}(r),\,\mu_{0}(r)\}$ are the unperturbed solutions for the
EH action with cosmological constant, \emph{i.e}., S(A)dS solution, given by:
\begin{eqnarray}
\mu_{0}(r)\,=\,1+\frac{C_1}{r^{D-3}}-\frac{\Lambda_{D}}{(D-1)}r^2\;\;\;,\;\;\;
\lambda_{0}(r)\,=\,-C_{2}(D-2)(D-1)\,\mu_{0}(r)
\label{mu0_lambda0}
\end{eqnarray}
which are the standard BH solutions in a $D$-dimensional Schwarzschild--(A)dS
space-time. The factor $C_2$ can be chosen by
performing a coordinate $t$ reparametrization so that both
functions in Equation~\eqref{mu0_lambda0} can be identified. For the moment, the
background solutions are kept as given in Equation~\eqref{mu0_lambda0} since the
possibility of getting $\lambda(r)=\mu(r)$  in the
perturbative expansion will be discussed later on.

The perturbative procedure is as follows: by inserting  Equations~\eqref{expansion_en_alpha_fR} and
\eqref{expansion_en_alpha_lambda&mu} into the general Einstein modified Equations
\eqref{eqn_lambda} and \eqref{eqn_mu}, equations for each order in $\alpha$ parameter powers are obtained.
Please refer to \cite{Dombriz_PRD_BH,Dombriz_PRD_BH1} for details.

Notice that from the obtained results up to second order in $\alpha$,
the corresponding metric has constant scalar curvature for any
value of the integration constants. As a matter of fact,
this metric is nothing but the standard Schwarzschild--(A)dS geometry,
and can be easily recast in the usual form by making a trivial reparametrization in the time coordinate.

Therefore, this analysis proved that at least up to second order, the only static,
spherically symmetric solutions which are analytical
in $\alpha$ are the standard Schwarzschild--AdS space-times irrespectively of the $f(R)$ model under consideration. This result is valid
provided that both the solutions and the gravitational Lagrangian are analytic
in the $\alpha$ parameter and that $g(R)$ corresponds to a small deviation from the usual EH gravitational Lagrangian.
Further orders in $\alpha^{3,4,...}$ can be obtained by inserting
previous results in the equations of order $3,4,...$ and so on. This way one may get the solutions $\{\lambda_{3,4,...}(r),\mu_{3,4,...}(r) \}$ with the added difficulty that the corresponding equations become more and more complicated.


To conclude Sections \ref{Static_BH} and \ref{Perturbations}, we can summarize by saying that
in the context of $f(R)$ gravities the only  spherically symmetric and
static solutions of negative constant curvature are the standard BH in
AdS space. The same result applies in the general case (without imposing
constant curvature) in perturbation theory up to second order. However,
the possibility of having static and spherically symmetric solutions
with non-constant curvature cannot be excluded in the case of
$f(R)$ functions, which are not analytical in $\alpha$ parameter.

\section{Kerr--Newman Black Holes in $f(R)$ Theories}
\label{Section_KN_BH}

In this section, we shall focus our attention in obtaining
constant curvature $R_0$
vacuum solutions for fields generated by massive charged
objects in the frame of $f(R)$ gravity theories. Hence, the appropriate action to derive the
field equations will be now Equation~\eqref{Action_RN} and $D=4$ will be assumed throughout this section. The obtained equations (in $G_N=c=\hbar=k_B=1$ units) are:
%
%
\begin{eqnarray}
 R_{\mu\nu}\,(1+f'(R_0))-\frac{1}{2}\,g_{\mu\nu}\,(R_0+f(R_0))
-2\left( F_{\mu\alpha}F_{\,\,\,\nu}^\alpha-\frac{1}{4}g_{\mu\nu}F_{\alpha\beta}F^{\alpha\beta}\right) \,=\,0\,
\label{ec_tensorial}
\end{eqnarray}

In the studied case $D=4$, the trace of the previous equation leads again to expression Equation~\eqref{const_R0}
due to the conformal character of the Faraday tensor, \emph{i.e}., $F^{\mu}\,_{\mu}=0$.

For these $f(R)$ field equations the axisymmetric, stationary and constant curvature $R_0$ solution that describes a BH with mass, electric charge and angular momentum is analogous to the one found by Carter and published for the first time in 1973 \cite{Carter}. In Boyer--Lindquist coordinates, the metric
takes the form \cite{Jimeno_BH,Cembranos:2012ji}:
\begin{eqnarray}
\text{d}s^2\,=\,\frac{\rho^2}{\Delta_r}\,\text{d}r^2+\frac{\rho^2}{\Delta_\theta}\,\text{d}\theta^2+\frac{\Delta_\theta\,\sin^2{\theta}}{\rho^2}\,\left[a\,\frac{\text{d}t}{\Xi}-\left(r^2+a^2\right)\,\frac{\text{d}\phi}{\Xi}\right]^2-\frac{\Delta_r}{\rho^2}\,\left(\frac{\text{d}t}{\Xi}-a\sin^2{\theta}\frac{\text{d}\phi}{\Xi}\right)^2\,
\label{metrica}
\end{eqnarray}

with
\begin{eqnarray}
\Delta_r\,&:=&\,\left(r^2+a^2\right)\left(1-\frac{R_0}{12}\,r^2\right)-2Mr+\frac{Q^2}{\left( 1+f'(R_0 )\right)}\,\nonumber
\end{eqnarray}
\begin{eqnarray}
\rho^2\,:=\,r^2+a^2\cos^2{\theta}\;\;;\;\;
\Delta_\theta\,:=\,1+\frac{R_0}{12}\,a^2\cos^2{\theta}\;\;;\;\;
\Xi\,:=\,1+\frac{R_0}{12}\,a^2\,
\label{definiciones}
\end{eqnarray}
where $M$, $a$ and $Q$ denote the mass, spin and electric charge parameters respectively. Notice that, unlike in the GR case, the contribution of the charge of the BH to the metric is corrected by a $\left(1+f'(R_0) \right)^{-1/2}$ factor. This feature was already obtained for Reissner--Nordstr\"{o}m
BH presented in expression Equation~\eqref{RN_simple_solution} and originally in \cite{Dombriz_PRD_BH,Dombriz_PRD_BH1}.
On the other hand, the required potential vector and electromagnetic field tensor in Equation \eqref{ec_tensorial} solutions
for metric Equation~\eqref{metrica} are respectively:
\begin{eqnarray}
A&=&-\frac{Q\,r}{\rho^2}\left(\frac{\text{d}t}{\Xi}-a\sin^2{\theta}\frac{\text{d}\phi}{\Xi}\right),\nonumber\\[0.15cm]
F&=&-\frac{Q\,(r^2-a^2\cos^2{\theta})}{\rho^4}\left(\frac{\text{d}t}{\Xi}-a\sin^2{\theta}\frac{\text{d}\phi}{\Xi}\right)\wedge\text{d}r
-\frac{2\,Q\,r\,a\cos{\theta}\sin{\theta}}{\rho^4}\,\text{d}\theta\wedge\left[a\,\frac{\text{d}t}{\Xi}-(r^2+a^2)\,\frac{\text{d}\,\phi}{\Xi}\right]\,\nonumber\\
&&
\label{EM}
\end{eqnarray}

To lighten notation, from now on let us use  the notation $\bar{Q}^2\equiv Q^2\, /\,\left(1+f'(R_0) \right)$ as a normalized electric charge parameter of the BH. 

\subsection{Event Horizons}

\begin{figure}[h]
	\centering
		\includegraphics[width=0.38\textwidth]{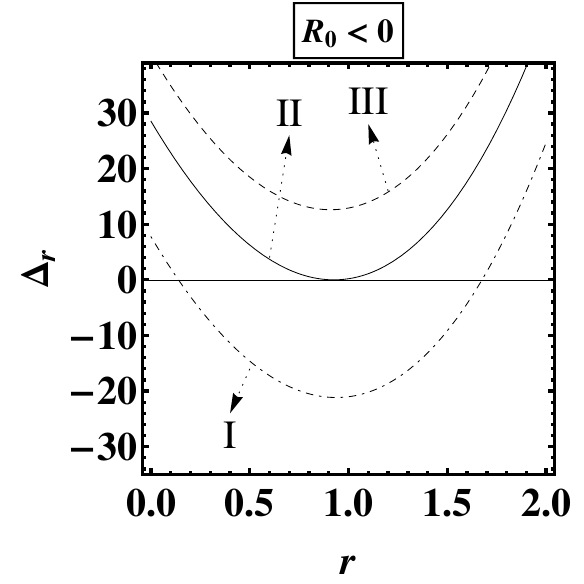}\,\,\,\,\,\,\,\,\,\,\,\,\,\,
		\includegraphics[width=0.38\textwidth]{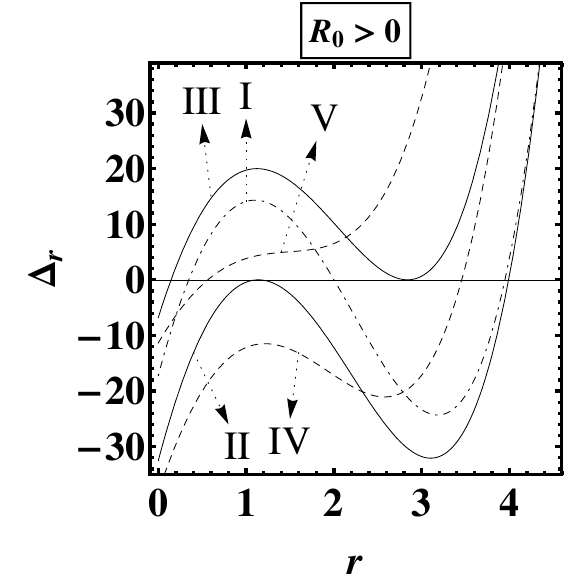}
		\caption{
		Graphics showing horizons positions as solutions of the equation $\Delta_r=0$ taken from \cite{Jimeno_BH,Cembranos:2012ji}. On the left panel ($R_0<0$) the presented cases are: $h>0$ ({\bf I}), BH with well-defined horizons, dashed with dots), $h=0$ ({\bf II}), {\it extremal} BH, continuous line) and $h<0$ ({\bf III}), {\it naked singularity}, dashed). On the right panel ($R_0>0$) the represented cases are: $h<0$ ({\bf I}, BH with well-defined horizons, dashed with dots), $h=0$ ({\bf II}, $extremal$ BH and {\bf III}, $extremal$ $marginal$ BH, continuous line), and $h>0$ ({\bf IV}, {\it naked singularity} and {\bf V}, {\it naked marginal singularity}, dashed).}
	\label{fig:graficashor}
\end{figure}

Let us now briefly study the horizon structure of these BH: according to the event horizon definition, \emph{i.e}., $g^{rr}=0$, these hypersurfaces are found as the roots of the equation $\Delta_r=0$, \emph{i.e}.,
\begin{eqnarray}
r^4+\left(a^2-\frac{12}{R_0}\right)\,r^2+\frac{24\,M}{R_0}\,r-\frac{12}{R_0}\,\left(a^2+\bar{Q}^2\right)\,=\,0\,
\label{poli_hori}
\end{eqnarray}
This is an algebraic fourth order equation that, following \cite{Jimeno_BH,Cembranos:2012ji}, can be rewritten as:
\begin{eqnarray}
(r-r_{-})(r-r_{int})(r-r_{ext})(r-r_{cosm})\,=\,0\,
\label{cuartic}
\end{eqnarray}
where
\begin{itemize}
\item $r_-$ is always a negative solution with no physical meaning,
\item $r_{int}$ and $r_{ext}$ are the interior and exterior horizon respectively, and
\item $r_{cosm}$ represents---provided that it arises---the cosmological event horizon for observers between $r_{ext}$ and $r_{cosm}$.
\end{itemize}
The latter divides the accessible region from the hidden region for an exterior observer. The Equation~\eqref{cuartic}
can be solved by using L. Ferrari's method  for quartic equations. The existence 
of real solutions for this equation is given by a factor $h$, which we shall denote as {\it horizon parameter}:
\begin{eqnarray}
h\equiv&&\left[\frac{4}{R_0}\left(1-\frac{R_0}{12}\,a^2\right)^2-4\,\left(a^2+\bar{Q}^2\right)\right]^3\nonumber\\
&&+\frac{4}{R_0}\left\{\left(1-\frac{R_0}{12}\,a^2\right)\left[\frac{4}{R_0}\left(1-\frac{R_0}{12}\,a^2\right)^2+12\,\left(a^2+\bar{Q}^2\right)\right] -18\,M^2\right\}^2\,.\nonumber\\
\label{ecu_D}
\end{eqnarray}
In Figure \ref{fig:graficashor} we show the zeros of the fourth order polynomial $\Delta_r$ that lead to different astrophysical configurations, depending on the values and signs of $R_0$ and $h$.%

Let us stress at this stage that, from a certain positive value of the curvature $R_{0}^{\,crit}$ onward, the $h$ factor goes to zero for two values of $a$. They can be understood as follows:
\begin{itemize}
\item Upper spin bound, $a=a_{max}$ for which the BH turns $extremal$---the interior and exterior horizons have merged into a single horizon with a null surface gravity. This is the usual configuration  for the BH to become extremal.
\item Lower spin bound, $a=a_{min}$, below which the BH turns into a {\it marginal extremal} BH. This value can be understood as the cosmological limit for which a BH preserves its exterior horizon without being ``torn apart" due to the relative recession speed between two radially separated points induced by the cosmic expansion \cite{Jimeno_BH,Cembranos:2012ji}.
\end{itemize}
%

%

\begin{figure}[h]
	\centering
		\includegraphics[width=0.38\textwidth]{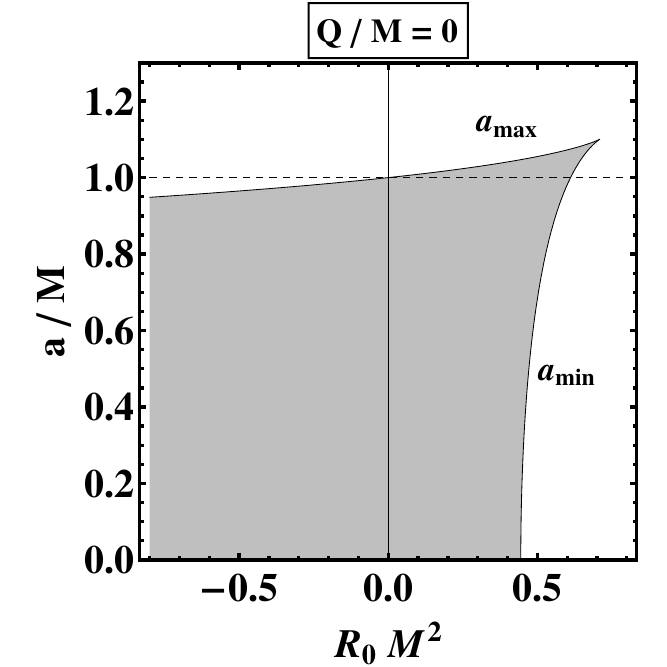}\,\,\,\,\,\,\,\,\,\,\,\,\,\,
		\includegraphics[width=0.38\textwidth]{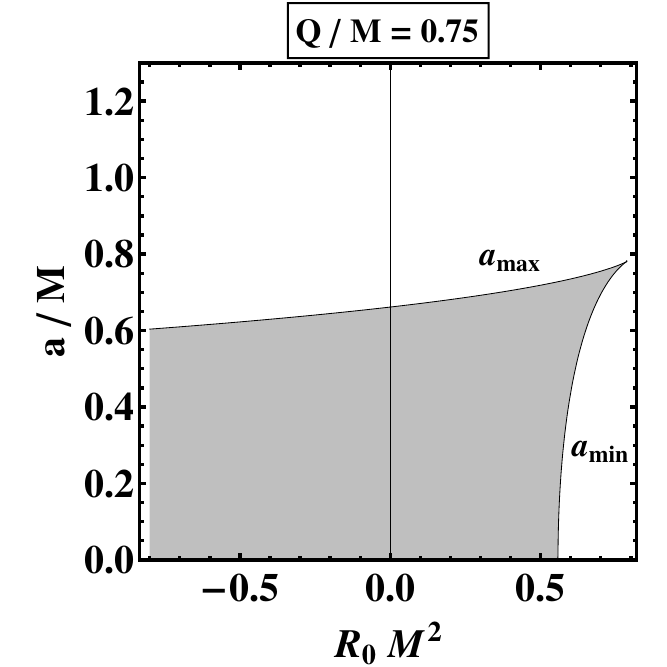}
		\caption{The shaded regions, delimited by the upper $a_{max}$ and lower $a_{min}$ curves, represent the values of $a/M$
		for which the existence of BH is possible once $R_0\,M^2$ value is fixed. Panels show for $\bar{Q}\,/\,M=0$ (left) and $\bar{Q}\,/\,M=0.75$ (right) on the left and right panels respectively. Note that $R_0$ has dimensions of [length]$^{-2}$ when normalizing. Original plots at~\cite{Jimeno_BH,Cembranos:2012ji}.}
	\label{fig:rangoespin}
\end{figure}

In Figure \ref{fig:rangoespin} we revisit the original plot shown in \cite{Jimeno_BH,Cembranos:2012ji}, where for certain values of the electric charge parameter $\bar{Q}$ the allowed ranges of the spin $a$ parameter values are presented.

\section{Black Holes Thermodynamics in $f(R)$ Theories}
\label{Thermodynamics}

As mentioned in the Introduction, the study of BH thermodynamics has drawn a lot of attention when studying these configurations in alternative theories of gravity. We shall divide this section in two main parts: the study of thermodynamics in AdS case and then the study of KN case as the natural generalization of the former. This approach will enable to introduce first the rudiments of the Euclidean method so the reader can familiarize with it, and then extend the formulation for more general objects such as the KN BH configurations. Finally, we will provide some examples by considering $f(R)$ models that illustrate the procedure for both configurations.
%
In the AdS case, the dimension of the space-time will be arbitrary and then fixed to several values in order to illustrate the found results. On the contrary, for the KN study, dimension will be fixed from the very beginning to $D=4$ for the sake of simplicity.


\subsection{BH Thermodynamics for AdS Configuration}
\label{AdS_Subsection}

In order to study the different thermodynamic quantities for
the $f(R)$ BH in AdS space, the temperature is usually considered as the
basic 
quantity from which the rest of the thermodynamic quantities are derived. In principle,
there are two different approaches of introducing this quantity for the
AdS solutions. 
Firstly, one of the possible definitions is the one
provided by the Euclidean quantum gravity \cite{HGG,HGG1,HGG2,HGG3} that was revised in \cite{Dombriz_PRD_BH,Dombriz_PRD_BH1} yielding:

\begin{equation}
T_E=\frac{1}{4\pi} e^{-\Phi(r_H)}A'(r_H)
\end{equation}

Another possible definition of temperature was firstly proposed in
\cite{Hawking1974} stating that temperature can be given in terms of
the horizon gravity $\mathcal{K}$ as
$T_{\mathcal{K}}\equiv\frac{\mathcal{K}}{4\pi}$.
Then, by reminding the metric definition Equation~\eqref{metric_D_v1}, it is straightforward to obtain that
$T_{\mathcal{K}}= T_E$.
In conclusion, both definitions lead to the same result for this kind of
solution. Notice also that in any case the temperature depends
only on the behavior of the metric near the horizon and it is
independent of the underlying gravitational action. This last statement can be understood
by remarking that either different actions or
different field equations giving rise to the same solutions of the form Equation~\eqref{metric_D_v1} will
also present the same temperature. This is not the case for other thermodynamic
quantities as will be seen later. Taking into account the results
in previous sections, the usual analyses for
$f(R)$ theories \cite{Dombriz_PRD_BH,Dombriz_PRD_BH1} have focused on
the
constant curvature AdS BH solutions with $\Phi=0$ as a natural
choice, and $A(r)$ as given by expression Equation~\eqref{A_Witten}.
Then, both definitions of temperature lead to:
\begin{equation}
\beta\equiv \frac{1}{T}\,=\,\frac{4 \pi l^2 r_H}{(D-1)r_H^2+(D-3)l^2}
\label{T_AdS}
\end{equation}
Notice that for a given dimension $D$, the temperature is a function of $r_H$ only, \emph{i.e}., it
depends only on the BH size. In the limit $r_H$ going to zero the
temperature diverges as $T \sim 1/r_H$ and for $r_H$ going to
infinite $T$ grows linearly with $r_H$. Consequently $T$ has a
minimum 
%
%
corresponding to a temperature:
\begin{equation}
T_0=\frac{\sqrt{(D-1)(D-3)}}{2 \pi l}
\end{equation}
The existence of this minimum was established in \cite{Hawking&Page}
for $D=4$ by Hawking and Page.
More recently, Witten extended this result to higher dimensions
\cite{Witten}. This minimum will be relevant in order to set the regions
with different thermodynamic behaviors and stability properties.

Explicit solutions for  $A(r_H)=0$ in Equation~\eqref{A_Witten} can be found for $D=4, 5$ and were found \linebreak in \cite{Dombriz_PRD_BH,Dombriz_PRD_BH1}.
%
%
%
%
%

Another important temperature value is the one corresponding to a
horizon radius $r_H=l$. Let us denote it as $T_1$ and from Equation~\eqref{T_AdS}
it is given by:
\begin{equation}
T_1=\frac{D-2}{2\pi l}
\end{equation}
Notice that for $D>2$ we have $T_0<T_1$.


In order to compute the remaining thermodynamic quantities, the standard
procedure is to consider the $D$-dimensional
Euclidean action defined as follows:
\begin{equation}
S_E=-\frac{1}{16 \pi G_D}\int \text{d}^{D}x\sqrt{g_E}\,(R+f(R))
\label{Euclidean_action}
\end{equation}
When the previous expression is evaluated on some
metric with a periodic Euclidean time of period $\beta$, it equals
$\beta$ times the free energy $F$ associated to this metric.
The computation by Hawking and
Page \cite{Hawking&Page}, generalized to higher dimensions by Witten
\cite{Witten}, was extended to $f(R)$ theories in \cite{Briscese}
and \cite{Dombriz_PRD_BH,Dombriz_PRD_BH1}. In these
references the difference of this action, when it is evaluated on
the BH and on the AdS metric, was computed leading to:
%
%
%
%
%
\begin{equation}
\Delta S_E=-\frac{(R_0+f(R_0))\beta \mu_{D-2}}{32 \pi (D-1)
G_D}(l^2r^{D-3}_H-r^{D-1}_H)=\beta F
\label{F_AdS}
\end{equation}
where $R_0=-D(D-1)/l^2$. From this expression it is straightforward to obtain the
free energy as $F=\Delta S_E /\beta$. One can also see that
provided that the condition $-(R_0+f(R_0))>0$ is fulfilled---which is the usual
case in EH gravity---we have
$F>0$ for $r_H<l$ and $F<0$ for $r_H>l$.

Once that both temperature and free energy have been calculated,  the total thermodynamical
energy may now be obtained as:
\begin{equation}
   E\,\equiv\,\frac{\partial \Delta S_E}{\partial \beta}\,=\,-\frac{(R_0+f(R_0))M l^2}{2(D-1)}
\label{Energy}
\end{equation}
where $M$ is the mass defined in Equation~\eqref{BHmass}. This is one of the possible
definitions for the BH energy for $f(R)$ theories but other definitions are available in the literature, see for instance \cite{Multamaki2007}
for a more general discussion. For the EH
action with non-vanishing cosmological constant, we have \linebreak $f(R)=-(D-2)\Lambda_D$ and then it is immediate to
find $E=M$. However this is not the case for general $f(R)$
gravitational Lagrangians as seen in Equation~\eqref{Energy}.  Notice that positive energy in AdS space-time requires
$R_0+f(R_0)<0$.

Now the entropy $S$ can be obtained from the well-known relation
$S\,\equiv\,\beta E- \beta F$.
Then one gets:
\begin{equation}
S\,=\,-\frac{(R_0+f(R_0))l^2 A_{D-2}(r_H)}{8 (D-1)G_D}
\end{equation}
where $A_{D-2}(r_H)$ is the horizon area given by
$A_{D-2}(r_H)\equiv r_H^{D-2}\mu_{D-2}$. Notice that once again
positive entropy requires $R_0+f(R_0)<0$.
For the EH action non-vanishing cosmological constant, we
have $R_0+f(R_0)=-2(D-1)/l^2$ and then the famous
Hawking--Bekenstein result \cite{Bekenstein} is recovered:
\begin{equation}
S=\frac{ A_{D-2}(r_H)}{4G_D}
\end{equation}

Finally we can compute the heat capacity $C$ which can be written
as:
\begin{equation}
C\,\equiv\,\frac{\partial E}{\partial T}=\frac{\partial E}{\partial
r_H}\frac{\partial r_H}{\partial T}
\end{equation}
Then it is easy to find:
\begin{equation}
C=\frac{-(R_0+f(R_0))(D-2)\mu_{D-2}r^{D-2}_Hl^2}{8G_D(D-1)}
\frac{(D-1)r^2_H+(D-3)l^2}{(D-1)r^2_H-(D-3)l^2}
\end{equation}
For the particular case of the EH action with cosmological constant we find:
\begin{equation}
C=\frac{(D-2)\mu_{D-2}r^{D-2}_H}{4G_D}\frac{(D-1)r^2_H+(D-3)l^2}{(D-1)r^2_H-(D-3)l^2}
\end{equation}
In the Schwarzschild limit, \emph{i.e}., $l$ going to infinity, this formula turns into:
\begin{equation}
C=-\frac{(D-2)\mu_{D-2}r^{D-2}_H}{4G_D}< 0
\end{equation}
which is the well-known result for standard Schwarzschild BH, \emph{i.e}., $C<0$.

Provided that the condition $(R_0+f(R_0))<0$ is accomplished,  the $f(R)$
general case  lead, like in the EH case, to
$C>0$ for $r_H > r_{H0}$ (the large BH region) and $C<0$ for
$r_H < r_{H0}$ (the small BH region). For $r_H \sim r_{H0}$ ($T$
close to $T_0$)   $C$ is divergent. Notice that in EH gravity, $C<0$ necessarily
implies $F>0$ since $T_0<T_1$.

In summary, $f(R)$ theories accomplishing the condition:
\begin{eqnarray}
R_0+f(R_0)<0
\label{condition}
\end{eqnarray}
provide a
scenario analogous to the one described in full detail by Hawking and Page in \cite{Hawking&Page}  for the EH case. This
analysis was originally shown in \cite{Dombriz_PRD_BH,Dombriz_PRD_BH1} and then widely used in literature.

In principle, particular $f(R)$ models violating
the condition $R_0+f(R_0)<0$ can be naively considered. However, in such cases,
 the mass and the entropy would be negative and therefore
the AdS BH solutions for these models would be unphysical.
Therefore $R_0+f(R_0)<0$ can be regarded  as a necessary condition
for $f(R)$ theories in order to support the usual AdS BH solutions. Using
Equation~\eqref{const_R0}, this condition is fully equivalent to $1+f'(R_0)>0$. This last
condition has a clear physical interpretation in $f(R)$ gravity theories
(see \cite{silvestri} and references therein). Indeed, it can be
interpreted as the condition for the effective Newton constant
$G_{eff}=G_{D}/(1+f'(R_0))$ to be positive. It can also be
interpreted from the quantum point of view as the condition which
prevents the graviton from becoming a ghost. Therefore
the requirement Equation~\eqref{condition} can be also regarded as a thermodynamical viability condition for $f(R)$
gravity theories when AdS configurations are studied \cite{Dombriz_PRD_BH,Dombriz_PRD_BH1}.

Once that the main thermodynamical quantities have been derived, let us now revise the
local and global stabilities phenomenology:
For $T<T_0$, the only possible state of thermal equilibrium in an
AdS space is pure radiation with negative free energy and hence there is
no stable BH solutions. For $T>T_0$ there exist two possible BH
solutions; the small (and light) BH and the large (heavy) BH. The
small one has negative heat capacity and positive free energy as the
standard Schwarzschild BH. Therefore it is unstable under Hawking
radiation decay. For the large BH there appear two possibilities; if
$T_0<T<T_1$ then both the heat capacity and the free energy are
positive and the BH will decay by tunneling into radiation, but if
$T>T_1$ then the heat capacity is still positive but the free energy
becomes negative. In this case the free energy of the heavy BH will
be less than that of pure radiation. Then pure radiation will tend
to tunnel or to collapse to the BH configuration in equilibrium with
thermal radiation.
This phenomenology will be applied to different $f(R)$ models in Section \ref{AdS_Thermodynamics}.

\subsection{BH Thermodynamics for KN Configuration}

Following the reasoning at the beginning of this section, 
the research concerning the KN configuration will be also done for
BH with a well-defined horizon structure for negative values of $R_0$.
%
%
In order to study the different thermodynamical properties of Kerr--Newman BH in $f(R)$ theories, the usual approach is to start studying the temperature of the exterior horizon $r_{ext}\equiv r_{ext}\left(R_0,\,a,\,\bar{Q},\,M\right)$.  For that purpose, the Euclidean action method \cite{HGG,HGG1,HGG2,HGG3} is again considered. The procedure consists now of performing the change of variables $t\rightarrow -i\tau$, $a\rightarrow ia$ on the metric \eqref{metrica} in order to obtain the Euclidean section \linebreak(see \cite{Jimeno_BH,Cembranos:2012ji} for details). After some calculations, the inverse Hawking temperature now becomes:
%
%
\begin{eqnarray}
\beta_{\text{KN}}\equiv\frac{1}{T_{E,\,\text{KN}}}\,=\,\frac{\displaystyle 4\pi\left(r_{ext}^2+a^2\right)}{\displaystyle r_{ext}\left[1-\frac{R_0\,a^2}{12}-\frac{R_0\,r_{ext}^2}{4}-\frac{\left(a^2+\bar{Q}^2\right)}{r_{ext}^2}\right]}
\label{inversatemp}
\end{eqnarray}
For null spin and charge, this expression becomes naturally Equation~\eqref{T_AdS}.
%
%
\smallskip
\smallskip

BH horizon temperature can also be obtained through Killing vectors, as is explained in \cite{Hawking1974} where temperature is defined as: $T_{\kappa,\,\text{KN}}\equiv \kappa / 4\,\pi$,
with $\kappa$ the surface gravity defined now as $\chi^\mu\,\nabla_\mu\chi_\nu=\kappa\,\chi_\nu$. It can be verified that $\kappa$ is the same at any horizon point and consequently $T_{\kappa,\,\text{KN}}=T_{E,\,\text{KN}}$ as obtained in~\cite{Bardeen1973}.

Now that we know the expression for the temperature for KN configurations, the natural step in order to obtain the remaining thermodynamical quantities is to consider the corresponding extension of the Euclidean action with respect to the previously considered Equation~\eqref{Euclidean_action}. It now becomes:
\begin{eqnarray}
\Delta S_{E,\,\text{KN}}\,=\,\frac{1}{16 \pi}\int _{\cal Y}\text{d}^{4}x\sqrt{\mid g\mid}\,\left(R_0+f(R_0)-F_{\mu\nu}F^{\mu\nu}\right)\,
\label{Euclidean_action_KN}
\end{eqnarray}
with ${\cal Y}$ the integration region. As is described in \cite{Hawking&Page},
and fully derived in \cite{Jimeno_BH,Cembranos:2012ji} the result for $\Delta S_{E, \,\text{KN}}$ becomes:
%
%
%
\begin{eqnarray}
\Delta S_{E\,\text{KN}}\,=\,\frac{ \beta\,\left(R_0+f(R_0)\right)}{24\,\Xi}
\left[
r_{ext}^3+\left(a^2+\frac{12}{R_0}\right)\,r_{ext}+\frac{12\,a^2}{R_0\,r_{ext}}
\right]
+
\frac{\beta}{2}\,\Phi_e\,{\cal{Q}}\,\left(\frac{r_{ext}^2+a^2}{2\,r_{ext}^2}+1\right)
\label{accion}
\end{eqnarray}
%
where $\Phi_e$ is the electric potential of the horizon as seen from infinity:
%
\begin{eqnarray}
\Phi_e \equiv \frac{Q\,r_{ext}}{r_{ext}^2+a^2}=\frac{\bar{Q}\,r_{ext}}{r_{ext}^2+a^2}\,\left(1+f'(R_0) \right)^{1/2}\,
\label{Phi_e}
\end{eqnarray}
and ${\cal{Q}}$ is the physical electric charge of the BH, obtained by integrating the flux of the electromagnetic field tensor at infinity, which happens to be:
\begin{eqnarray}
{\cal{Q}} \equiv \frac{Q}{\Xi}= \frac{\bar{Q}}{\Xi}\,\left(1+f'(R_0) \right)^{1/2}\,
\label{cargafisica}
\end{eqnarray}
From expressions Equations~\eqref{Phi_e} and \eqref{cargafisica}, it can be seen that $\Phi_e$ and ${\cal{Q}}$ definitions are $f(R)$-model independent.
The reason for this lies in the fact that
these calculations involve the vector potential and the electromagnetic field tensor, which are independent of the considered $f(R)$ model.
Further analysis of the action revealed \cite{Jimeno_BH,Cembranos:2012ji} that it became singular for $h=0$, as could be expected from {\it extremal} BH, whose temperature $T_{E,\,\text{KN}}=0$ makes the $\beta_{\text{KN}}$ factor diverge. Since thermodynamical potentials are obtained by dividing the action by the
$\beta$ factor, they still remain well-defined at $T_{E,\,\text{KN}}=0$.

%

By using the expression Equation~\eqref{accion} we can immediately obtain Helmholtz free energy $F$. For rotating configurations, the correct definition is now:
\begin{eqnarray}
F=\frac{\Delta S_E}{\beta}+\Omega_H\,J
\end{eqnarray}
where the term $\Omega_H\,J$ comes from the required Legendre transformation to fix angular momentum, being $J$ the angular momentum of the BH and $\Omega_H$ the angular velocity of the rotating horizon. For the explicit expressions, we refer the reader to \cite{Jimeno_BH,Cembranos:2012ji}. Thus, performing some calculations one gets:
%
%
\begin{eqnarray}
F\,=\,(1+f'(R_0))\frac{\displaystyle \left[36\,\bar{Q}^2+12\,r_{ext}^2+r_{ext}^4\,R_0+a^2\,(36-r_{ext}^2\,R_0)\right]}{\displaystyle  24\,r_{ext}\,\Xi}\,
\label{F_KN}
\end{eqnarray}
%
%

%
Once again, expression Equation~\eqref{F_KN} reduces to Equation~\eqref{F_AdS} for a non-rotating uncharged BH.
Provided that the condition $1+f'(R_0)>0$ is required to hold in order to obtain positive values of the mass, the inspection of the numerator of $F$ renders that $F>0$ for values of the horizon below $r_{ext}^{\,\text{\it limit}}$ (with an associated mass $M^{\,\text{\it limit}}$ through Equation \eqref{poli_hori}), and $F<0$ for larger values.

Using the appropriate thermodynamical relations \cite{Caldarelli}, we can derive the entropy $S$ of the KN BH, which reads:
\begin{eqnarray}
S\,=\,\beta\,({\cal{M}}-\Omega_H\,J)-\Delta S_E 
\,=\,(1+f'(R_0))\,\frac{\pi\,(r_{ext}^2+a^2)}{\displaystyle \Xi}\,
\label{entropia}
\end{eqnarray}

If the area ${\cal{A}}_H$ of the exterior horizon is now computed, one gets:
\begin{eqnarray}
{\cal{A}}_H=4\pi\,(r_{ext}^2+a^2)\,/\,\Xi
\label{A_KN}
\end{eqnarray}
%
%
Thus, if expression Equation~\eqref{A_KN} is substituted into the entropy expression Equation~\eqref{entropia}, the entropy can be expressed as:
\begin{eqnarray}
S=(1+f'(R_0))\,\frac{ {\cal A}_H }{4}\,
\end{eqnarray}
Consequently $1+f'(R_0)>0$ is again a mandatory condition to obtain a positive entropy, as was shown in the AdS as well. Note that if for example we want to recover GR with a cosmological constant taking $f(R_0)=-2\Lambda$, this relation turns into:
$S={\cal A}_{H}/4$, \emph{i.e}., the famous Bekenstein result \cite{Bekenstein}.

Once that the temperature $T$ and the entropy $S$ of the KN BH are known, the natural step further can be taken and study the heat capacity at constant scalar curvature $R_0$ and at fixed spin $a$ and charge $\bar{Q}$ parameters. From the definition:
\begin{eqnarray}
C=T \left.\frac{\partial S}{\partial T}\right|_{R_0,a,Q}\,
\end{eqnarray}
we obtain the expression:
%
\begin{eqnarray}
C\,=\,(1+f'(R_0))\,\frac{\displaystyle 2\pi\, r_{ext}^2\,(a^2+ r_{ext}^2)\left[a^2\,(12+ r_{ext}^2\, R_0)+3\,(4\,\bar{Q}^2-4\, r_{ext}^2+ r_{ext}^4 \,R_0)\right]}{
Den(\bar{Q},\,r_{ext},\, R_0,\,a)}
\end{eqnarray}
%
with $Den(\bar{Q},\,r_{ext},\, R_0,\,a)\,\equiv\,Den$ defined as:
\begin{eqnarray}
Den\,\equiv\,
 \Xi \left[-36\,\bar{Q}^2\,r_{ext}^2+a^4\,(-12+r_{ext}^2\, R_0)+3\,r_{ext}^4\,(4+r_{ext}^2\,R_0)-4a^2\,(3\,\bar{Q}^2+12\,r_{ext}^2-2\,r_{ext}^4\, R_0)\right]\nonumber
 \end{eqnarray}

%
At this stage, it is interesting to find out for which values of $R_0$, $a$, $\bar{Q}$ and $M$ the denominator of the thermal capacity becomes
zero, \emph{i.e}., the thermal capacity goes through an infinite discontinuity, \emph{i.e}., the BH experiences a phase transition. Following the analysis presented
in \cite{Jimeno_BH,Cembranos:2012ji}, two kinds of BH on this subject can be distinguished depending on the values of the $a$, $\bar{Q}$ and $M$ parameters and scalar curvature $R_0$:
\begin{itemize}
\item
{\it fast} BH, without phase transitions and always positive heat capacity $C>0$.
\item
{\it slow} BH, presents two phase transitions for two determined values of $r_{ext}$.
\end{itemize}

\begin{figure}[h]
	\centering
		\includegraphics[width=0.375\textwidth]{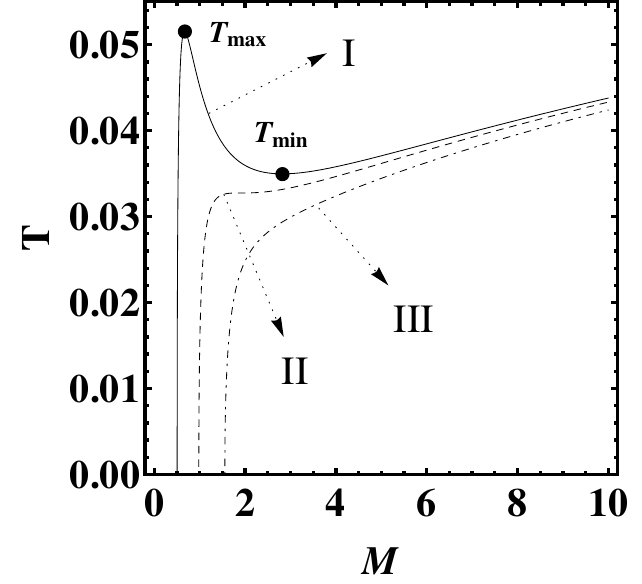}\,\,\,\,\,\,\,\,\,\,\,\,\,\,
		\includegraphics[width=0.40\textwidth]{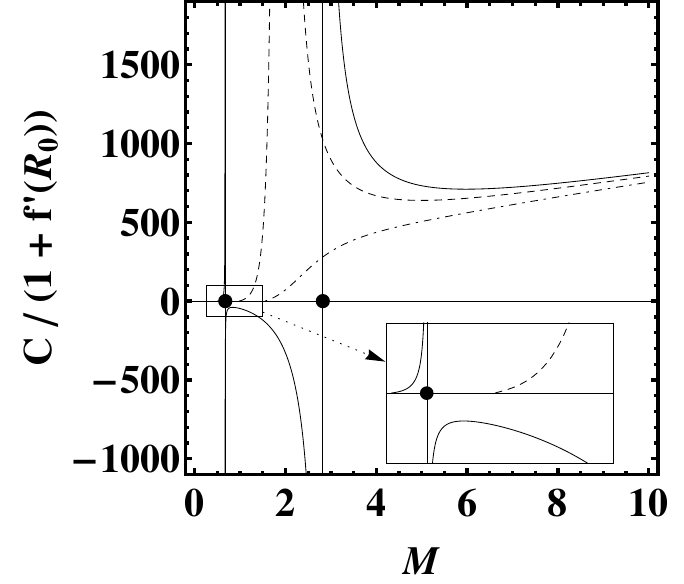}
		\caption{For $R_0=-0.2$, we graphically display temperature (left) and heat capacity (right)
of a BH as functions of the mass parameter $M$ for the cases: {(\textbf{I})} $a=0.5$ y $Q=0$ : ``slow'' BH that shows a local maximum temperature $T_{max}$ and a local minimum temperature $T_{min}$ at the points where the heat capacity diverges, taking the latter negative values between $T_{max}$ y $T_{min}$; {(\textbf{II})} $a\approx0.965$ y $Q=0$ : limit case where $T_{max}$ and $T_{min}$ merge, hence resulting in an inflection point in the temperature and an always positive heat capacity; {(\textbf{III})} $a=1.5$ y $Q=0$ : ``fast'' BH with both temperature and heat capacity monotonously growing (always positive too).
}
	\label{fig:temperatura}
\end{figure}

In Figure \ref{fig:temperatura} we have visualized the behavior of the temperature $T$ and the heat capacity $C$ of a BH for different values of mass parameter $M$, with fixed $a$, $\bar{Q}$ and $R_0$ values.
%
\smallskip
\smallskip

As was remarked in \cite{Jimeno_BH,Cembranos:2012ji}, KN--AdS BH, unlike Schwarzschild--AdS BH case \cite{Hawking&Page},
are allowed for any value of the temperature $T$.
Hence stability of each BH will be exclusively given by the corresponding values of heat capacity $C$ and free energy $F$ as functions of the $a$, $\bar{Q}$, $M$ and $R_0$ parameters, that ultimately define the BH.
However, for a set of fixed values of $a$, $\bar{Q}$ and $R_0$, the mass parameter must be bigger than a minimum $M^{\text{\it min}}$ (characterized by $T=0$) to host BH configuration, otherwise radiation is the only possible equilibrium state up to such a minimum mass. For bigger masses, we shall distinguish between the {\it fast} and the {\it slow} BH.
\begin{itemize}
\item  {\it Fast} BH, with bigger values of the spin and the electric charge than the {\it slow} ones, shows a heat capacity always positive and a positive free energy up to a $M=M^{\,\text{\it limit}}$ value, and negative onwards. Thus, this BH is unstable against {\it tunneling} decay into radiation for mass parameter values of $M<M^{\,\text{\it limit}}$. For $M>M^{\,\text{\it limit}}$, free energy becomes negative, therefore smaller than that of pure radiation, that will tend to collapse to the BH configuration in equilibrium with thermal~radiation.
\item  
{\it Slow} BH shows a more complex thermodynamics, being necessary to distinguish between four regions delimited by the mass parameter values:  $M^{min}<M^{\textnormal {I}}<M^{\textnormal {II}}<M^{\,\text{\it limit}}$, as follows:
\begin{itemize}

\item For $M^{min}<M<M^{\textnormal {I}}$ and for $M^{\textnormal {II}}<M<M^{\,\text{\it limit}}$, both the heat capacity and the free energy are positive, which means that the BH is unstable and decays into radiation by {\it tunneling}.

\item If $M^{\textnormal {I}}<M<M^{\textnormal {II}}$, the heat capacity becomes negative but free energy remains positive, being therefore unstable and decays into pure thermal radiation or to larger values of mass.
\item Finally, for $M>M^{\,\text{\it limit}}$ the heat capacity is positive whereas the free energy is now negative, thus tending pure radiation to tunnel to the BH configuration in equilibrium with thermal radiation.
\end{itemize}
\end{itemize}
\smallskip
\smallskip

Let us conclude this section as we did the previous one, by emphasizing that, although not quantitatively, the thermodynamical behavior of KN--AdS BH in $f(R)$ modified gravity theories is qualitatively analogous to that of GR \cite{Caldarelli}. This was an original important result presented for the first time in \cite{Jimeno_BH,Cembranos:2012ji}. In that investigation, authors proved that some thermodynamical quantities like the physical mass ${\cal{M}}$, the angular momentum $J$ and the entropy $S$ differ from those of GR by a $(1+f'(R_0))$ factor. One might think that the free energy $F$ and the heat capacity $C$ also differ in that same multiplicative way, but because of the dependence with the normalized charge parameter $\bar{Q}$, this same term $(1+f'(R_0))$ leads to a more complex deviation from the GR behavior.


\section{Thermodynamics in A${\text d}$S and Kerr--Newman:  Particular Examples}
\label{AdS_Thermodynamics}
\vspace{-12pt}
\subsection{Thermodynamics in A${\text d}$S }

In order to illustrate the calculations sketched in Section \ref{AdS_Subsection},
let us consider some particular $f(R)$ models in
whose heat capacity $C$ and the free energy $F$ will be determined that were originally studied in \cite{Dombriz_PRD_BH,Dombriz_PRD_BH1}. As
was pointed out above, those two quantities provide the
%
%
local and global stability of AdS BH. For the particular models that are studied below, the constant scalar curvature $R_0$ could be
calculated exactly by using Equation~\eqref{const_R0}. For the sake of
simplicity the $D$-dimensional Schwarzschild radius in
Equation~(\ref{BHmass}) was taken as $R_S^{D-3}\equiv 2$. The models we have considered are the following.

\subsubsection{. Model I: $f(R)\,=\,\alpha (-R)^{\beta}$ }

This model has been widely studied because the $\alpha R^2$ term with $\alpha >0$ can account for the accelerated expansion of the Universe \cite{Starow}. This model can also explain the observed temperature anisotropies observed in the CMB, and can become a viable alternative to scalar field inflationary models, since reheating after inflation would have its origin on the production of particles during the oscillation phase of the Ricci scalar \cite{Mijic}. For $\beta=2$, this model was recently pointed out  \cite{varios4,varios6} as a possible explanation to dark matter.

Substituting this model in Equation \eqref{const_R0} for arbitrary dimension 
and by only considering  non-vanishing curvature solutions, the found solution is:
\begin{eqnarray}
R_{0}\,=\,-\left[\frac{2-D}{(2\beta-D)\alpha}\right]^{1/(\beta-1)}
\label{R0_Model_I}
\end{eqnarray}
For $D>2$,
the condition $(2\beta-D)\alpha<0$ provides well-defined scalar curvatures $R_{0}$.
Two separated regions have thus to be studied: Region $1$ $\{\alpha<0,\,\beta>D/2\}$
and region $2$ $\{\alpha>0,\, \beta<D/2\}$.
For this model we also get:
\begin{eqnarray}
1+f'(R_{0})\,=\,\frac{D(\beta-1)}{2\beta-D}
\end{eqnarray}
Notice that in region $1$, $1+f'(R_0)>0$ for $D>2$, since in this case $\beta>1$
is straightforwardly accomplished.
In region $2$, we find that for $D>2$, the requirement $R_0+f(R_0)<0$, \emph{i.e}., $1+f'(R_0)>0$,
fixes $\beta<1$, since this is the most stringent constraint over the
parameter $\beta$ in this region. Therefore the physical space of
parameters in region $2$ is restricted to be $\{\alpha>0,\,\beta<1\}$.

\begin{figure}[h]
\centering
\subfigure{
\begin{overpic}[width=7.0cm]{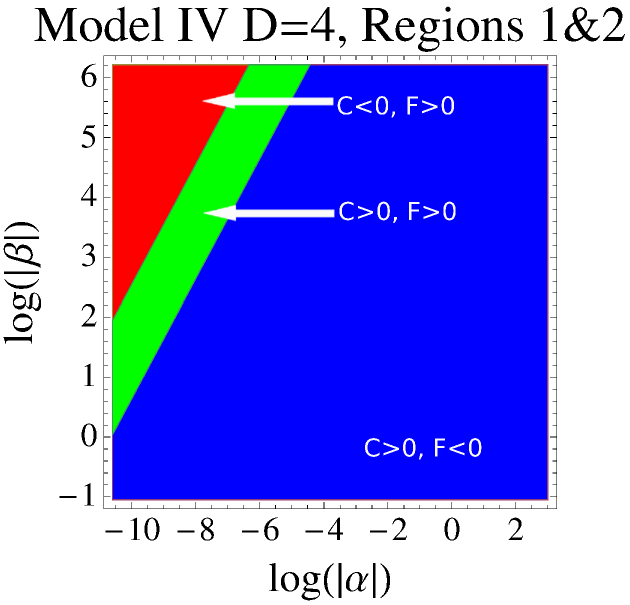}
\end{overpic}
}
\subfigure{
\begin{overpic}[width=7.0cm]{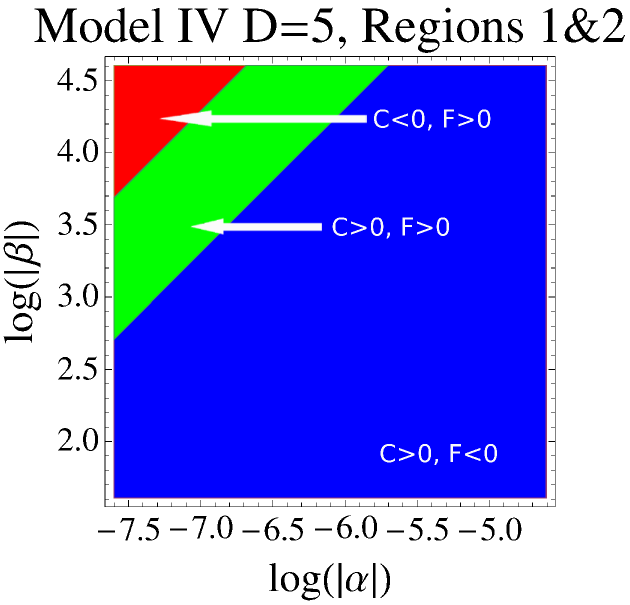}
\end{overpic}
}
\caption{Thermodynamical regions in the $(|\alpha|,|\beta|)$ plane
 for Model I in $D=4$  (left) and  $D=5$ (right).}
 \label{D4and5}
\end{figure}

\begin{figure}[h]
	\centering
		\includegraphics[width=7.0cm]{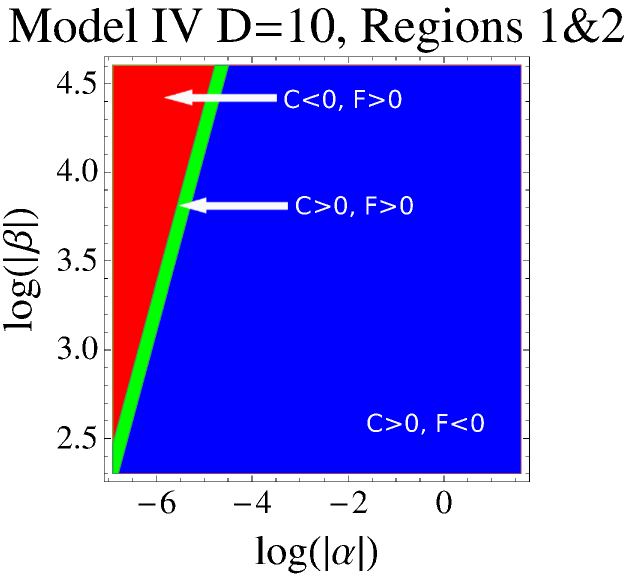}
		\caption{Thermodynamical regions in the $(|\alpha|,|\beta|)$ plane
 for Model I in $D=10$.
		}
	\label{fig:termoregionesIV}
\end{figure}

In Figures \ref{D4and5}  and \ref{fig:termoregionesIV}, taken from \cite{Dombriz_PRD_BH,Dombriz_PRD_BH1}, one can see the physical regions in the parameter space $(\alpha,\beta)$
corresponding to the different signs of $(C,F)$.
%

\subsubsection{. Model II:  $f(R)\,=\,-\alpha\,\displaystyle \frac{\displaystyle \kappa\left(\frac{R}{\alpha}\right)^n}{\displaystyle 1+\beta\left(\frac{R}{\alpha}\right)^n}$}
This model has been proposed in \cite{Hu&Sawicki2007} as cosmologically viable and has attracted a lot of attention in order to constrain its validity in different astrophysical and cosmological scenarios.
Throughout this study, let us consider $n=1$ for the sake of simplicity. A vanishing curvature solution also appears in this model and the two nontrivial
curvature solutions are given by \cite{Dombriz_PRD_BH,Dombriz_PRD_BH1}:
\begin{eqnarray}
R_{0}^{\pm}\,=\,\frac{\alpha \left[(\kappa-2) D+4\pm\sqrt{\kappa} \sqrt{\kappa D^2-8 D+16}\right]}{2 \beta (D-2)}
\label{R0_model_IV}
\end{eqnarray}
where $\kappa>0$ and $\kappa>(8D-16)/D^2$ are required for real $R_{0}$ solutions.
The corresponding $1+f'(R_0)$ values for Equation~\eqref{R0_model_IV} are:
\begin{eqnarray}
1+f'(R_{0}^{\pm})\,=\,1-\frac{4(D-2)^2}{\left(\sqrt{\kappa D^2-8D+16}\pm\sqrt{\kappa} D\right)^2}
\label{1+f'(R0)_model_IV}
\end{eqnarray}
Since $1+f'(R_{0})>0$ is required, one has
$\text{sign}(R_{0}^{\pm})=\text{sign}(\alpha\beta)$. In fact it turns out that $1+f'(R_{0}^{-})$ is not
positive for any allowed value of $\kappa$ and therefore this curvature solution $R_{0}^{-}$ is excluded for our study (see \cite{Dombriz_PRD_BH,Dombriz_PRD_BH1} for details). On the other hand,
$1+f'(R_{0}^+)>0$ only requires $\kappa>0$ for dimension $D\geq4$ and
therefore $f'(R_0)<0$.
Therefore only two accessible regions need to be studied:  region $1$
$\{\alpha>0,\, \beta<0\}$ and region $2$, $\{\alpha<0,\,\beta>0\}$.



\subsection{KN--AdS Thermodynamics}
\label{KN_Thermodynamics}

In this section we will study the thermodynamics of the two $f(R)$ models presented in the last section but when a Kerr--Newman configuration is considered.

First, as was shown above, the condition $1+f'(R_0)>0$ must be satisfied in order to obtained physical results, so first of all the viability of each model depending upon the values of the parameters that define them will be checked out.
Thus, for each model, we will study the range of parameters that allows the existence of Kerr--Newman BH, \emph{i.e}., we will graphically display $a_{max}$ and $a_{min}$ (if present) in terms of those parameters, and study the region of $a$ confined between these two surfaces. This phenomenology was absent in the SAdS study since the configuration was spherically symmetric and therefore $a=0$.
Finally we will focus on the thermodynamical quantities that define BH stability depending again upon the model range of parameters. In order to sketch graphically the thermodynamical behavior, 
let us then consider the BH parameter values: $M=1$, $a=0.4$ and $\bar{Q}=0.2$. Let us remind at this stage that this analysis is restricted to $R_0<0$ condition, as explained in the beginning of Section \ref{Thermodynamics}.

For the sake of simplicity, let us introduce the dimensionless variables notation \cite{Jimeno_BH,Cembranos:2012ji}
that we will use in the graphical visualization of the results to be presented:
\begin{eqnarray}
\frac{r}{M}\rightarrow r,\,\,\,\,\frac{a}{M}\rightarrow a,\,\,\,\,\frac{\bar{Q}}{M}\rightarrow \bar{Q},\,\,\,\,R_0\,M^2\rightarrow R_0
\end{eqnarray}
where the previous symbols keep their original meaning as defined before. The considered models are~as follows.

\subsubsection{. Model I: $f(R)\,=\,\alpha (-R)^{\beta}$}

Let us remind the solution Equation~\eqref{R0_Model_I} found above for this model.
%
%
%
%
%
%
%
The viability condition \linebreak $1+f'(R_0^{})>0$ restricts the range of parameters that define this $f(R)$ model.

\begin{figure}[h]
	\centering
		\includegraphics[width=0.40\textwidth]{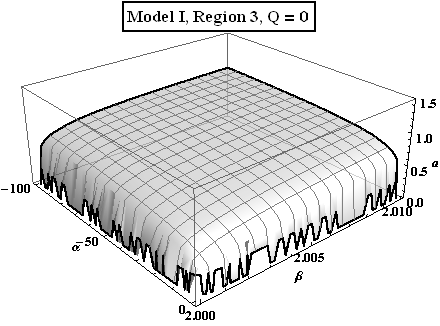}\,\,\,\,\,\,\,\,\,\,\,\,\,\,
		\includegraphics[width=0.40\textwidth]{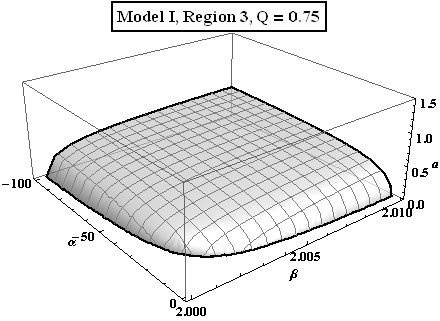}\\[0.5cm]
		\includegraphics[width=0.40\textwidth]{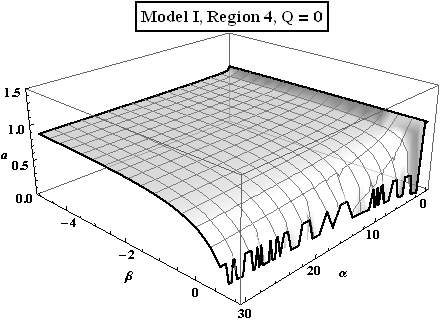}\,\,\,\,\,\,\,\,\,\,\,\,\,\,
		\includegraphics[width=0.40\textwidth]{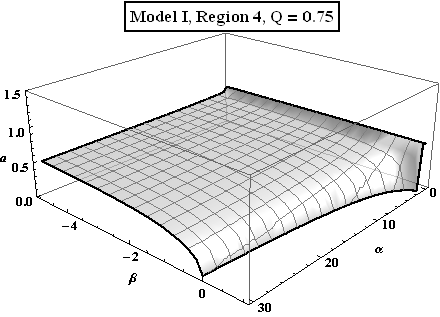}
		\caption{
		\textbf{Model I}: 
		Region 3: $\left\{ \alpha<0,\,\beta>2\right\}$, and Region 4: $\left\{ \alpha>0,\,\beta<1\right\}$. BH with a well defined horizon structure will only exist if they have a spin parameter below the upper surface $a_{max}$. For the presented regions 3 and 4, the surface $a_{min}$ does not exist. 
		}
	\label{fig6}
\end{figure}

In Figure \ref{fig6}, we show the range of the spin parameter $a$ for which BH are allowed, depending on the parameters $\alpha$ and $\beta$ and for charge parameters values $Q=0$ and $Q=0.75$.
%
%
For the solution $R_0<0$ there exist two possible regions:

\begin{itemize}

\item Region 3 $\left\{ \alpha<0,\,\beta>2\right\}$. For this region, the value of $a_{max}$ decreases suddenly from its normal value to 0 near the frontier of the region, \emph{i.e}., $\alpha\approx 0$ and $\beta \approx 2$; other values of $\alpha$ and $\beta$ display a relatively low curvature and the existence of BH is assured.

\item Region 4 $\left\{ \alpha>0,\,\beta<1\right\}$. This region only shows problems when $0<\beta <1$, where $a_{max} \rightarrow 0$, but, as $\beta$ becomes more negative, the surface of $a_{max}$ slowly acquires higher values, recovering its usual value for $\beta=-2$.

\end{itemize}

\begin{figure}[h]
	\centering
		\includegraphics[width=0.45\textwidth]{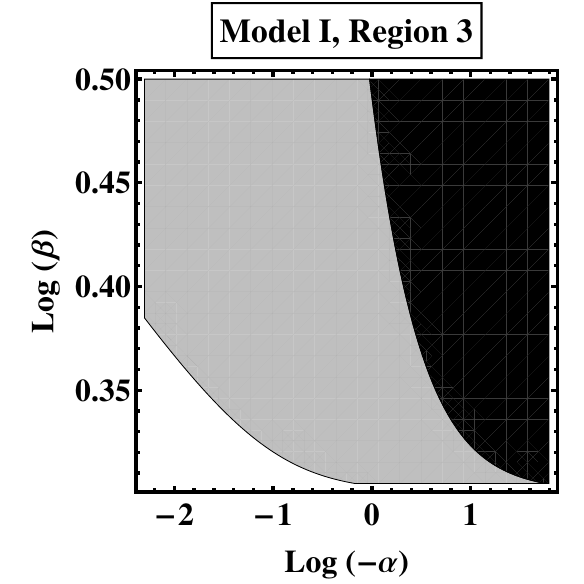}\,\,\,\,\,\,\,\,\,\,\,\,\,\,
		\includegraphics[width=0.45\textwidth]{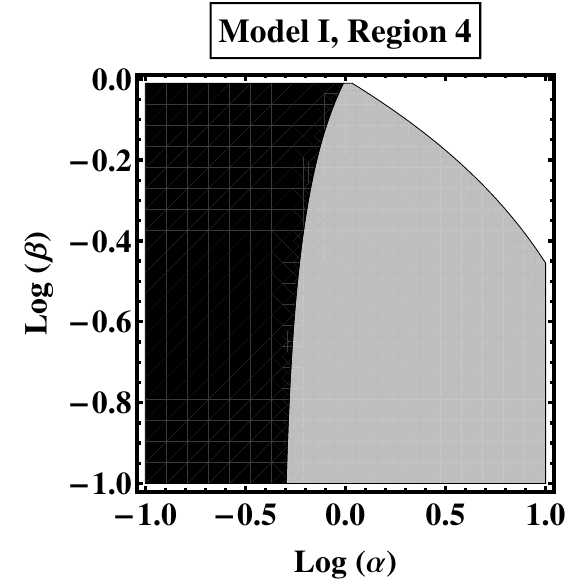}		 		
		\caption{Thermodynamical regions for Model I. We distinguish between three different regions: {(\textbf{i})} $C<0$ and $F>0$, in black; {(\textbf{ii})} $C>0$ and $F>0$, in gray; {(\textbf{iii})} $C>0$ and $F<0$, in white.
		}
	\label{fig7}
\end{figure}

We have graphically schematized in Figure \ref{fig7} the possible thermodynamical configurations of a BH for Regions 3 and 4, \emph{i.e}., $R_0<0$ following the results presented in \cite{Jimeno_BH,Cembranos:2012ji}. As we can see in Region 3, only for values close to the region boundary ($\alpha \approx 0$ and $\beta \approx 2$) is BH stable both locally and globally ($C>0$ and $F<0$). Also, as the absolute value of $\alpha$ increases, it is more probable that the BH shows global instability ($C>0$ and $F>0$) and finally global and local instability ($C<0$ and $F>0$). In Region 4, however, we find the opposite behavior: near the limit of the region values ($\alpha \approx 0$ and $\beta \approx 1$) we have global and local instability, and, as $\alpha$ gets bigger, the BH becomes more stable both locally and globally.

\subsubsection{. Model II:  $f(R)\,=\,-\alpha\,\displaystyle \frac{\displaystyle \kappa\left(\frac{R}{\alpha}\right)^n}{\displaystyle 1+\beta\left(\frac{R}{\alpha}\right)^n}$ }
%
 %

\begin{figure}[h]
	\centering
		\includegraphics[width=0.40\textwidth]{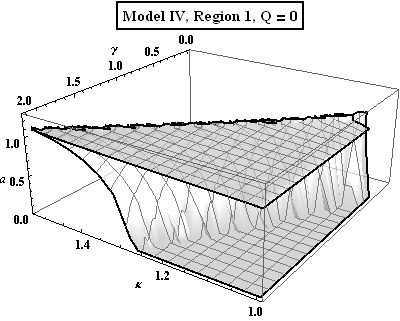}\,\,\,\,\,\,\,\,\,\,\,\,\,\,
		\includegraphics[width=0.40\textwidth]{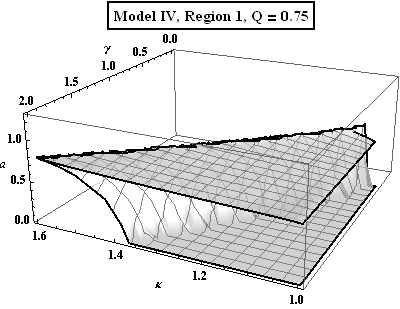}\\[0.5cm]
		\includegraphics[width=0.40\textwidth]{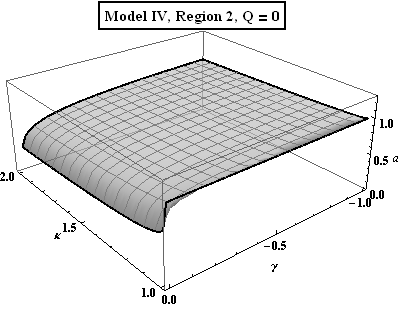}\,\,\,\,\,\,\,\,\,\,\,\,\,\,
		\includegraphics[width=0.40\textwidth]{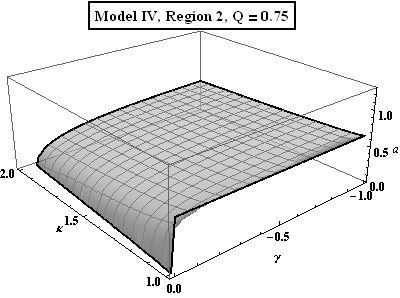}
		\caption{
		\textbf{Model II}. Region 1: $\left\{ \kappa>1,\,\gamma>0 \right\}$, and Region 2: $\left\{ \kappa>1,\,\gamma<0 \right\}$. BH with a well-defined horizon structure will only exist if they have a spin parameter below the upper surface $a_{max}$, and above a second surface $a_{min}$ (in case it exists, only in region 1 for this model) for certain values of $\kappa$ and $\gamma$.
		}
	\label{fig8}
\end{figure}

Let us consider for the sake of simplicity again 
$n=1$, thus having a two-parameters model, since we can define $\gamma=\beta/\alpha$ and then obtain:
\begin{eqnarray}
f(R)\,=\,-\frac{\kappa\,R}{1+\gamma\,R}
\end{eqnarray}
Replacing the latter in Equation~\eqref{const_R0}, one gets Equation~\eqref{R0_model_IV}, that for $D=4$,  and focusing on the negative constant curvature solution, yields:
\begin{eqnarray}
R_0^{}=-\frac{1-\kappa}
{\gamma}-\sqrt{\frac{-\kappa\,(1-\kappa)}{\gamma^2}},
\label{R0_model_IV_neg_D4}
\end{eqnarray}
Keeping in mind that we have to satisfy $1+f'(R_0)>0$, $\kappa$ happens to be restricted to values of $\kappa>1$. On the other hand, computation of $1+f'(R_0)$ reveals that 
$R_0$ is only a valid solution for values of $\kappa$ and $\gamma$ in the Region 2: $\{ \kappa>1,\,\gamma<0\}$,
being as mentioned 
$R_0<0$. 
In Figure \ref{fig8} we show the range of the spin parameter $a$ for which BH are allowed, depending on the
parameters $\kappa$ and $\gamma$ for certain values of the charge $\bar{Q}$.
Region 2 displays an $a_{max}$ surface that descends slightly as the value of $\kappa$ increases, and more significantly when $\gamma \approx 0$, becoming 0 when this limit is reached.

\begin{figure}[h]
	\centering
		\includegraphics[width=0.43\textwidth]{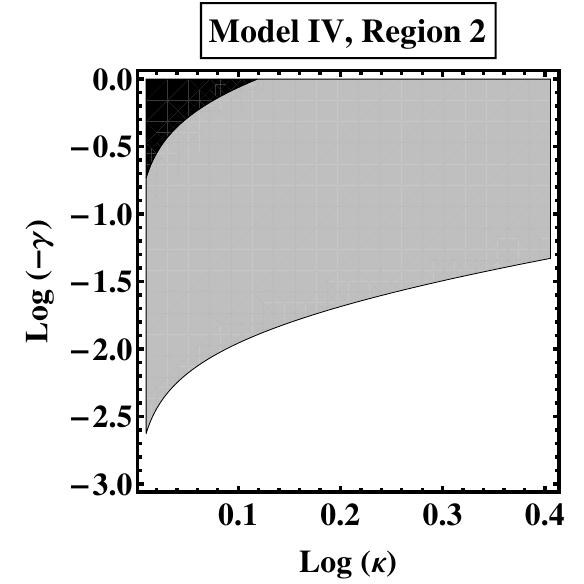}
		\caption{Thermodynamical regions with negative scalar curvature $R_0<0$ of model I. We distinguish between three different regions: {(\textbf{i})} $C<0$ and $F>0$, in black; {(\textbf{ii})} $C>0$ and $F>0$, in gray; {(\textbf{iii})} $C>0$ and $F<0$, in white.
		}
	\label{fig9}
\end{figure}

 We graphically schematize in Figure \ref{fig9} the different possible thermodynamical configurations for the region in which $R_0<0$, in this case the region 2. In this parameter space region, a combination of high values of $\kappa$ and small absolute values of $\gamma$ favor BH local and global stability, and, conversely, a combination of small values of $\kappa$ and high absolute values of $\gamma$ lead to BH instability, first globally and then also locally.

\section{Cosmological Solutions in Modified Gravity}
\label{CosmoSol}

Modified gravity has become very popular in recent years because it is capable to reproduce the dark energy epoch, and even the inflationary phase, with no need to introduce an additional component in the universe. Hence, the reconstruction of appropriate theories that contain some particular cosmological solutions have involved an important effort (see~\cite{varia,varia1,varia2,varia3,varia4,ScalarFR,reconstruction1,reconstruction11,reconstruction12,reconstruction13,reconstruction14,reconstruction15,reconstruction16,reconstruction17,reconstruction18,reconstruction19,reconstruction110,reconstruction111,reconstruction112,reconstruction114,reconstruction2,reconstruction3,Dunsby1,Nojiri:2009kx,Nojiri:2011kd}). The same is true for the reconstruction of so-called viable modified gravities, which can recover GR in some limit while they introduce additional effects at cosmological scales (see~\cite{Hu&Sawicki2007,Viable,Viable1,Viable2,Viable3}). However, as in every dark energy model, the required fine tuning on the form of the action as on free parameters of the model have resulted in a large effort in searching additional and independent predictions in the frame of modified gravities, including other aspects different than cosmology as the ones studied in the above sections (see also the reviews \cite{Reviews,Reviews1,Reviews2,Reviews3,Reviews4,Reviews5,Reviews6,Reviews7} and references therein). In this section we are interested to make a short review on cosmological solutions in the frame of modified gravities, and in particular in $F(R)$ gravity, while the thermodynamical aspects are studied in the following sections.  As it is widely accepted, our universe seems to be well described by FLRW metrics:
\be
ds^2=-dt^2+a^2(t)\left[\frac{1}{1-kr^2}dr^2+r^2 d\theta^2+r^2\sin^2\theta\ d\varphi^2\right]\
\label{D1}
\ee
and specifically the universe seems to be spatially flat, $k=0$. Hence, by introducing the metric in $F(R)$ field Equation~\eqref{In.2}, modified FLRW equations are obtained:
\bea
H^2=\frac{1}{3F_R}\left[\kappa^2 \rho_m +\frac{RF_R-F}{2}-3H\dot{R}F_{RR}\right]\  \nn
-3H^2-2\dot{H}=\frac{1}{F_R}\left[\kappa^2p_m+\dot{R}^2F_{RRR}+2H\dot{R}F_{RR}+\ddot{R}F_{RR}+\frac{1}{2}(F-RF_R)\right]\
\label{D2}
\eea
Recent astronomical data indicates that the well-known $\Lambda$CDM model where DE is represented by an
effective cosmological constant fits quite well the observational
data. Nevertheless, it is shown that $F(R)$ gravity can even reproduce exactly the $\Lambda$CDM model in absence of a cosmological constant s shown below (see~\cite{reconstruction3,Dunsby_PRD}). Let us consider $\Lambda$CDM cosmology:
\be
\label{D3}
\frac{3}{\kappa^2} H^2 = \frac{3}{\kappa^2} H_0^2 + \rho_0 a^{-3}
= \frac{3}{\kappa^2} H_0^2 + \rho_0 a_0^{-3} \e^{-3N} \
\ee
where we have introduced the number of e-foldings $N=ln\frac{a}{a_0}$ instead of the cosmological time $t$, and we have assumed a pressureless fluid. We can easily solve the continuity equation for $\rho_m$:
 \be
 \dot{\rho}+3H(1+w_m)\rho=0\
 \label{D3a}
 \ee
while the first FLRW equation in Equation~\eqref{D2} can be rewritten in terms of the number of e-foldings, which yields \cite{Nojiri:2009kx}:
\bea
\label{D4}
0 &=& -9 G\left(N\left(R\right)\right)\left(4 G'\left(N\left(R\right)\right)
+ G''\left(N\left(R\right)\right)\right) \frac{d^2 F(R)}{dR^2}
+ \left( 3 G\left(N\left(R\right)\right)
+ \frac{3}{2} G'\left(N\left(R\right)\right) \right) \frac{dF(R)}{dR} \nn
&& - \frac{F(R)}{2}
+ \sum_i \rho_{i0} a_0^{-3(1+w_i)} \e^{-3(1+w_i)N(R)}\
\eea
where $G(N)=H^2$. Hence, for $\Lambda$CDM cosmology Equation~\eqref{D3}, the scalar curvature is given by $R=3 G'(N) + 12 G(N) = 12 H_0^2 + \kappa^2\rho_0 a_0^{-3} \e^{-3N}$, and the Equation \eqref{D4} gives a hypergeometric equation \cite{reconstruction3}:
\be
\label{D5}
0=3\left(R - 9H_0^2\right)\left(R - 12H_0^2\right) \frac{d^2 F(R)}{d^2 R}
 - \left( \frac{1}{2} R - 9 H_0^2 \right) \frac{d F(R)}{dR} - \frac{1}{2} F(R)\
\ee
whose solution is given by:
\be
\label{D6}
F(x) = A F(\alpha,\beta,\gamma;x) + B x^{1-\gamma} F(\alpha - \gamma + 1, \beta - \gamma + 1,
2-\gamma;x)\
\ee
where $x=\frac{R}{3H_0^2} - 3$, $\{A,\ B\}$ are integration constants, and the parameters are:
\be
\label{D7}
\gamma = - \frac{1}{2}\ ,\alpha + \beta = - \frac{1}{6}\ ,\quad \alpha\beta = - \frac{1}{6}\
\ee
Hence, the action Equation~\eqref{D6} is capable to reproduce $\Lambda$CDM model with no cosmological constant. Nevertheless, the action Equation~\eqref{D6} is composed by hypergeometric functions, which do not seem to provide a successful explanation for dark energy, since the theory turns out much complicated than the usual GR with a cosmological constant. However, other viable models of $F(R)$ gravity capable to mimic a cosmological constant at the present time have been suggested in the literature (see~\cite{Hu&Sawicki2007,Viable,Viable1,Viable2,Viable3}), which in general seem to cross the phantom barrier \cite{NuevoSaez}. Then, let us now study a model where the dominant component is phantom-like. Such kind of system can be easily expressed in the standard General Relativity when a phantom fluid is considered, where the FLRW equation reads $H^2(t)=\frac{\kappa^2}{3}\rho_\mathrm{ph}$  (the subscript $ph$ denotes the phantom nature of the fluid). As the EoS for the fluid is given by $p_\mathrm{ph}=w_\mathrm{ph}\rho_\mathrm{ph}$ with $w_\mathrm{ph}<-1$, by using the conservation equation, the solution yields $a(t)=a_0(t_s-t)^{-H_0}$, where $a_0$ is a constant, $H_0=-\frac{1}{3(1+w_\mathrm{ph})}$ and $t_s$ is the so-called Rip time. Then, the solution describes a universe that ends in a Big Rip singularity at the time $t_s$. The same behavior can be achieved in $F(R)$ theory with no need to introduce a phantom fluid. The expression for the Hubble parameter as a function of the number of e-foldings is given by $H^2(N)=H^2_0 \e^{2N/H_0}$. Then, the Equation (\ref{D4}), with no matter contribution, takes the form:
\be
R^2\frac{d^2F(R)}{dR^2}+AR\frac{dF(R)}{dR}+BF(R)=0\
\label{D8}
\ee
where $A=-H_0(1+H_0)$ and $B=\frac{(1+2H_0)}{2}$. This equation is an Euler
equation whose solution yields:
\be
F(R)=C_1R^{m_+}+C_2R^{m_-}\ , \quad \text{where} \quad m_{\pm}=\frac{1-A\pm\sqrt{(A-1)^2-4B}}{2}\
\label{D9}
\ee
Thus, the phantom cosmology $a(t)=a_0(t_s-t)^{-H_0}$ can be
also obtained in the frame of $F(R)$ theory and no phantom fluid is needed.

In addition, the equivalence between $f(R)$ gravity and scalar-tensor theories Equation~\eqref{In.3} can be used to  easily obtain cosmological solutions in $f(R)$ gravity. For the metric Equation~(\ref{D1}),  Friedmann equations read:
\bea
3H\frac{dP(\phi)}{dt}+3H^2P(\phi)+\frac{1}{2}Q(\phi)-\rho_m=0\  \nn
\frac{d^2P(\phi)}{dt^2}+2H\frac{dP(\phi)}{dt}+(2\dot{H}+3H^2)P(\phi)+\frac{1}{2}Q(\phi)+p_m=0\
\label{D10}
\eea
We redefine the scalar field such that it is chosen to be the time coordinate $\phi=t$. Hence, taking into account the Equations (\ref{D10}) and the continuity equation for the perfect fluid $\rho_m$, the Hubble parameter may be calculated as a function of the scalar field $\phi$, $H=g(\phi)$. By combining the Equation (\ref{D10}), the function $Q(\phi)$ is deleted, and it yields:
\be
\frac{d^2P(\phi)}{d\phi^2}-g(\phi)\frac{dP(\phi)}{d\phi}+2g'(\phi)P(\phi)+(1+w_m)\rho_{m0}\exp\left[-3(1+w_m)\int d\phi\ g(\phi) \right] =0\
\label{D11}
\ee
By solving this equation for a given function $P(\phi)$, a cosmological solution $H(t)$ is found, and the function $Q(\phi)$ is obtained by means of equations in Equation~(\ref{D10}):
\be
Q(\phi)=-6(g(\phi))^2P(\phi)-6g(\phi)\frac{dP(\phi)}{d\phi}+2\rho_{m0}\exp\left[-3(1+w_m)\int d\phi\ g(\phi) \right]\
\label{D12}
\ee
If we neglect the contribution of matter, then the Equation (\ref{D11}) is a first order differential equation on $g(\phi)$, and it can be easily solved. The solution found is the following \cite{reconstruction2}:
\be
g(\phi)=C\sqrt{P(\phi)}-\sqrt{P(\phi)}\int d\phi \frac{P''(\phi)}{2P^{3/2}(\phi)}\
\label{D13}
\ee
where $C$ is an integration constant. Hence, we have shown that $F(R)$ gravity represents a serious candidate to explain the phenomena of late-time acceleration in the expansion of the universe. In the following sections, we explore the thermodynamical properties of FLRW metrics in the frame of modified gravity as well as an attempt to generalize Cardy--Verlinde formula.


\section{First Law of Thermodynamics and FLRW Equations}
\label{ThermFLRW}

In this section, we review some of the most important results obtained in the study of FLRW metrics and their thermodynamics, specifically the relation between the first law of thermodynamics and FLRW equations, which can be deduced by assuming the well known relation between the entropy and the horizon area, as well as the relation between the gravity surface and the temperature~\cite{Cai:2005ra,Cai:2005ra1,Cai:2005ra2}. Such study has been extended not only to $f(R)$ gravity, as is also shown below (see ~\cite{Akbar:2006er}), but also to many other different scenarios \cite{Wu:2007se,Wu:2007se1,Wu:2007se2,Wu:2007se3,Wu:2007se4,Wu:2007se5,Wu:2007se6,Wu:2007se7,Wu:2007se8,Wu:2007se9,Wu:2007se10}. In fact, it was found in~\cite{Jacobson:1995ab} and extended in ~\cite{Elizalde:2008pv} for $f(R)$ gravities, a direct deduction of gravity field equations by assuming the relation of the horizon area and the entropy. Nevertheless, here we are only interested to show the deduction of FLRW equation from the thermodynamics defined in the corresponding horizons of a particular spacetime. Let us consider a $n+1$ FLRW metric Equation~\eqref{D1}, with arbitrary spatial curvature $k=-1,0,1$. This metric may be rewritten~as:
\be
ds^2=h_{ab}dx^adx^b+\tilde{r}^2d\Omega^2_{n-1}\
\label{D14}
\ee
where the indexes $a,b=0,1$, being $x^0=t$, $x^1=r$, and $\tilde{r}=a(t) r$, while $d\Omega^2_{n-1}$ expresses the line element of an $n-1$ unit sphere. Then, the apparent horizon for this general kind of spacetimes is defined as $h^{ab}\partial_a\tilde{r}\partial_b\tilde{r}=0$, whose radius yields:
\be
\tilde{r}_A=\frac{1}{\sqrt{H^2+k/a^2}}\
\label{D15}
\ee
As will be shown below, one can assign an entropy proportional to the area of the apparent horizon as well as an energy flux through the horizon. This is the reason for considering the apparent horizon instead of other horizons defined in FLRW spacetimes. We may define the work density and the energy flux vector as \cite{Cai:2005ra,Cai:2005ra1,Cai:2005ra2}:
\bea
w=-h_{ab}T^{ab}\ , \nn \\
\Psi_a=T^{b}_a\partial_b\tilde{r}+w\partial_a\tilde{r}\
\label{D16}
\eea
where the work $w$ refers to the change experimented by the apparent horizon, while the vector $\Psi_a$ refers to the flux of energy through the horizon. Then, the variation of the internal energy can be written as:
\be
\vec{\nabla} E=A\vec{\Psi}+w\vec{\nabla} V\
\label{D17}
\ee
where $A$ is the area and $V$  is the volume of an n-dimensional space with radius $\tilde{r}$. This equation represents the unified first law according to~\cite{Hayward:1997jp}. The first term in the r.h.s. in Equation~\eqref{D17} takes the form:
\be
A\vec{\Psi}=\frac{\kappa}{8\pi G}\vec{\nabla}A+\tilde{r}^{n-2}\vec{\nabla}\left(\frac{E}{\tilde{r}^{n-2}}\right)
\label{D18}
\ee
where $\kappa$ is the surface gravity:
\be
\kappa=\frac{1}{\sqrt{-h}}\partial_a\left(\sqrt{-h}h^{ab}\partial_b\tilde{r}\right)\
\label{D19}
\ee
Note that the energy is defined as:
\be
E=\frac{n(n-1)\Omega_n}{16\pi G}\tilde{r}^{n-2}(1-h^{ab}\partial_a\tilde{r}\partial_b\tilde{r})\
\label{D20}
\ee
Hence at the apparent horizon, the second term in the r.h.s. in Equation~\eqref{D18} vanishes, as on the black hole event horizon, so that the expression Equation~\eqref{D18} can be identified with the thermodynamical relation $dE=-TdS$ if the surface gravity is assumed to be the temperature $\kappa/2\pi$, and the entropy yields the usual expression, proportional to $A/4 G$. Now if we assume a perfect fluid filling the universe, with an energy-momentum tensor given by $T_{\mu\nu}=(\rho+p)u_{\mu}u_{\nu}+pg_{\mu\nu}$, the energy flux vector yields:
\be
\vec{\Psi}=\left(-\frac{1}{2}(\rho+p)H\tilde{r},\frac{1}{2}(\rho+p)a\right)\
\label{D21}
\ee
Then, the change of the internal energy  through the apparent horizon Equation~\eqref{D18} becomes:
\be
dE=A\Psi=-A(\rho+p)H\tilde{r}_Adt\
\label{D22}
\ee
Identifying the entropy and temperature associated to the apparent horizon,
\be
S=\frac{A}{4G}\ , \quad T=\frac{1}{2\pi \tilde{r}_A}\
\label{D23}
\ee
and using the first law of thermodynamics $dE=-TdS$, the second  FLRW equation is obtained:
\be
\dot{H}=-\frac{8\pi G}{n-1}(\rho+p)+\frac{k}{a^2}\
\label{D24}
\ee
By combining it with the continuity Equation \eqref{D3a},  the first FLRW equation is also obtained. Hence, FLRW equations can be deduced by using only thermodynamical arguments and identifying the entropy of the apparent horizon and its temperature, which is an identity that supports the idea of the relation between gravitational field equations and thermodynamical ones, as suggested by Jacobson in~\cite{Jacobson:1995ab}.

In  the same way, this analysis can be easily extended to $f(R)$ gravity as well as to other modified theories of gravity. Specifically, for $F(R)$ gravity, the entropy relation is defined (see previous sections) as~follows:
\be
S=\frac{AF'(R)}{4G}\
\label{D25}
\ee
Hence, following the same steps performed above for GR, the FLRW equations for $f(R)$ gravity Equation~\eqref{D2} are recovered \cite{Akbar:2006er}. Then, a strong relation between thermodynamics of the horizons and gravitational field equations is established. In the next section, another aspect of the relation of FLRW equations and thermodynamics is explored, in that case following the holographic principle through the entropy bounds and the Cardy--Verlinde formula.


\section{Generalization of Cardy--Verlinde Formula}
\label{CVformula}

This section is devoted to consideration of the Cardy--Verlinde
formula for general scenarios, and in particular for modified gravity. The Cardy--Verlinde formula was established by Verlinde in \cite{Verlinde} as a thermodynamical relation between FLRW equations  with a conformal fluid (a radiation-like fluid) and the Cardy formula (see \cite{Cardy}), and it was extended for a general perfect fluid in~\cite{Youm}. In~\cite{Brevik:2010jv}, attempts to reproduce the Cardy--Verlinde formula was suggested for different scenarios, including modified gravity.
Other aspects of the Cardy--Verlinde formula such as the presence of a viscous fluid have been studied \linebreak in~\cite{SDO-IB1,SDO-IB1a,SDO-IB1b}.
In this section, we mainly review those results. We consider a
$(n+1)$-dimensional spacetime described by the FLRW metric Equation~\eqref{D1}. By inserting the
metric Equation~(\ref{D1}) in the Einstein field equations, the FLRW equations for GR are derived:
\be
H^2 = \frac{16\pi G}{n(n-1)}\sum^m_{i=1}\rho_i-\frac{k}{a^2}\, ,\quad \dot{H} =
-\frac{8\pi G}{n-1}\sum^m_{i=1}(\rho_i + p_i) + \frac{k}{a^2}\
\label{3} \ee
Here $\rho_i={E_i}/{V}$ and $p_i$ are the
energy-density and pressure of the matter component $i$ that fills
the Universe. In this section, we consider only the $k=1$ closed
Universe. Moreover, we assume an equation of state (EoS) of the
form $p_i=w_i\rho_i$ with $w_i$ constant for each fluid, and
assume no interaction between the different components.
Then, the continuity Equation \eqref{D3a} has the form now $
\dot{\rho_i}+nH \left(\rho_i+p_i\right)=0$ and by
solving it, we find that the $i$ fluid depends on the scale factor as:
\be \rho_i\propto a^{-n(1+w_i)}\ \label{5} \ee
Let us now review the case of~\cite{Youm}, where just one fluid with
EoS $p=w\rho$ and $w= \text{constant}$ is considered. The total
energy inside the comoving volume $V$, $E=\rho V$, can be written
as the sum of an extensive part $E_\mathrm{E}$ and a subextensive
part $E_\mathrm{C}$, called the Casimir energy (for other ways to account
the Casimir energy in a FLRW universe, see~\cite{Casimir,Casimir1}), and  takes the
form:
\be E(S,V)=E_\mathrm{E}(S,V)+\frac{1}{2}E_\mathrm{C}(S,V)\
\label{6}
\ee
Under a rescaling of the entropy
($S\rightarrow\lambda S$) and the volume ($V\rightarrow\lambda
V$), the extensive and subextensive parts of the total energy
transform as:
\be E_\mathrm{E}(\lambda S,\lambda V) = \lambda
E_\mathrm{E}(S,V)\ , \quad E_\mathrm{C}(\lambda S,\lambda V) =
\lambda^{1-2/n} E_\mathrm{C}(S,V)\  \label{7} \ee
Hence, by assuming that the Universe satisfies the first law of
thermodynamics, the term corresponding to the Casimir energy
$E_\mathrm{C}$ can be seen as a violation of the Euler identity
according to the definition in~\cite{Verlinde}:
\be
E_\mathrm{C}=n(E+pV-TS) \  \label{8}
\ee
Since the total energy behaves as $E\sim a^{-nw}$ and by the definition Equation~(\ref{6}), the
Casimir energy also goes as $E_\mathrm{C}\sim a^{-nw}$. The FLRW
universe expands adiabatically, $dS=0$, so the products
$E_\mathrm{C}a^{nw}$ and $E_\mathrm{E}a^{nw}$ should be
independent of the volume $V$, and be just a function of the
entropy. Then, by the rescaling properties Equation~(\ref{7}), the
extensive and subextensive part of the total energy can be written
as functions of the entropy only \cite{Youm}:
\be E_\mathrm{E}=
\frac{\alpha}{4\pi a^{nw}}S^{w+1}\ , \quad
E_\mathrm{C}=\frac{\beta}{2\pi a^{nw}}S^{w+1-2/n}\  \label{9}
\ee
Here $\alpha$ and $\beta$ are undetermined constants. By combining
these expressions with Equation~(\ref{6}), the entropy of the Universe is
written as a function of the total energy $E$ and the Casimir
energy $E_\mathrm{C}$~\cite{Youm}:
\be S=\left(\frac{2\pi
a^{nw}}{\sqrt{\alpha\beta}}\sqrt{E_\mathrm{C}(2E-E_\mathrm{C})}
\right)^{\frac{n}{n(w+1)-1}}\  \label{10} \ee
which for $w=1/n$ (radiation-like fluid) reduces to \cite{Verlinde}:
\be
S=\frac{2\pi
a}{\sqrt{\alpha\beta}}\sqrt{E_\mathrm{C}(2E-E_\mathrm{C})}\
\label{19} \ee
which has the same form as the Cardy formula given
in~\cite{Cardy}. The first FLRW Equation (\ref{3}) can be
rewritten as a relation between thermodynamics variables, and
yields:
\be
S_\mathrm{H}=\frac{2\pi}{n}a\sqrt{E_\mathrm{BH}(2E-E_\mathrm{BH})}\
, \quad \mbox{where} \quad S_\mathrm{H}=(n-1)\frac{HV}{4G}\, ,
\quad E_\mathrm{BH}=n(n-1)\frac{V}{8\pi Ga^2}\ . \label{20} \ee
It is easy to check that for the bound proposed in~\cite{Verlinde}, $E_\mathrm{C}\leq E_\mathrm{BH}$, the
equation for the entropy Equation~(\ref{19}) coincides with the first FLRW
Equation (\ref{20}) when the bound is reached. We will see below
that when there are several fluid components, the same kind of
expression as in~\cite{Verlinde} cannot be found as well as its extension to modified gravity,
 nor is there the same correspondence with the FLRW equation when the bound is
saturated.

\subsection{Multicomponent Universe}

If $m$ fluids are considered with arbitrary EoS, $p_i=w_i\rho_i$,
the expression for the total entropy is simple to derive just by
following the same method as above. The total entropy is given by
the sum of the entropies for each fluid \cite{Brevik:2010jv}:
\be S=\sum_{i=1}^m S_i =
\sum_{i=1}^m \left(\frac{2\pi
a^{nw_i}}{\sqrt{\alpha\beta}}\sqrt{E_{iC}(2E_{i}-E_{iC})}
\right)^{\frac{n}{n(w_i+1)-1}}\  \label{11} \ee
This expression cannot be reduced to one depending only on the total energy unless
very special conditions on the nature of the fluids are assumed.
Let us   for simplicity  assume that  there are only two fluids with EoS given by $p_1=w_1\rho_1$ and
$p_2=w_2\rho_2$, $w_1$ and $w_2$ being constants. We can substitute the fluids by an effective fluid described by the EoS:
\be p_\mathrm{eff}=w_\mathrm{eff}\rho_\mathrm{eff}\ , \quad
\text{where} \quad w_\mathrm{eff}=
\frac{p_1+p_2}{\rho_1+\rho_2}=w_1+\frac{w_2-w_1}{1+\rho_1/\rho_2}\
\label{12} \ee
and $p_\mathrm{eff}=\frac{1}{2}(p_1+p_2)$, $\rho_\mathrm{eff}=\frac{1}{2}(\rho_1+\rho_2)$. Then, by using the
energy conservation Equation \eqref{D3a}, we find $\rho_1\sim
(a/a_0)^{-n(1+w_1)}$ and $\rho_2\sim (a/a_0)^{-n(1+w_2)}$, where $a_0$ is assumed to be the value of the scale factor at the time $t_0$. The effective
EoS parameter $w_\mathrm{eff}$ can be expressed as a function of
the scale factor $a(t)$:
\be
w_\mathrm{eff}=w_1+\frac{w_2-w_1}{1+(a/a_0)^{n(w_2-w_1)}}\  \label{13}
\ee
The total energy inside a volume $V$ becomes: \be E_T=E_1+E_2
\propto (a/a_0)^{-nw_1}+(a/a_0)^{-nw_2}\  \label{16} \ee
As the energy is proportional to two different powers of the scale factor
$a$, it is not possible to write it as a function of
the total entropy only. As a special case, if the EoS parameters are $w_1=w_2=w_\mathrm{eff}$, the formula for the entropy
reduces to Equation~(\ref{10}), and coincides with the CV formula  when $w_\mathrm{eff}=1/n$.

As another  case, one might consider that for some epoch of the cosmic history,  $w_1 \gg w_2$. Taking also   $a>>a_0$, we could then approximate  the total energy  by the
function $E_T \propto a^{-nw_2}$. From  Equation~(\ref{6}) the
Casimir energy would also depend on the same power of $a$,
$E_\mathrm{C}\propto a^{-nw_2}$. The expression Equation~(\ref{10}) is again recovered with $w=w_2$.

Thus in general, when a multicomponent FLRW Universe is assumed, the formula for the total entropy does
not resemble the Cardy formula, nor does it correspond to the FLRW
equation when the Casimir bound is reached.  It becomes possible to
reconstruct the formula Equation~(\ref{19}), and establish the correspondence with the
Cardy formula, only if we make specific choices for the EoS of the fluids.

\subsection{Inhomogeneous EoS Fluid and Modified Gravity}

Let us now explore the case of an $n+1$-dimensional Universe filled with
a fluid satisfying an inhomogeneous EoS. This kind of EoS,  generalizing  the perfect fluid model, has been
considered in several papers as a way to describe effectively the
DE (see \cite{InhEoS,InhEoS1}).
We assume an EoS expressed as a function of the scale factor:
\be
p=w(a)\rho+g(a)\
\label{28}
\ee
This EoS fluid could be taken to correspond to  modified gravity,
or to bulk viscosity \cite{InhEoS,InhEoS1}. By introducing Equation~(\ref{28})
in the energy conservation Equation~(\ref{D3a}) we obtain:
\be
\rho'(a)+\frac{n(1+w(a))}{a}\rho(a)=-n\frac{g(a)}{a}\
\label{29}
\ee
Here we have performed a variable change $t=t(a)$ such that
the prime over $\rho$ denotes derivative with respect to the
scale factor $a$. The general solution of this equation is:
\be
\rho(a)=\e^{-F(a)}\left(C -n\int\e^{F(a)}\frac{g(a)}{a}da\right)
\quad \text{where} \quad F(a)=\int^a \frac{1+w(a')}{a'}da'\
\label{30}
\ee
and $C$ is an integration constant. As shown above,
only for some special choices of the functions $w(a)$ and $g(a)$ can
the formula Equation~(\ref{19}) be recovered. Let us assume, as  an example,  that
$w(a)=-1$ and $g(a)=-a^m$, with $m$ being constant.
Then, the energy density behaves as $\rho\propto a^m$. Hence, by following the same steps as described above, the
extensive and subextensive energy go as $a^{m+n}$, and by imposing
conformal invariance and the rescaling properties Equation~(\ref{7}), we
calculate the dependence on the entropy to be:
\be
E_\mathrm{E}=
\frac{\alpha}{4\pi na^{-(m+n)}}S^{-m/n}\, , \quad
E_\mathrm{C}=\frac{\beta}{4\pi na^{-(m+n)}}S^{-(2n+m/n)}\
\label{31}
\ee
The expression for the entropy is easily
constructed by combining these two expressions and substituting
the extensive part by the total energy. This gives us the same expression as in Equation~(\ref{10}) with $w=-(n+m)/n$. Note that for
$m=-(1+n)$, the formula Equation~(\ref{19}) is recovered and also its
correspondence with the CFT formula. However for a generic power
$m$, the CV formula cannot be reconstructed, like
the cases studied above. Only for some special choices does the correspondence
work, leading to the identification between the FLRW equation and
the Cardy formula.

In the same way, we may reconstruct Cardy--Verlinde formula for modified gravity, and specifically for $f(R)$ gravity. Note that $f(R)$ gravity can be seen as GR plus an inhomogeneous fluid. For a closed
$3+1$ FLRW Universe, the modified FLRW equations are expressed now as:
\bea
\frac{1}{2}F(R)-3(H^2+\dot{H})F'(R)+3HF''(R)\dot{R}=\kappa^2
\rho_{m}\  \nn
-\frac{1}{2}F(R)+\left[3H^2+\dot{H}+\frac{2}{a^2}\right] F'(R)-[(\partial_{tt}F'(R))+2H(\partial_t F'(R))]=\kappa^2 p_m\
\label{F3} \eea
where  primes denote derivatives respect to $R$ and  dots with
respect to $t$. These equations can be rewritten in order to be
comparable with those of standard GR, similarly as in Equation~\eqref{D2}, by shifting terms to the right side of Equation~\eqref{F3}. For such a propose, the
geometric terms can be presented as an effective energy-density
$\rho_{F(R)}$ and a pressure $p_{F(R)}$:
\bea
\rho_{F(R)}=\frac{1}{\kappa^2 F'(R)}\left[\frac{RF'(R)-F(R)}{2}-3H\dot{R}F''(R)\right]
\  \nn
p_{F(R)}\frac{1}{\kappa^2 F'(R)}\left[\dot{R}^2F'''(R)+2H\dot{R}F''(R)+\ddot{R}F''(R)
+\frac{1}{2}(F(R)-RF'(R)) \right]\  \label{F4} \eea
Then, an EoS
for the geometric terms can be defined as
$p_{F(R)}=w_{F(R)}\rho_{F(R)}$. We can define an effective
energy-density $\rho=\rho_m/F'(R)+\rho_{F(R)}$ and pressure
$p=p_m/F'(R)+p_{F(R)}$. Hence, for some special cases the formula
for the entropy developed above can be obtained in
$F(R)$-gravity (for an early attempt deriving a  CV formula in a
specific version of $F(R)$-gravity, see \cite{vanzo}). For
example, for an $F(R)$ whose solution gives $\rho\propto
a^{-3(1+w_\mathrm{eff})}$, the formula for the entropy Equation~(\ref{10})
is recovered although in general, as in the cases studied above,
no such expression can be given. On the other hand, one could
assume that the geometric terms do not contribute to the matter
sector. Supposing a constant EoS matter fluid, the expression for
the entropy is given by Equation~(\ref{10}), although the cosmic Cardy
formula Equation~(\ref{20}) does not have the same form, and in virtue of the
modified first FLRW Equation (\ref{F3}) the form of the Hubble
entropy $S_\mathrm{H}$, the total energy $E$ and the Bekenstein
energy $E_\mathrm{BH}$ will be very different. It is not easy to
establish correspondence between two such approaches. Note that
using the effective fluid representation the generalized CV
formula may be constructed for any modified gravity.

Let us now consider de Sitter space solution in $F(R)$-gravity
(for review of CV formula in dS or AdS spaces, see \cite{cai,cai1}). As
was pointed in~\cite{Cognola}, almost every function
$F(R)$ admits a de Sitter solution. This can be easily seen from
the first FLRW equation in Equation~(\ref{F3}). A de Sitter solution is
given by a constant Hubble parameter $H(t)=H_0$; then by inserting
in Equation~(\ref{F3}) we obtain the following algebraic equation:
 \be
3H^2_0=\frac{F(R_0)}{2F'(R_0)}\  \label{F5}
\ee
Here $R_0=12H^2_0$ and the contribution of matter is neglected. Then,
for positive roots  $H_0$ of this equation, the corresponding
$F(R)$ leads to the de Sitter solution which may describe
inflation or DE. In this case the formula for the entropy
Equation~(\ref{10}) can be reproduced for $w=-1$, and even the universal
bound Equation~(\ref{B1}) can hold by taking a critical size of the
Universe. The formula that relates the cosmic bounds in
Equation~(\ref{B3a}) is easily obtained also in $F(R)$-gravity for a de
Sitter solution. In such a case one can identify:
\be
S_\mathrm{H}=\frac{H_0V}{2G}\ , \quad S_\mathrm{B}=\frac{a
V}{24G}\frac{F(R_0)}{F'(R_0)}\ , \quad
S_\mathrm{BH}=\frac{V}{2Ga}\  \label{F6}
\ee
which corresponds to the first FLRW equation written as
$S^2_\mathrm{H}+(S_\mathrm{B}-S_\mathrm{BH})^2=S^2_\mathrm{B}$.
Thus, one can conclude that dynamical entropy bounds are not
violated for modified gravity with de Sitter solutions. Note that
quantum gravity effects may be presented also as an effective
fluid contribution. In case when de Sitter space turns out to be
the solution, even with the account of quantum gravity the above
results indicate that dynamical cosmological/entropy bounds are
valid. In other words, the argument indicates the universality of
dynamical bounds. It seems that their violation is caused only by
future singularities if they are not cured by quantum gravity
effects. Note that a large number of modified gravity theories do
not contain  future singularities; they are  cured by higher
derivatives terms.


\section{On the Cosmological Bounds Near Future Singularities}
\label{Singularities}

In~\cite{Verlinde}, Verlinde proposed a new universal bound on
cosmology based on a restriction of the Casimir energy $E_C$; cf.
 his entropy formula Equation~(\ref{19}). This new bound postulated was:
\be
E_\mathrm{C}\leq E_\mathrm{BH}\
\label{B1}
\ee
where $E_\mathrm{BH}=n(n-1)\frac{V}{8\pi Ga^2}$. It was
deduced by the fact that in the limit when the Universe passes
between strongly and weakly self-gravitating regimes, the
Bekenstein entropy $S_\mathrm{B}=\frac{2\pi a}{n}E$ and the
Bekenstein--Hawking entropy $S_\mathrm{BH}=(n-1)\frac{V}{4Ga}$,
which define each regime, are equal. This bound could be
interpreted to mean that the Casimir energy  never  becomes  able
to reach sufficient energy, $E_\mathrm{BH}$, to form a black hole
of the size of the Universe. It is easy to verify that the strong
($Ha\geq 1$) and weak ($Ha\leq 1$) self-gravity regimes have the
following restrictions on the total energy:
\bea
E &\leq &
E_\mathrm{BH}\ \quad \text{for} \quad Ha\leq 1 \  \nn
E &\geq &
E_\mathrm{BH}\ \quad \text{for} \quad Ha\geq 1\
\label{B2}
\eea
From here it is easy to calculate the bounds on the entropy of
the Universe in the case when the Verlinde formula Equation~(\ref{19}) is
valid; this is (as  shown in the above sections) for an
effective radiation dominated Universe $w_\mathrm{eff}\sim 1/n$.
The bounds for the entropy deduced in~\cite{Verlinde} for
$k=1$ are:
\bea
S &\leq& S_\mathrm{B}\ \quad \text{for} \quad Ha\leq 1 \  \nn
S &\geq& S_\mathrm{H}\ \quad \text{for} \quad Ha\geq 1\
\label{B3}
\eea
where $S_\mathrm{B}$ is the Bekenstein entropy defined
above, and $S_\mathrm{H}$ is the Hubble entropy given by
Equation~(\ref{20}).
Note that for the strong self-gravity regime, $Ha\geq 1$,
the energy range is $E_\mathrm{C}\leq E_\mathrm{BH}\leq E$.
According to the formula Equation~(\ref{19}) the maximum entropy is
reached when the bound is saturated, $E_\mathrm{C}=E_\mathrm{BH}$.
Then $S=S_\mathrm{H}$, such that the FLRW equation coincides
with the CV formula, thus indicating  a connection with
CFT.  For the weak regime, $Ha\leq 1$, the range of energies
goes as $E_\mathrm{C}\leq E\leq E_\mathrm{BH}$ and the maximum
entropy is reached earlier, when $E_\mathrm{C}=E$, yielding the
result $S=S_\mathrm{B}$. The entropy bounds can be extended to more
general cases, corresponding to an arbitrary EoS parameter $w$. By taking
the bound Equation~(\ref{B1}) to be universally valid, one can easily
deduce the new entropy bounds for each regime, from the expression
of the entropy Equation~(\ref{10}). These
new bounds, discussed in~\cite{Youm}, differ from the ones given in Equation~(\ref{B3}), but
still establish a bound on the entropy as long as the bound on
$E_\mathrm{C}$ expressed in Equation~(\ref{B1}) is taken to be valid. The
entropy bounds can be related through  the first FLRW equation,
yielding the following quadratic expression (for $k=1$):
\be
S^2_\mathrm{H}+(S_\mathrm{B}-S_\mathrm{BH})^2=S^2_\mathrm{B}\
\label{B3a}
\ee
Here, we would like to review
what happens to the bounds,  particularly to the fundamental
bound Equation~(\ref{B1}), when the cosmic evolution is close to a future
singularity (see~\cite{Brevik:2010jv}); then the effective fluid  dominating the
cosmic evolution could have an unusual EoS. As  shown below,
for some class of future singularities, such a bound could soften
the singularities in order to avoid  violation of the
\textit{universal} bound Equation~(\ref{B1}). It  could be interpreted
to mean that  quantum effects become important when the bound is reached.
 However, as the violation of the bound could happen long before the singularity
even in the presence of  quantum effects, it could be a signal
of  breaking of the universality of the bound Equation~(\ref{B1}). Let
us first of all give a list of the possible future cosmic
singularities, which can be classified according to~\cite{ConformalAnomaly1} as:
\begin{itemize}
\item Type I (``Big Rip''): For $t\rightarrow t_s$, $a\rightarrow
\infty$ and $\rho\rightarrow \infty$, $|p|\rightarrow \infty$.
\item Type II (``Sudden''): For $t\rightarrow t_s$, $a\rightarrow
a_s$ and $\rho\rightarrow \rho_s$, $|p|\rightarrow \infty$.
\item Type III: For $t\rightarrow t_s$, $a\rightarrow a_s$
and $\rho\rightarrow \infty$, $|p|\rightarrow \infty$.
\item Type IV: For $t\rightarrow t_s$, $a\rightarrow a_s$ and
$\rho\rightarrow \rho_s$, $p \rightarrow p_s$
but higher derivatives of Hubble parameter diverge (see~\cite{Shtanov:2002ek}).
\end{itemize}
Note that the above list was suggested for the case of a flat FLRW
Universe (see~\cite{singularity, singularity1, singularity2, singularity3, singularity4}).
As we consider in this section a closed Universe ($k=1$), we should
make an analysis to see
if the list of singularities given above is also valid in this case.
It is straightforward to see that all the singularities listed above
can be reproduced for a particular choice of the effective EoS.
To show how the cosmic bounds behave for each type of singularity, we
could write an explicit
solution of the FLRW equations, expressed as a function of time
depending on free parameters
that will be fixed for each kind of singularity. Then, the Hubble
parameter may be written as follows:
\be
H(t)=\sqrt{\frac{16\pi G}{n(n-1)}\rho -\frac{1}{a^2}}=H_1(t_s-t)^m
+H_0\
\label{B4}
\ee
where $m$ is a constant properly chosen for each type of singularity.
Note that this is just a solution that ends in the singularities
mentioned above, but there are other solutions which also reproduce
such singularities.  We will study how the cosmic bounds behave near
each singularity listed above. As pointed out in~\cite{Elizalde:2004mq,phantom1,phantom2,phantom3,phantom4,phantom5,phantom6,phantom7,phantom8,phantom9,phantom10,phantom11,phantom12,phantom13,phantom14,phantom15,phantom16,phantom17,phantom18,phantom19,phantom20,ConformalAnomaly1}, around a singularity
 quantum effects could become important as the curvature of the
Universe grows and diverges in some of the cases. In other words, approaching the finite-time future singularity the curvature grows and universe reminds the early universe where quantum gravity effects are dominant ones because of extreme conditions.
Then one has to take into account the role of such quantum gravity effects which should define the behavior of the universe just before the singularity. Moreover, they may act so as to prevent the singularity occurrence. In a sense, one sees the return of quantum gravity era.
However, the consistent quantum gravity theory does not exist so far.
Then, in order to estimate the influence of quantum effects to universe near to singularity, one can use the effective action formulation.
We will apply the effective action produced by conformal anomaly (equivalently, the effective fluid with pressure/energy-density corresponding to conformal anomaly ones) because of several reasons.
It is known that at high energy region (large curvature) the conformal invariance is restored so one can neglect the masses. Moreover, one can use large N approximation to justify why large number of quantum fields may be considered as effective quantum gravity. Finally, in the account of quantum effects via conformal anomaly we keep explicitly the graviton (spin 2) contribution. The conformal anomaly $T_A$ has the
following well-known form:
\be
T_A=b\left(F+\frac{2}{3}\square R\right) +b'G+b''\square R\
\label{B4a}
\ee
Here we assume for simplicity a $3+1$ dimensional spacetime. Then, $F$
is the square of a 4D Weyl tensor and $G$ is the Gauss--Bonnet
invariant:
\be
F=\frac{1}{3}R^2-2R_{ij}R^{ij}+R_{ijkl}R^{ijkl}\ ,\quad
G=R^2-4R_{ij}R^{ij}+R_{ijkl}R^{ijkl}\,
\label{B4b}
\ee
The coefficients $b$ and $b'$ in Equation~(\ref{B4a}) are described by the
number of $N$ scalars, $N_{1/2}$ spinors, $N_{1}$ vector fields,
$N_2$ gravitons and $N_\mathrm{HD}$ higher derivative conformal
scalars.
They can be written as:
\be
b=\frac{N+6N_{1/2}+12N_{1}+611N_{2}-8N_\mathrm{HD}}{120(4\pi)^2}\
,\quad
b'=-\frac{N+11N_{1/2}+62N_{1}+1411N_{2}-28N_\mathrm{HD}}{360(4\pi)^2}\
\label{B4c}
\ee
As $b''$ is arbitrary it can be shifted by a finite
renormalization of the local counterterm. The conformal anomaly $T_A$
can be written as $T_A=-\rho_A+3p_A$, where $\rho_A$ and $p_A$ are
the energy and pressure densities respectively. By using Equation~(\ref{B4a})
and the energy conservation equation $\rho_A+3H(\rho_A+p_A)=0$, one
obtains the following expression for $\rho_A$
\cite{ConformalAnomaly1,ConformalAnomaly2}:
\begin{eqnarray}
\rho_A &=& -\frac{1}{a^4}\int dt a^4 HT_A \nn
&=& -\frac{1}{a^4}\int dt
a^4H\left[-12b\dot{H}^2+24b'(-\dot{H}^2+H^2\dot{H}+H^4)
� \right.\nonumber\\
&&\left.-(4b+6b'')(\dddot{H}+7H\ddot{H}+4\dot{H}^2+12H^2\dot{H}) \right]
\label{B4d}
\end{eqnarray}
The quantum corrected FLRW equation is given by:
\be
H^2=\frac{8\pi G}{3}(\rho+\rho_A)-\frac{1}{a^2}\
\label{B4e}
\ee
We study now how the bounds behave around the singularity in
the classical case when no quantum effects are added, and then
include the conformal anomaly Equation~(\ref{B4a}) quantum effects in the
FLRW equations. We will see that for some cases the violation of the
cosmic bound can be avoided.

\subsection{Big Rip Singularity}

This type of singularity has been very well studied and has become very popular
as it is a direct consequence in the majority of the cases when the
effective EoS parameter is less than $-1$, the so-called phantom case \cite{Elizalde:2004mq,phantom1,phantom2,phantom3,phantom4,phantom5,phantom6,phantom7,phantom8,phantom9,phantom10,phantom11,phantom12,phantom13,phantom14,phantom15,phantom16,phantom17,phantom18,phantom19,phantom20}.
Observations currently indicate that the phantom barrier could have
already been   or it will be crossed in the near future, so a lot of attention has
been paid to this case. It can be characterized by the solution
Equation~(\ref{B4}) with $m\leq -1$, and this yields
 the following dependence of the total energy density on the scale
factor near the singularity, when $a\gg 1$, for a closed Universe ($k=1$):
\be
\rho=\frac{n(n-1)}{16\pi G}H^2+\frac{1}{a^2}\sim a^{-n(1+w)} \quad
\text{for} \quad t\rightarrow t_s\
\label{B5}
\ee
where we have chosen $H_1=2/n|1+w|$ with $w<-1$, $m=-1$ and $H_0=0$
for clarity. This solution drives the Universe to a Big Rip singularity for
$t\rightarrow t_s$, where the scale factor diverges. If the singularity takes place, the
bound Equation~(\ref{B1}) has to be violated before this happens. This can be seen from
Equation (\ref{B5}), as the Casimir energy behaves as $E_\mathrm{C}\propto a^{n|w|}$ while
the Bekenstein--Hawking energy goes as $E_\mathrm{BH}\propto a^{n-2}$.
Then, as $w<-1$, the Casimir energy grows faster than the BH energy. Therefore,
 close to the singularity where the scale factor becomes very big, the value of
 $E_\mathrm{C}$ will be much larger than $E_\mathrm{B}$, thus violating
the bound Equation~(\ref{B1}).
Following the postulate from~\cite{Verlinde} one could interpret
the bound Equation~(\ref{B1})
as the limit where General Relativity and Quantum Field Theory
converge, such that when the bound is saturated quantum gravity effects should
become important. QG corrections could help to avoid the violation of the bound and may
be the Big Rip singularity occurrence.
As this is just a postulate based on the CV formula, which is only
valid for special cases as  shown in the sections above, the bound on $E_\mathrm{C}$
could not be valid for any kind of fluid.

Let us now include the conformal anomaly Equation~(\ref{B4a}) as a quantum effect that
 becomes important around the Big Rip. In such a case there is a phase
transition and the Hubble evolution will be given by the solution of
the FLRW Equation (\ref{B4e}). Let us  approximate to get some
qualitative results,  assuming $3+1$ dimensions.
Around $t_s$ the curvature is large, and
$|\rho_A|>>(3/\kappa^2)H^2+k/a^2$. Then  $\rho\sim-\rho_A$, and from
Equation~(\ref{B4d})we get:
\be
\dot{\rho}+4H\rho=H\left[-12b\dot{H}^2+24b'(-\dot{H}^2+H^2\dot{H}+H^4)
�-(4b+6b'')(\dddot{H}+7H\ddot{H}+4\dot{H}^2+12H^2\dot{H}) \right]\
\label{B5a}
\ee
We assume that the energy density, which diverges in the
classical case, behaves now as:
\be
\rho\sim (t_s-t)^{\lambda}\
\label{B5b}
\ee
where $\lambda$ is some negative number. By using the energy
conservation equation $\dot{\rho}+3H(1+w)\rho=0$, the Hubble
parameter goes as $H\sim 1/(t_s-t)$. We can check if this
assumption is correct in the presence of  quantum effects by
inserting both results in Equation~(\ref{B5a}). We get:
\be
\rho\sim 3H^4(-13b+24b')\
\label{B5c}
\ee
Hence as $b>0$ and $b'<0$,  $\rho$ becomes negative,
which is an unphysical result. Thus  $\rho$ should not
go to infinity in the presence of the quantum correction. This
is the same result as obtained in~\cite{ConformalAnomaly1} where
numerical analysis showed that the singularity is moderated by the
conformal anomaly, so that the violation of the bound that naturally
occurs in the classical case can be avoided/postponed when quantum
effects are included.

Now we consider the case where $f(R)$ has the form:
\be
\label{FBB1}
f(R) \sim R^\alpha \
\ee
when the curvature is small or large.
Then if the matter has an EoS parameter $w>-1$, by solving
(\ref{F3}) we find:
\be
\label{FBB2}
a \sim \left\{ \begin{array}{cl}
t^{\frac{2\alpha}{n(1+w)}} & \mbox{when}\ \frac{2\alpha}{n(1+w)}> 0
\\
(t_s - t)^{\frac{2\alpha}{n(1+w)}} & \mbox{when}\
\frac{2\alpha}{n(1+w)}< 0
\end{array} \right.
\ee
Then there may appear a singularity at $t=0$, which corresponds to
the Big Bang singularity,
or at $t=t_s$, which corresponds to the Big Rip singularity.
Since the Casimir energy behaves as $E_\mathrm{C} \sim a^{-nw}$ but
$E_\mathrm{BH} \sim a^{n-2}$,
only when $t\to 0$, $E_\mathrm{C}$ dominates in case that $n> 2$ and
$w\geq 0$ or in case that $n \geq 2$ and $w>0$.
Even in the phantom phase where $\frac{2\alpha}{n(1+w)}< 0$, the
bound Equation~(\ref{B1}) is not violated.

\subsection{Sudden Singularity}

This kind of singularity is also problematic with respect to the bounds.
Nevertheless, as long as the energy density $\rho$ does not diverge,
the violation of the bound may be avoided for some special choices.
The sudden singularity can be described
by the solution Equation~(\ref{B4}) with $0<m<1$, and constants $H_{0,1}>0$.
Then the scale factor goes as:
\be
a(t)\propto \exp\left[ -\frac{H_1}{m+1}(t_s-t)^{m+1}+H_0t\right] \
\label{B6}
\ee
which gives $a(t)\sim \e^{h_0t}$ (de Sitter) close to $t_s$. From the
first FLRW equation the total energy density~becomes:
\be
\rho=H^2(t)+\frac{1}{a^2}=\left[H_1(t_s-t)^m+H_0 \right]^2
+ \exp\left[ 2\frac{H_1}{m+1}(t_s-t)^{m+1}-2H_0t\right]\
\label{B6a}
\ee
which tends to a constant $\rho\sim H_0^2+\e^{-2H_0t_s}$ for
$t\rightarrow t_s$.
Then the Casimir energy grows as $E_\mathrm{C}\propto H^2_0
a^{n}+a^{n-2}$, while $E_\mathrm{BH}\propto a^{n-2}$ close to $t_s$.
The BH energy grows slower than the Casimir energy, and the bound is
violated for a finite $t$.
However, by a specific choice of the coefficients, the violation of
the bound Equation~(\ref{B1}) could be avoided.
For $H_0=0$, and by some specific coefficients, the bound could be
obeyed.
In general, it is very possible that $E_\mathrm{C}$ exceeds
its bound. In the presence of quantum corrections, the singularity
can be avoided but the bound can still be violated, depending on the
free parameters for each model. We may assume that in the presence of
the conformal anomaly for $n=3$, the energy density grows as
\cite{ConformalAnomaly1}:
\be
\rho=\rho_0+\rho_1(t_s-t)^{\lambda}\
\label{B6b}
\ee
where $\rho_0$ and $\rho_1$ are constants, and $\lambda$ is now a
positive number. Then the divergences on the higher derivatives of
the Hubble parameter can be avoided, as  is shown in~\cite{ConformalAnomaly1}. Nevertheless, $E_\mathrm{C}$ still
grows faster than $E_\mathrm{BH}$, such that the Universe has to be
smaller than a critical size in order to hold the bound Equation~(\ref{B1}) as
 is pointed in~\cite{Youm} for the case of a vacuum dominated
universe.

\subsection{Type III Singularity}

This type of singularity is very similar to the Big Rip, in spite of
the scale factor $a(t)$ being finite at the singularity.
The solution Equation~(\ref{B4}) reproduces this singularity by taking
$-1<m<0$. The scale factor goes as:
\be
a(t)= a_s\exp \left[-\frac{H_1}{m+1}(t_s-t)^{m+1} \right]\
\label{B7}
\ee
where for simplicity we take $H_0=0$. Then, for $t\rightarrow t_s$,
the scale factor $a(t)\rightarrow a_s$.
To see how $E_\mathrm{C}$ behaves near the singularity, let us write
it in terms of the time instead of the scale factor:
\be
E_\mathrm{C}\propto a_s^nH_1^2 (t_s-t)^{2m}+ a_s^{n-2}\
\label{B8}
\ee
where $m<0$.
Hence, the Casimir energy diverges at the singularity, while
$E_\mathrm{BH}\propto a_s^{n-2}$ takes a finite value
for the singularity time $t_s$, so the bound is clearly violated long
before the singularity.
Then, in order to maintain the validity of the bound Equation~(\ref{B1}), one
might assume, as in the Big Rip case,
that GR is not valid near or at the bound. Even if  quantum
effects are included, as was pointed in~\cite{ConformalAnomaly1},
 for this type of singularity the energy density diverges more
rapidly than in the classical case, so that the bound is also
violated in the presence of  quantum effects.

\subsection{Type IV Singularity}

For this singularity, the Hubble rate behaves as:
\be
\label{III_1}
H = H_1(t) + \left(t_s - t\right)^\alpha H_2(t)\
\ee
Here $H_1(t)$ and $H_2(t)$ are regular function and do not vanish at
$t=t_s$. The constant $\alpha$ is not integer and larger than $1$.
Then the scale factor behaves as
\be
\label{III_2}
\ln a(t) \sim \int dt H_1(t) + \int dt \left(t_s - t\right)^\alpha
H_2(t)\
\ee
Near $t=t_s$, the first term dominates and all quantities like
$\rho$, $p$, and $a$ \emph{etc}. are finite and
therefore the bound Equation~(\ref{B1}) would not be violated near the
singularity.

\subsection{Big Bang Singularity}

When the matter with $w\geq 0$ coupled with gravity and dominates,
the scale factor behaves as
\be
\label{BB1}
a \sim t^{\frac{2}{n(1+w)}}\
\ee
Then there appears a singularity at $t=0$, which may be a Big Bang
singularity.
Although the Big Bang singularity is not a future singularity, we
may consider the bound Equation~(\ref{B1}) when $t\sim 0$.
Since $n(1+w)>2$, the energy density behaves as
$\rho \sim a^{-n(1+w)}$ and therefore the Casimir energy behaves as
$E_\mathrm{C} \sim a^{-nw}$. On the other hand, we find
$E_\mathrm{BH} \sim a^{n-2}$.
Then when $n> 2$ or when $n \geq 2$ and $w>0$, $E_\mathrm{C}$
dominates when $a\to 0$, that is, when $t\to 0$,
and the bound Equation~(\ref{B1}) is violated. This tells us, as expected, that
 quantum effects become important in the early universe.

Thus, in the above we have explored what happens near the future cosmic
singularities. We have seen that in general, and with some very special exceptions
in the case of Type II and Type IV,
the bound will be violated if one assumes the validity of GR close to
the singularity. Even if  quantum corrections are assumed, it seems that the bound will  be violated, although in the Big Rip case the singularity may be avoided when quantum effects are incorporated.
It is natural to suggest, in accordance with Verlinde, that
the bound on the Casimir energy means a finite range for the validity of the classical theory. When this kind of theory becomes saturated,
 some other new quantum gravity effects
have to be taken into account. Hence, the universality of the bound Equation~(\ref{B1}) is
not clear and may hold just
for some specific cases, like the radiation dominated Universe. Furthermore, even in absence of a future singularity, a possibility allowed also for the case that $w<-1$ (see~\cite{Sahni:2002dx}),  the bound may be violated, specially for the case of a ``Little Rip'' (see~\cite{Frampton:2011sp}), a stage of the universe when the expansion would be strong enough to break some binding systems such as Solar System. That scenario, absence of singular points, can be very possible for a lot of DE models, and also in modified gravity (see~\cite{Nojiri:2011kd,LopezRevelles:2012cg,12071646})

\section{Conclusions}
\label{Conclusions}

In this work we have considered both static spherically symmetric
solutions and massive charged and spinning  solutions in
$f(R)$ theories of gravity in a vacuum situation within the metric formalism.
The first case was analyzed in arbitrary dimensions, whereas for the second
we focused on four dimensions.

We first discussed the constant curvature case (including charged
black holes in four dimensions) in static and spherically symmetric configurations.
Then we studied the general case without imposing, a priori, the condition of constant curvature. For the uncharged case, we tackled this calculation through a
perturbative analysis around the EH case, which
made it possible to study those solutions being regular in the
perturbative parameter. We have reproduced the explicit expressions
up to second order for the metric coefficients, which only give rise to
constant curvature (Schwarzschild AdS) solutions as in the EH
case \cite{Dombriz_PRD_BH,Dombriz_PRD_BH1}.

With regards to the massive, charged and spinning BH for $f(R)$ gravity, we have derived
the metric tensor that describes such objects. We presented how this solution differs from
that found by Carter~\cite{Carter} by a proper redefinition of the spin, electric charge and vacuum scalar curvature and therefore that  
this type of solutions are supported in this kind of modified gravity theories.
Further study of the \linebreak Kerr--Newman metric allowed us to describe
the different black-hole-type configurations derived from the presence
of different horizons: usual, {\it extremal} and {\it marginal extremal} black holes, {\it naked singularities} and {\it naked extremal singularities}.
%


On the other hand, we have also calculated thermodynamical
quantities for the AdS as well as for
KN configurations. We paid special attention to
the issue of the stability of these kinds of solutions.
As a very remarkable result, we have presented that the
condition for a $f(R)$ theory of gravity to support these kinds of
black holes is given by $R_0+f(R_0)<0$ where $R_0$ is the constant
curvature of the AdS space-time. In fact, this condition also implies that the effective
Newton constant is positive and consequently
the graviton does not become a ghost.
By employing the Euclidean action method, we revised how
the mass, the energy or the entropy of these BH (AdS and KN) differ from those predicted in
GR just by a multiplicative factor $1+f'(R_0)$. This factor has to be positive in order to guarantee a positive mass and entropy for these kinds of BH.

For the AdS configuration, we sketched the fact that
the qualitative thermodynamic behavior of the AdS BH in $f(R)$ gravity theories
is the same as the one found by Hawking and Page.
On the other hand, with regards to the KN configurations, the analysis of the BH heat capacity revealed how two different
types of BH can be distinguished: {\it fast} and {\it slow}, showing the latter two phase transitions.
With respect to the horizon structure in the KN configuration,
we have shown how horizons can only exist for values of the spin lower than a maximum value
$a_{max}$, and that from a certain positive value of the curvature onward, only above a minimum value $a_{min}$.

Then, for two paradigmatic $f(R)$ models, we have investigated the stability of the different possible configurations
that arise from the values of the free energy and the heat capacity, quantities that depend on the particular $f(R)$ model
under study. We have analyzed explicitly the rich thermodynamical phenomenology that characterizes these gravitational
models with a complete set of different figures that were originally presented in \cite{Dombriz_PRD_BH,Dombriz_PRD_BH1} and \cite{Jimeno_BH,Cembranos:2012ji}.
For doing so, we have studied the parameter regions in which BH in such models are locally stable and globally preferred.

%

%
%
Furthermore, we have investigated the deep connection between FLRW equations and the first law of thermodynamics. The study of FLRW metrics is a very important issue as the universe seems to be well described by this kind of metrics, where eventually one should assume extra components to explain the cosmological evolution. However, as suggested above, modified gravity can even avoid the need to introduce dark energy, and even a kind of inflaton, to explain the whole cosmological evolution. Hence, FLRW metrics and their connection to thermodynamics turns out to be a fundamental issue. Then, as suggested in~\cite{Cai:2005ra,Cai:2005ra1,Cai:2005ra2}, FLRW equations can be derived by assuming the entropy relation with the area of the apparent horizon. Even more, such a relation can be extended to $f(R)$ gravity when one assumes the modified relation for the entropy of the horizon.

%
%

In addition, we have also reviewed the possible extension of generalized CV formula for
multicomponent fluids, generalized in the sense that
an inhomogeneous EoS (including modified $f(R)$ gravity) was assumed. For some special cases the formula
is reduced to the standard CV formula expressing the
correspondence with 2d CFT theory. The dynamical entropy bound for
all above cases was found. The universality of dynamical entropy
bound near  all four types of the future singularity, as well as
the initial Big Bang singularity, was investigated. Note that the study of future singularities is a crucial step since a lot of dark energy models contain some kind of singularity. In fact, so-called viable modified gravities usually cross the phantom barrier $(w<-1)$ (see~\cite{Bamba:2011dk}), and although it seems that future singularities do not occur,  the ``Little Rip'' event may take place (see \cite{NuevoSaez}).
 It was proved that except for some special cases of Type II and Type IV
singularity, the dynamical entropy bound is violated near the
singularity. Taking into account  quantum effects of conformally
invariant matter does not improve the situation. Hence, the dynamical
entropy bound may not be universal and  its
violation simply indicates that the situation will be changed with the
introduction of  quantum gravity effects.


This deep connection between gravitation and thermodynamics in so many scenarios, as the ones reviewed here in extended gravitational theories and others (see~\cite{Padmanabhan:2009vy}), may suggest that gravity could be just an emergent force as suggested in~\cite{Verlinde:2010hp}, which may imply the birth of new physics that may lead to the search of a more fundamental theory.

Experimental checks to test the validity of a particular $f(R)$ model can be applied in multiple realms of astrophysics. Among others, future data from CMB tensor perturbations  \cite{CMB_Dombriz} as well as gravitational collapse predictions \cite{Collapse_Dombriz_1, Collapse_Dombriz_2, Collapse_Dombriz_3} and the growth of cosmological perturbations (and subsequent structure formation)
\cite{Dombriz_perturbaciones_PRD,Dombriz_perturbaciones_PRD1,Amare,Comment} may constrain the available range of parameters for viable models \cite{Oyaizu:2008tb,Schmidt:2008tn}. In addition, the study of the weak lensing in $f(R)$ gravity can also provide complementary constraints on the models and free parameters as pointed out in~\cite{Schmidt:2008hc}.
The general viability conditions for $f(R)$ theories prior to model specifications are summarized in reference \cite{silvestri}.
Furthermore, the study of star structure and evolution in modified gravity theories may lead additional and independent tests for this class of fourth order theories of gravity (see~\cite{Capozziello:2011nr}). Finally, the aforementioned constraints 
may be enlarged not only by studying astrophysical BH stability but also if quantum gravity scale is accessible at particle accelerators. For example, it has been shown that LHC could produce about one {\it microBH} per second \cite{microBH,microBH1}. Stability and thermodynamical properties of the produced BH might shed some light about the underlying theory of gravity.


\section*{Acknowledgments}
\vspace{12pt}

AdlCD acknowledges financial support from URC and NRF (South Africa), MICINN (Spain) project numbers FIS2011-23000, FPA2011-27853-C02-01 and Consolider-Ingenio MULTIDARK \linebreak CSD2009-00064.

DSG acknowledges support from  a postdoctoral contract from the University of the Basque Country (UPV/EHU). DSG is also supported by the research project FIS2010-15492, and also by the Basque Government through the special research action KATEA and UPV/EHU under program UFI 11/55.


\begin{thebibliography}{99}

\bibitem{SIa} Riess,  A.G.;  Filippenko, A.V.;  Challis, P.; Clocchiattia, A.;  Diercks, A.;  Garnavich, P.M.;  Gilliland, R.L.;  Hogan, C.J.;  Jha, S.;  Kirshner, R.P.; {\it et al}. {Observational evidence from supernovae for an accelerating universe and a cosmological constant}. {\it Astron. J.} \textbf{1998}, \emph{116}, 1009--1039.
\bibitem{SIa1}Perlmutter, S.;  Aldering, G.;  Goldhaber, G.;  Knop, R.A.;  Nugent, P.;  Castro, P.G.;  Deustua,~S.;  Fabbro, S.;  Goobar, A.;  Groom, D.E.; {\it et al}. {Measurements of Omega and Lambda from 42 high redshift supernovae}. {\it Astrophys. J.} \textbf{1999}, \emph{517}, 565--586.
\bibitem{SIa2} Tonry, J.L.;  Schmidt, B.P.;  Barris, B.;  Candia, P.;  Challis, P.;  Clocchiatti, A.;  Coil, A.L.;  Filippenko, A.V.;  Garnavich, P.;  Hogan, C.; {\it et al}. {Cosmological results from high-z supernovae}. {\it Astrophys. J.} \textbf{2003}, \emph{594}, 1--24.
\bibitem{inflac}Guth,  A.H. {The inflationary universe: A possible solution to the horizon and flatness problems}. {\it Phys.\ Rev.\ D\ }  \textbf{1981}, \emph {23}, 347--356.
\bibitem{Peebles}  Peebles, P.J.E. {\it Principles of Physical Cosmology}; Princeton University Press: Princeton, NJ, USA, 1993.
\bibitem{Komatsu}  Komatsu, E.;  Smith, K.M.;  Dunkley, J.;  Bennett, C.L.; Gold, B.;  Hinshaw, G.;  Jarosik, N.;  Larson,~D.;  Nolta, M.R.;  Page, L.; {\it et al}. {Seven-Year Wilkinson Microwave Anisotropy Probe (WMAP) observations: Cosmological interpretation} . {\it Astrophys.\ J.\ Suppl.\ } \textbf{2011}, \emph {192},  arXiv:1001.4538[gr-qc].
\bibitem{Percival}   Percival, W.J.;  Reid, B.A.;  Eisenstein, D.J.; Bahcall, N.A.;  Budavari, T.;  Frieman, J.A.; Fukugita,~M.; Gunn, J.E.; Ivezic, Z.; Knapp, G.R.; {\it et al}. {Baryon acoustic oscillations in the sloan digital sky survey data release 7 galaxy sample}. {\it Mon. Not. Roy. Astron. Soc.} \textbf{2010}, \emph{401}, 2148--2168.
\bibitem{Riess}  Riess,  A.G.;  Macri, L.;  Casertano, S.;  Sosey, M.;  Lampeitl, H.;  Ferguson, H.C.;  Filippenko, A.V.;  Jha, S.W.;  Li, W.;  Chornock, R.; {\it et al}. { A redetermination of the hubble constant with the hubble space telescope from a differential distance ladder}. {\it Astrophys. J.} \textbf{2009}, {\emph 699}, 539--563.
\bibitem{DE}
 Biswas, T.; Cembranos, J.A.R.;  Kapusta, J.I. { 	
Thermal duality and hagedorn transition from p-adic strings}. {\it Phys.\ Rev.\ Lett.}  \textbf{2010}, \emph {104}, 021601--021604. 
\bibitem{DE1} Biswas, T.;  Cembranos, J.A.R.;  Kapusta, J.I. { 	
Thermodynamics and cosmological constant of non-local field theories from p-adic strings}. \emph{J. High Energy Phys.} \textbf{2010}, {\it 1010}, doi:10.1007/JHEP10(2010)048.
\bibitem{DE2}  Biswas, T.;  Cembranos, J.A.R.;  Kapusta, J.I. { 	
Finite temperature solitons in non-local field theories from p-adic strings}. {\it Phys.\ Rev.\  D} \textbf{2010}, \emph {82}, arXiv:1006.4098[gr-qc]. 
\bibitem{cosmoproblema} Weinberg, S. {The cosmological constant problem.} {\it Rev.\ Mod.\ Phys.} \textbf{1989}, \emph {61}, 1--23.
\bibitem{varios}
Nojiri, S.;  Odintsov, S.D. { 	
Modified gravity with ln R terms and cosmic acceleration}. {\it Gen. Rel. Grav.} \textbf{2004}, \emph {36}, 1765--1780.
\bibitem{varios1}
 Carroll, S.M.;  Duvvuri, V.;  Trodden, M.;  Turner, M.S. { 	
Is cosmic speed-up due to new gravitational physics?} {\it Phys. Rev. D
} \textbf{2004}, \emph {70}, doi:10.1103/PhysRevD.70.043528.
\bibitem{varios3}
 Dvali, G.;  Gabadadze, G.;  Porrati, M. {
4-D gravity on a brane in 5-D Minkowski space}. \emph{Phys. Lett. B} \textbf{2000}, \emph {485}, 208--214. \bibitem{varios4}
Cembranos, J.A.R. {Dark matter from R2-gravity}. \emph{Phys.\ Rev.\ Lett.} \textbf{2009}, {\it 102}, doi:10.1103/PhysRevLett.102.141301. 
\bibitem{varios5}Cembranos, J.A.R.
{QCD effects in cosmology}. \emph{AIP Conf.\ Proc}. \textbf{2009}, {\it 1182}, 288--291. 
\bibitem{varios6}Cembranos, J.A.R.
{$R^2$ dark matter}. \emph{J.\ Phys.\ Conf.\ Ser.} \textbf{2011}, {\it 315}, doi:10.1088/1742-6596/315/1/012004. 
\bibitem{varios7}
Cembranos, J.A.R. {
The Newtonian limit at intermediate energies}. \emph{Phys. Rev. D} \textbf{2006}, {\it 73}, doi:10.1103/PhysRevD.73.064029. 
\bibitem{varios8}
Cembranos, J.A.R.; Olive, K.A.; Peloso, M.; Uzan, J.P. {Quantum corrections to the cosmological evolution of conformally coupled fields}. \emph{J. Cosmol. Astropart. Phys.} \textbf{2009}, {\it 0907}, doi:10.1088/1475-7516/2009/07/025. 
\bibitem{varios9}
 Beltr\'an, J.;  Maroto, A.L. {Cosmic vector for dark energy}. \emph{Phys. Rev. D} \textbf{2008}, {\it 78}, doi:10.1103/PhysRevD.78.063005.
\bibitem{varios10} Beltr\'an, J.; Maroto, A.L. {
Cosmological electromagnetic fields and dark energy}. \emph{J. Cosmol. Astropart. Phys.} \textbf{2009}, {\it 0903}, doi:10.1088/1475-7516/2009/03/016.
\bibitem{varios11} Beltr\'an, J.;  Maroto, A.L. { 	
Cosmological evolution in vector-tensor theories of gravity}. \emph{Phys. Rev. D} \textbf{2009}, {\it 80}, 063512--063533.
\bibitem{varios12} Beltr\'an, J.;  Maroto, A.L. {
Dark energy: The Absolute electric potential of the universe}. \emph{Int. J. Mod. Phys. D} \textbf{2009}, {\it 18}, 2243--2248.

\bibitem{Dombriz-Saez}  De la Cruz-Dombriz, A.; S\'aez-G\'omez, D. {On the stability of the cosmological solutions in $f(R,G)$ gravity}. \textbf{2011},
  arXiv:1112.4481 [gr-qc].

  \bibitem{Reviews}   Clifton, T.; Ferreira, P.G.; Padilla, A.; Skordis, C. { 	
Modified gravity and cosmology}. \emph{Phys.\ Rept.} \textbf{2012}, {\it 513}, 1--189.   
\bibitem{Reviews1} Nojiri, S.;  Odintsov, S.D. { 	
 Introduction to modified gravity and gravitational alternative for dark energy}. \emph{Int. J. Geom. Meth. Mod. Phys.} \textbf{2007}, \emph{4}, 115--146.  
\bibitem{Reviews2} Nojiri, S.;  Odintsov, S.D. { 	
Dark energy, inflation and dark matter from modified F(R) gravity}. \textbf{2008}, arXiv: 0807.0685 [gr-qc].
\bibitem{Reviews3} Capozziello, S.;  Francaviglia, M. { 	
Extended theories of gravity and their cosmological and astrophysical applications}. \emph{Gen. Rel. Grav.} \textbf{2008}, \emph{40}, 357--420.
\bibitem{Reviews4} Sotiriou, T.P.;  Faraoni, V. { 	
f(R) theories of gravity}. \textbf{2010}, arXiv: 0805.1726 [gr-qc].
\bibitem{Reviews5} Lobo, F.S.N. { 	
The dark side of gravity: Modified theories of gravity}. \textbf{2008}, arXiv: 0807.1640 [gr-qc].
\bibitem{Reviews6} Capozziello, S.; Faraoni, V. {Beyond Einstein Gravity}. In \emph{Fundamental Theories of Physics Volume 170}; Springer: Dordrecht, The Netherlands, 2011.
\bibitem{Reviews7}S\'aez-G\'omez, D. {On Friedmann-Lema\^{\i}tre-Robertson-Walker cosmologies in non-standard gravity}. PhD Thesis, University of Barcelona, Barcelona, Spain, 2011, arXiv:1104.0813 [hep-th].
\bibitem{Capozziello:2002rd}
  Capozziello, S. {Curvature quintessence}. \emph{Int.\ J.\ Mod.\ Phys.\ D} \textbf{2002}, {\it 11}, doi:10.1142/S0218271802002025.
\bibitem{Capozziello:2002rd1}  Capozziello, S.; Carloni, S.; Troisi, A. {Quintessence without scalar fields}. \emph{Recent Res.\ Dev.\ Astron.\ Astrophys.} \textbf{2003}, {\it 1}, arXiv:astro-ph/0303041.



\bibitem{Odintsov} Nojiri, S.; Odintsov, D. {Unified cosmic history in modified gravity: From F(R) theory to Lorentz non-invariant models}. \emph{Phys. Rept.} \textbf{2011}, {\it 505}, 59--144.

\bibitem{reconstruction3}
De la Cruz-Dombriz, A.; Dobado, A.
{A $f(R)$ gravity without cosmological constant}.
\emph{Phys.\ Rev. D} \textbf{2006}, {\it 74}, 087501--087504. 

\bibitem{Dunsby1}Goheer, N.; Larena, J.; Dunsby, P.K.S. { 	
Power-law cosmic expansion in f(R) gravity models}. \emph{Phys.\ Rev.\ D} \textbf{2009}, {\it 80}, 061301--061304.

\bibitem{Tsujikawa}
Abdelwahab, M.; Goswami, R.; Dunsby, P.K.S. { 	
Cosmological dynamics of fourth order gravity: A compact view}.  \emph{Phys.\ Rev.\ D} \textbf{2012}, {\it 85}, 083511--083517. 
\bibitem{Tsujikawa1}
Carloni, S.; Goswami, R.; Dunsby, P.K.S. {	
A new approach to reconstruction methods in f(R) gravity}. \emph{Class.\ Quant.\ Grav.} \textbf{2012}, {\it 29}, doi:10.1088/0264-9381/29/13/135012. 
\bibitem{Tsujikawa2}
Capozziello, S.; de Laurentis, M. {Extended theories of gravity}.
  \emph{Phys.\ Rept.} \textbf{2011}, {\it 509}, 167--321.
\bibitem{varia}
Nzioki, A.M.; Dunsby, P.K.S.; Goswami, R.; Carloni, S. {Geometrical approach to strong gravitational lensing in f(R) gravity}. \emph{Phys.\ Rev.\ D} \textbf{2011}, {\it 83}, 024030--024039.
\bibitem{varia1}
Abebe, A.; Goswami, R.; Dunsby, P.K.S. { 	
On shear-free perturbations of f(R) gravity}. \emph{Phys.\ Rev.\ D} \textbf{2011}, {\it 84}, 124027--124033.  
\bibitem{varia2}
Sotiriou, T.P. {The Nearly Newtonian regime in non-linear theories of gravity}. \emph{Gen.\ Rel.\ Grav.} \textbf{2006}, {\it 38}, 1407--1417.
\bibitem{varia3}
Faraoni, V. { 	
Solar system experiments do not yet veto modified gravity models}. \emph{Phys.\ Rev.\  D} \textbf{2006}, {\it 74}, 023529--023535.
\bibitem{varia4}
Nojiri, S.; Odintsov, S.D. { 	
Modified f(R) gravity consistent with realistic cosmology: From matter dominated epoch to dark energy universe}. \emph{Phys.\ Rev. D} \textbf{2006}, {\it 74}, 086005--086017.


\bibitem{Dombriz_perturbaciones_PRD} De la Cruz-Dombriz A., Dobado A. and Maroto, A.L. { 	
On the evolution of density perturbations in f(R) theories of gravity Cosmological density perturbations in modified gravity theories}. \emph{Phys.\ Rev. D} \textbf{2008}, {\it 77}, 123515--123523.
 \bibitem{Dombriz_perturbaciones_PRD1}
De la Cruz-Dombriz, A.; Dobado, A.; Maroto, A.L. {Cosmological Density Perturbations in Modified Gravity Theories}.  In Proceedings of the AIP Conference, Salamanca, Spain, September 2008; Volume 1122, p.~252.

\bibitem{Amare} Abebe, A.; Abdelwahab, M.; de la Cruz-Dombriz, A.; Dunsby, P.K.S. {Covariant gauge-invariant perturbations in multifluid f(R) gravity}.
  \emph{Class.\ Quant.\ Grav.} \textbf{2012}, {\it 29}, doi:10.1088/0264-9381/29/13/135011.   

\bibitem{Comment} De la Cruz-Dombriz, A.; Dobado, A.;  Maroto, A.L. {Comment on `Viable singularity-free f(R) gravity without a cosmological constant'}. \emph{Phys.\ Rev.\ Lett.\ D} \textbf{2009}, {\it 103}, 179001.

\bibitem{cvetic}
  Cvetic, M.; Nojiri, S.; Odintsov, S.D. {Black hole thermodynamics and negative entropy in de Sitter and anti-de Sitter Einstein-Gauss-Bonnet gravity}.
  \emph{Nucl.\ Phys. B} \textbf{2002}, {\it 628}, 295--330.

\bibitem{Cai_GaussBonet_AdS}  Cai,  R.G. Gauss-Bonnet black holes in AdS spaces. \emph{Phys.\ Rev. D} \textbf{2002}, {\it 65},  084014--084022.

\bibitem{Cho} Cho, Y.M.; Neupane, I.P. AntiÐde Sitter black holes, thermal phase transition, and holography in higher curvature gravity. \emph{Phys.\ Rev.\  D} \textbf{2002}, {\it 66}, 024044--024059.

\bibitem{Matyjasek}
Cai, R.G. {A note on thermodynamics of black holes in Lovelock gravity}. \emph{Phys.\ Lett. B } \textbf{2004}, {\it 582}, 237--242.
\bibitem{Matyjasek1}
Matyjasek, J.; Telecka, M.; Tryniecki, D. {Higher dimensional black holes with a generalized gravitational action}. \emph{Phys.\ Rev. D} \textbf{2006}, {\it 73},  124016--124024.


\bibitem{Horava}Park, M. {The black hole and cosmological solutions in IR modified Ho\v{r}ava gravity}. \emph{J. High. Energy Phys.} \textbf{2009}, {\it 9}, doi:10.1088/1126-6708/2009/09/123.
\bibitem{Horava1} Lee, H.W.;  Kim, Y.W.;  Myung, Y.S. {Extremal black holes in the Horava-Lifshitz gravity}. \emph{Eur.\ Phys.\ J.\ C} \textbf{2010}, {\it 68}, 255--263.
\bibitem{Horava2} Castillo, A.;  Larranaga, A. { 	
Entropy for black holes in the deformed Ho\v{r}ava-lifshitz gravity}. \emph{Electron.\ J.\ Theor.\ Phys.} \textbf{2011}, {\it 8}, 1--10.

\bibitem{Wang:2011xf}
  Wang, T.
  {Static solutions with spherical symmetry in f(T) theories}.
  \emph{Phys.\ Rev.\ D} \textbf{2011}, {\it 84}, 024042--024051.

\bibitem{Whitt} Whitt, B. {Fourth order gravity as general relativity plus matter}.\emph{  Phys.\ Lett. B} \textbf{1984}, {\it 145}, 176--178.

\bibitem{Mignemi} Mignemi, S.; Wiltshire, D.L. {Black holes in higher derivative gravity theories}. \emph{Phys.\ Rev. D} \textbf{1992}, {\it 46}, 1475--1506.

\bibitem{Multamaki} Multamaki, T.; Vilja, I. { 	
Spherically symmetric solutions of modified field equations in f(R) theories of gravity}. \emph{Phys.\ Rev. D} \textbf{2006}, {\it 74}, 064022--064026.

\bibitem{olmo} Olmo, G.J. {Limit to general relativity in f(R) theories of gravity}. \emph{Phys.\ Rev. D } \textbf{2007}, {\it 75}, 023511--023518.

\bibitem{Nzioki:2009av} Nzioki, A.M.; Carloni, S.; Goswami, R.; Dunsby, P.K.S.
{A New framework for studying spherically symmetric static solutions in f(R) gravity}.
\emph{Phys.\ Rev.\ D} \textbf{2010}, {\it 81}, 084028--084038. 

\bibitem{Taeyoon1} Moon, T.; Myung, Y.S.; Son, E.J.
  {f(R) black holes}.
  \emph{Gen.\ Rel.\ Grav.} \textbf{2011}, {\it 43}, arXiv:1101.1153 [gr-qc]. 
\bibitem{Capozziello:2009jg} Capozziello, S.; de Laurentis, M.; Stabile, A. {Axially symmetric solutions in f(R)-gravity}.
\emph{Class.\ Quant.\ Grav.} \textbf{2010}, {\it 27}, doi:10.1088/0264-9381/27/16/165008. 
\bibitem{Myung}  Myung, Y.S. { 	
Instability of rotating black hole in a limited form of f(R) gravity}. \emph{Phys.\ Rev.\ D} \textbf{2011}, {\it 84}, 024048--024053.
\bibitem{Palatini_Noether}  Vollick, D.N. { 	
Noether charge and black hole entropy in modified theories of gravity}. \emph{Phys.\ Rev. D} \textbf{2007}, {\it 76}, 124001--124005.

\bibitem{Cognola} Cognola, G.; Elizalde, E.; Nojiri, S.; Odintsov, S.D.; Zerbini, S. {One-loop f(R) gravity in de Sitter universe}. \emph{J. Cosmol. Astropart. Phys.} \textbf{2005}, {\it 502}, doi:10.1088/1475-7516/2005/02/010.

\bibitem{Hawking&Page} Hawking, S.W.; Page, D.N. { 	
Thermodynamics of black holes in anti-de sitter space}.  \emph{Commun.\ Math.\ Phys.} \textbf{1983}, {\it 87},
 577--588.

\bibitem{Witten} Witten,  E. {Anti-de sitter space, thermal phase transition, and confinement in gauge theories}. \emph{Adv.\ Theor.\ Math.\ Phys.} \textbf{1998}, {\it 2}, 505--532.

\bibitem{Briscese} Briscese, F.; Elizalde, E. { 	
Black hole entropy in modified gravity models}.  \emph{Phys.\ Rev. D} \textbf{2008}, {\it 77}, 044009--011013.


\bibitem{Myung:2011ih}
  Myung, Y.S.; Moon, T.; Son, E.J.
 {Stability of f(R) black holes}.
  \emph{Phys.\ Rev.\ D} \textbf{2011}, {\it 83}, 124009--124015.


\bibitem{PerezBergliaffa:2011gj}   Perez Bergliaffa, S.E.; de Oliveira Nunes,  Y.E.C. {Static and spherically symmetric black holes in $f(R)$ theories}.
  \emph{Phys.\ Rev.\ D} \textbf{2011}, {\it 84}, 084006--084009.   


\bibitem{Moon:2011sz}
  Moon, T.; Myung, Y.S.; Son, E.J. {Stability analysis of f(R)-AdS black holes}.
  \emph{Eur.\ Phys.\ J.\ C} \textbf{2011}, {\it 71}, doi:10.1140/epjc/s10052-011-1777-0.

\bibitem{Nelson:2010ig}
  Nelson, W.
{Static Solutions for 4th order gravity}.
  \emph{Phys.\ Rev.\ D} \textbf{2010}, {\it 82}, 104026--104036.

  \bibitem{Larranaga:2011fv}
  Larranaga, A.
  {A rotating charged black hole solution in f(R) gravity}.
  \emph{Pramana} \textbf{2012}, {\it 78}, 697--703.


\bibitem{Myung:2011we}
  Myung, Y.S.
  {Instability of rotating black hole in a limited form of $f(R)$ gravity}.
  \emph{Phys.\ Rev.\ D} \textbf{2011}, {\it 84}, 024048--024053.


\bibitem{Hendi:2012nj}
  Hendi, S.H.; Momeni, D.
 {Black hole solutions in F(R) gravity with conformal anomaly}.
  \emph{Eur.\ Phys.\ J.\ C} \textbf{2011}, {\it 71}, 10.1140/epjc/s10052-011-1823-y.

 \bibitem{Jacobson:1995ab}
  Jacobson, T.
  {Thermodynamics of space-time: The Einstein equation of state}.
  \emph{Phys.\ Rev.\ Lett.} \textbf{1995}, {\it 75}, 1260--1263. 

\bibitem{Elizalde:2008pv}
  Elizalde, E.; Silva, P.J.
 {F(R) gravity equation of state}.
  \emph{Phys.\ Rev.\ D} \textbf{2008}, {\it 78}, 061501--061504. 

\bibitem{Cai:2005ra}
Hayward, S.A.; Mukohyama, S.; Ashworth, M.C.
 {Dynamic black hole entropy}.
  \emph{Phys.\ Lett.\ A} \textbf{1999}, {\it 256}, 347--350. 
%
\bibitem{Cai:2005ra1}
Bak, D.; Rey, S.J.
 {Cosmic holography}.
  \emph{Class.\ Quant.\ Grav.} \textbf{2000}, {\it 17}, doi:10.1088/0264-9381/17/15/101. 
%
\bibitem{Cai:2005ra2}
    Cai, R.G.; Kim, S.P. {First law of thermodynamics and Friedmann equations of Friedmann-Robertson-Walker universe}.
  \emph{J. High Energy Phys.} \textbf{2005}, {\it 0502}, doi:10.1088/1126-6708/2005/02/050. 
%
    \bibitem{Akbar:2006er}
  Akbar, M.; Cai, R.G.
  {Friedmann equations of FLRW universe in scalar-tensor gravity, f(R) gravity and first law of thermodynamics}.
  \emph{Phys.\ Lett.\ B} \textbf{2006}, {\it 635}, 7--10. 

  \bibitem{Wu:2007se}
  Wu, S.F.; Wang, B.; Yang, G.H.
  {Thermodynamics on the apparent horizon in generalized gravity theories}.
  \emph{Nucl.\ Phys.\ B} \textbf{2008}, {\it 799}, 330--344. 
  \bibitem{Wu:2007se1}
  Bamba, K.; Geng, C.Q.
 {Thermodynamics of cosmological horizons in $f(T)$ gravity}.
  \emph{J. Cosmol. Astropart. Phys.} \textbf{2011}, {\it 1111}, doi:10.1088/1475-7516/2011/11/008. 
  \bibitem{Wu:2007se2}
  Radicella, N.; Pavon, D.
 {The generalized second law in universes with quantum corrected entropy relations}.
  \emph{Phys.\ Lett.\ B} \textbf{2010}, {\it 691}, 121--126. 
  \bibitem{Wu:2007se3}
  Cao, Q.J.; Chen, Y.X.; Shao, K.N.
 {Clausius relation and Friedmann equation in FLRW universe model}.
 \emph{J. Cosmol. Astropart. Phys.} \textbf{2010}, {\it 1005}, doi:10.1088/1475-7516/2010/05/030. 
  \bibitem{Wu:2007se4}
  Cai, R.G.; Ohta, N.
  {Horizon thermodynamics and gravitational field equations in Ho\v{r}ava-lifshitz gravity}.
  \emph{Phys.\ Rev.\ D} \textbf{2010}, {\it 81}, 084061--084068. 
  \bibitem{Wu:2007se5}
  Bamba, K.; Geng, C.Q.
  {Thermodynamics in F(R) gravity with phantom crossing}.
  \emph{Phys.\ Lett.\ B} \textbf{2009}, {\it 679}, 282--287. 
  \bibitem{Wu:2007se6}
  Sheykhi, A.; Wang, B.
  {The Generalized second law of thermodynamics in Gauss-Bonnet braneworld}.
  \emph{Phys.\ Lett.\ B} \textbf{2009}, {\it 678}, 434--437. 
  \bibitem{Wu:2007se7}
  Zhu, T.; Ren, J.R.; Li, M.F.
 {Influence of generalized and extended uncertainty principle on thermodynamics of FLRW universe}.
  \emph{Phys.\ Lett.\ B } \textbf{2009}, {\it 674}, 204--209. 
  \bibitem{Wu:2007se8}
  Cai, R.G.
 {Thermodynamics of apparent horizon in brane world scenarios}.
  \emph{Prog.\ Theor.\ Phys.\ Suppl.} \textbf{2008}, {\it 172}, 100--109. 
  \bibitem{Wu:2007se9}
  Akbar, M.; Cai, R.G.
  {Friedmann equations of FLRW universe in scalar-tensor gravity, f(R) gravity and first law of thermodynamics}.
  \emph{Phys.\ Lett.\ B} \textbf{2006}, {\it 635}, 7--10. 
  \bibitem{Wu:2007se10}
  Cai, R.G.; Cao, L.-M.; Hu, Y.P.
 {Corrected entropy-area relation and modified friedmann equations}.
  \emph{J. High Energy Phys.} \textbf{2008}, {\it 808}, doi:10.1088/1126-6708/2008/08/090. 

  \bibitem{Cardy}
Cardy, J.L. {Operator content of two-dimensional conformally invariant}. \emph{Nucl.\ Phys.\ B} \textbf{1986}, {\it 270}, 186--204.

 \bibitem{Verlinde}Verlinde, E. {On the holographic principle in a radiation dominated universe}. \textbf{2000}, arXiv: hep-th/0008140.

\bibitem{Youm} 
Youm, D.
{A note on the Cardy-Verlinde formula}.
\emph{Phys.\ Lett.\ B} \textbf{2002}, {\it 531}, 276--280.

\bibitem{Brevik:2010jv}
  Brevik, I.; Nojiri, S.; Odintsov, S.D.; S\'aez-G\'omez, D.
  {Cardy-Verlinde formula in FLRW Universe with inhomogeneous generalized fluid and dynamical entropy bounds near the future singularity}.
  \emph{Eur.\ Phys.\ J.\ C} \textbf{2010}, {\it 69}, 563--574.

 \bibitem{ScalarFR}
Nojiri and, S.; Odintsov, S.D. { 	
Modified gravity with negative and positive powers of the curvature: Unification of the inflation and of the cosmic acceleration}.
 \emph{Phys.\ Rev.\ D} \textbf{2003}, {\it 68}, 123512--123521. 

\bibitem{Dombriz_thesis}
  De la Cruz Dombriz, A.
 {Some cosmological and astrophysical aspects of modified gravity theories}, PhD Thesis, Complutense University of Madrid, Madrid, Spain, 2010,
  arXiv:1004.5052 [gr-qc].

  \bibitem{khoury}
   Khoury, J.; Weltman, A.
   {Chameleon fields: Awaiting surprises for tests of gravity in space}.
   \emph{Phys.\ Rev.\ Lett.} \textbf{2004}, {\it 93}, 171104--171107.
\bibitem{Khoury:2003rn}
  Khoury, J.; Weltman, A.
   {Chameleon cosmology}.
   \emph{Phys.\ Rev. D} \textbf{2004}, {\it 69}, 044026--044040.

\bibitem{Nojiri:2007as}
  Nojiri, S.; Odintsov, S.D.
  {Unifying inflation with LambdaCDM epoch in modified f(R) gravity consistent with Solar System tests}.
  \emph{Phys.\ Lett.\ B} \textbf{2007}, {\it 657}, 238--245.


 \bibitem{Hu&Sawicki2007}Hu, W.; Sawicki, I. {Models of f(R) cosmic acceleration that evade solar-system tests}. \emph{Phys.\ Rev.\ D} \textbf{2007}, {\it 76}, 064004--064016.

 \bibitem{Viable}
   Nojiri, S.; Odintsov, S.D.
  {Modified f(R) gravity unifying R**m inflation with Lambda CDM epoch}.
  \emph{Phys.\ Rev.\ D} \textbf{2008}, {\it 77}, 026007--026013. 
 \bibitem{Viable1}
     Pogosian, L.; Silvestri, A.
  {The pattern of growth in viable f(R) cosmologies}.
  \emph{Phys.\ Rev.\ D} \textbf{2008}, {\it 77}, 023503--023517. 
 \bibitem{Viable2}
Capozziello, S.; Tsujikawa, S.
 {Solar system and equivalence principle constraints on f(R) gravity by chameleon approach}.
  \emph{Phys.\ Rev.\ D} \textbf{2008}, {\it 77}, 107501--107504. 
 \bibitem{Viable3}
Starobinsky, A.A. {Disappearing cosmological constant in f(R) gravity}.
  \emph{J. Exp. Theor. Phys. Lett.} \textbf{2007}, {\it 86}, 157--163. 

\bibitem{ortin} Ort\'in, T. \emph{Gravity and Strings}; Cambridge University Press: Cambridge, UK, 2003.

\bibitem{Dombriz_PRD_BH} De~la~Cruz-Dombriz, A.; Dobado, A.; Maroto, A.L. {Black Holes in f(R) theories}.  \emph{Phys.\ Rev. D} \textbf{2009}, {\it 80}, 124011--124023. [Erratum: \emph{Phys.\ Rev.\  D} \textbf{2011}, {\it 83}, 029903(E).]
\bibitem{Dombriz_PRD_BH1}
De~la~Cruz-Dombriz, A.; Dobado, A.; Maroto, A.L.
{Black holes in modified gravity theories}.
\emph{J. Phys. Conf. Ser.} \textbf{2010}, \emph{229}, doi:10.1088/1742-6596/229/1/012033.

\bibitem{silvestri}  Pogosian, L.; Silvestri, A. Pattern of growth in viable f(R) cosmologies. \emph{Phys.\ Rev. D} \textbf{2008}, {\it 77}, 023503--023517.

\bibitem{Birkhoff} Birkhoff, G.D. \textit{Relativity and Modern Physics}; Harvard University Press: Cambrigde, MA, USA, 1923.
\bibitem{Birkhoff2}
 Jebsen, J.T. {\"Uber die allgemeinen kugelsymmetrischen L\"osungen der Einsteinschen Gravitationsgleichungen im Vakuum}. \emph{Ark. Mat. Ast. Fys.} \textbf{1921}, \emph{15}, 18.

\bibitem{Capozziello:2011wg}
  Capozziello, S.; S\'aez-G\'omez, D.
  {Scalar-tensor representation of $f(R)$ gravity and Birkhoff's theorem}.
  \emph{Annalen Phys}. \textbf{2012}, {\it 524}, 279--285.
 \bibitem{Capozziello:2011wg1}
  Capozziello, S.; S\'aez-G\'omez, D.
    {Conformal frames and the validity of Birkhoff's theorem}.
   \emph{AIP Conf.\ Proc.} \textbf{2011}, {\it 1458}, 347--350. 
\bibitem{Carter} Carter, B. \textit{Les Astres Occlus}; DeWitt, C.M., Ed.; Gordon and Breach: New York, NY, USA, 1973.

\bibitem{Jimeno_BH}  Cembranos, J.A.R.; de~la~Cruz-Dombriz, A.; Jimeno-Romero, P.  {
Kerr-Newman black holes in f(R) theories}. \textbf{2011}, arXiv:1109.4519 [gr-qc].

\bibitem{Cembranos:2012ji}
  Cembranos, J.A.R.; de la Cruz-Dombriz, A.; Romero, P.J.
  {Modified spinning black holes}.
  \emph{AIP Conf.\ Proc.} \textbf{2011}, {\it 1458}, 439--442.

\bibitem{HGG}   Hartle, J.B.;  Hawking, S.W. {Path integral derivation of black hole radiance}. \emph{Phys.\ Rev. D} \textbf{1976}, {\it 13}, 2188--2203.
\bibitem{HGG1}  Gibbons, G.W.;  Perry, M.J. {Black holes and thermal green functions}. \emph{Proc.\ R.\ Soc.\ Lond.  A } \textbf{1978}, {\it 358}, 467--494.
\bibitem{HGG2}  Gibbons, G.W.;  Hawking, S.W. {Action integrals and partition functions in quantum gravity}. \emph{Phys.\ Rev. D} \textbf{1977}, {\it 15}, 2752--2756.
\bibitem{HGG3}  Gibbons, G.W.;  Hawking, S.W. \emph{Euclidean Quantum Gravity}; World Scientific Pub Co Inc: Singapore, 1993.

\bibitem{Hawking1974} Hawking, S.W. {Particle creation by black holes}. \emph{Commun.\ Math.\ Phys.} \textbf{1975}, {\it 43}, 199--220. [Erratum-ibid.\  \textbf{1976}, {\it 46}, 206].


\bibitem{Multamaki2007} Multamaki, T.; Putaja, A.; Vilja, I.; Vagenas, E.C. {Energy-momentum complexes in f(R) theories of gravity}. \emph{Class.\ Quant.\ Grav.} \textbf{2008}, {\it 25}, doi:10.1088/0264-9381/25/7/075017.

\bibitem{Bekenstein}  Bekenstein, J.D. {Black holes and entropy}. \emph{Phys.\ Rev.\  D} \textbf{1973}, {\it 7}, 2333--2346.

\bibitem{Bardeen1973} Bardeen, J.M.; Carter, B.; Hawking, S.W. {The four laws of Black Hole mechanics}. \emph{Commun.\ Math.\ Phys.} \textbf{1973}, {\it 31}, 161--170.

\bibitem{Caldarelli} Caldarelli, M.M.; Cognola, G.; Klemm, D. { 	
Thermodynamics of Kerr-Newman-AdS black holes and conformal field theories}. \emph{Class.\ Quant.\ Grav.} \textbf{2000}, {\it 17}, 399--420.

\bibitem{Starow}  Starobinsky, A.A. {A new type of isotropic cosmological models without singularity}. \emph{Phys. Lett. B} \textbf{1980}, {\it 91}, 99--102.

\bibitem{Mijic} Miji\'c, M.B.; Morris, M.S.; Suen, W.M. {The $R^2$ cosmology: Inflation without a phase transition
} \emph{Phys.\ Rev. D} \textbf{1986}, {\it 34}, 2934--2946.




 \bibitem{reconstruction1}
Nojiri, S.; Odintsov, S.D.
  {Modified gravity and its reconstruction from the universe expansion}.
  \emph{J.\ Phys.\ Conf.\ Ser.} \textbf{2007}, {\it 66}, doi:10.1088/1742-6596/66/1/012005. 
        \bibitem{reconstruction11}
Nojiri, S.; Odintsov, S.D.
  {Modified gravity as an alternative for Lambda-CDM cosmology}.
  \emph{J.\ Phys.\ A} \textbf{2007}, {\it 40}, doi:10.1088/1751-8113/40/25/S17.  

 \bibitem{reconstruction12}
Capozziello, S.; Nojiri, S.; Odintsov, S.D.; Troisi, A.
{Cosmological viability of f(R)-gravity as an ideal fluid and its
compatibility with a matter dominated phase}.
 \emph{Phys.\ Lett.\ B} \textbf{2006}, {\it 639}, 135--143. 
%
 \bibitem{reconstruction13}
Elizalde, E.; S\'aez-G\'omez, D.
{F(R) cosmology in presence of a phantom fluid and its scalar-tensor counterpart: Towards a unified precision model of the universe evolution}.
  \emph{Phys.\ Rev.\ D} \textbf{2009}, {\it 80}, 044030--044041.
  %
 \bibitem{reconstruction14}
  Brevik, I.H.
{Crossing of the w = -1 barrier in two-fluid viscous modified gravity}.
\emph{Gen.\ Rel.\ Grav.} \textbf{2006}, {\it 38}, 1317--1328.
%
 \bibitem{reconstruction15}
 Granda, L.N.
{Reconstructing the f(R) gravity from the holographic principle}. \textbf{2009}, arXiv:0812.1596 [hep-th].
%
 \bibitem{reconstruction16}
Setare, M.R.
{Holographic modified gravity}.
 \emph{Int.\ J.\ Mod.\ Phys. D} \textbf{2008}, {\it 17}, doi:10.1142/S0218271808013819.
%
 \bibitem{reconstruction17}
Wu, X.; Zhu, Z.H.
 {Reconstructing f(R) theory according to holographic dark energy}.
 \emph{Phys.\ Lett. B} \textbf{2008}, {\it 660}, 293--298.
%
 \bibitem{reconstruction18}
Bamba, K.; Geng, C.Q.; Nojiri, S.; Odintsov, S.D.
{Crossing of the phantom divide in modified gravity}.
\emph{Phys.\ Rev. D} \textbf{2009}, {\it 79}, 083014--083028.
%
%
 \bibitem{reconstruction19}
Elizalde, E.; Myrzakulov, R.; Obukhov, V.V.; S\'aez-G\'omez, D.
 {LambdaCDM epoch reconstruction from F(R,G) and modified Gauss-Bonnet gravities}.
  \emph{Class.\ Quant.\ Grav.} \textbf{2010}, {\it 27}, doi:10.1088/0264-9381/27/9/095007.
  %
 \bibitem{reconstruction110}
  Myrzakulov, R.; S\'aez-G\'omez, D.; Tureanu, A.
 {On the $\Lambda$CDM Universe in $f(G)$ gravity}.
  \emph{Gen.\ Rel.\ Grav.} \textbf{2011}, {\it 43}, arXiv:1009.0902.
 \bibitem{reconstruction111}
 Carroll, S.M.;  Duvvuri, V.;  Trodden, M.;  Turner, M.S. {Is cosmic speed-up due to new gravitational physics?}
 \emph{Phys. Rev. D} \textbf{2004}, {\it 70}, 043528--043532.
%
 \bibitem{reconstruction112}
 Dobado, A.;  Maroto, A.L. Inflatonless inflation.
  \emph{Phys.\ Rev. D} \textbf{1995}, {\it 52}, 1895--1901. 
%
%
 \bibitem{reconstruction114}
 Cembranos, J.A.R. {The Newtonian limit at intermediate energies}.
\emph{Phys.\ Rev.\ D} \textbf{2006}, {\it 73}, 064029--064033.
%
\bibitem{reconstruction2}
S\'aez-G\'omez, D.
  {Modified f(R) gravity from scalar-tensor theory and inhomogeneous EoS dark energy}.
  \emph{Gen.\ Rel.\ Grav.} \textbf{2009}, {\it 41}, 1527--1538.

\bibitem{Nojiri:2009kx}
  Nojiri, S.; Odintsov, S.D.; S\'aez-G\'omez, D.
 {Cosmological reconstruction of realistic modified F(R) gravities}.
  \emph{Phys.\ Lett. B} \textbf{2009}, {\it 681}, 74--80.

\bibitem{Nojiri:2011kd}
  Nojiri, S.; Odintsov, S.D.; S\'aez-G\'omez, D.
  {Cyclic, ekpyrotic and little rip universe in modified gravity}.
  \emph{AIP Conf.\ Proc.} \textbf{2011}, {\it 1458}, 207--221.
  \bibitem{Dunsby_PRD}
Dunsby, P.K.S.; Elizalde, E.; Goswami, R.; Odintsov, S.; S\'aez-G\'omez, D.
  {On the LCDM Universe in f(R) gravity}.
  \emph{Phys.\ Rev. D} \textbf{2010}, {\it 82}, 023519.

  \bibitem{NuevoSaez}
  S\'aez-G\'omez, D.
  {Cosmological evolution, future singularities and Little Rip in viable f(R) theories and their scalar-tensor counterpart}. \textbf{2012}, arXiv:1207.5472 [gr-qc].

   \bibitem{Hayward:1997jp}
  Hayward, S.A.
  {Unified first law of black hole dynamics and relativistic thermodynamics}.
  \emph{Class.\ Quant.\ Grav.} \textbf{1998}, {\it 15}, doi:10.1088/0264-9381/15/10/017.
  \bibitem{SDO-IB1}
Brevik, I.; Odintsov, S.D.
{On the Cardy-Verlinde entropy formula in viscous cosmology}.
\emph{Phys.\ Rev. D} \textbf{2002}, {\it 65}, 067302--067305. 
\bibitem{SDO-IB1a}
Brevik, I.
 {Cardy-verlinde entropy formula in the presence of a general
state equation}.
\emph{Phys.\ Rev. D} \textbf{2002}, {\it 65}, 127302--127305. 
\bibitem{SDO-IB1b}
Brevik, I.
{Viscous cosmology and the Cardy-Verlinde formula}.
\emph{Int.\ J.\ Mod.\ Phys. A } \textbf{2003}, {\it 18}, doi:10.1142/S0217751X03015593.
  \bibitem{Casimir}
  Brevik, I.; Gorbunova, O.; S\'aez-G\'omez, D.
  {Casimir effects near the big rip singularity in viscous cosmology}.
  \emph{Gen.\ Rel.\ Grav.} \textbf{2010}, {\it 42}, 1513--1522.

\bibitem{Casimir1}
Gorbunova, O.; S\'aez-G\'omez, D. {The Oscillating dark energy and cosmological Casimir effect}.
 \emph{ Open Astron.\ J.} \textbf{2010}, {\it 3}, 73--75.

 \bibitem{InhEoS}
Nojiri, S.; Odintsov, S.D.
{Inhomogeneous equation of state of the universe: Phantom era,
singularity and crossing the phantom barrier}.
\emph{Phys.\ Rev. D} \textbf{2005}, {\it 72}, 023003--023014. 

\bibitem{InhEoS1}
Nojiri, S.; Odintsov, S.D.
{The New form of the equation of state for dark energy fluid and accelerating universe}.
\emph{Phys.\ Lett. B} \textbf{2006}, {\it 639}, 144--150.


\bibitem{vanzo}
Brevik, I.; Nojiri, S.; Odintsov, S.D.; Vanzo, L.
{Entropy and universality of Cardy-Verlinde formula in dark energy universe}.
\emph{Phys.\ Rev. D} \textbf{2004}, {\it 70}, 043520--043533.

    \bibitem{cai}
Cai, R.G.
{Cardy-Verlinde formula and thermodynamics of black holes in de
spaces}.
\emph{Nucl.\ Phys. B} \textbf{2002}, {\it 628}, 375--386. 
\bibitem{cai1}
Cai, R.G.
{Cardy-Verlinde formula and asymptotically de Sitter spaces}.
\emph{Phys. Lett. B} \textbf{2002}, \emph{525}, 331--336.


\bibitem{ConformalAnomaly1}
Nojiri, S.;  Odintsov, S.D.;  Tsujikawa, S.
{Properties of singularities in (phantom) dark energy universe}.
\emph{Phys. Rev. D} \textbf{2005}, \emph{71}, 063004--063019.
   \bibitem{Shtanov:2002ek}
  Shtanov, Y.; Sahni, V.
  {Unusual cosmological singularities in brane world models}.
  \emph{Class.\ Quant.\ Grav.} \textbf{2002}, {\it 19}, L101--L107.

   \bibitem{singularity}
Nojiri, S.; Odintsov, S.D. {The future evolution and finite-time singularities in
unifying the inflation and cosmic acceleration}.
\emph{Phys.\ Rev.\ D} \textbf{2008}, {\it 78}, 046006--046017. 

\bibitem{singularity1}
Bamba, K.; Nojiri, S.; Odintsov, S.D.
{The future of universe  in modified gravity theories: Approaching the
finite-time future singularity}.
\emph{J. Cosmol. Astropart. Phys.} \textbf{2008}, {\it 0810}, doi:10.1088/1475-7516/2008/10/045. 

\bibitem{singularity2}
Capozziello, S.; de Laurentis, M.; Nojiri, S.; Odintsov, S.D.
{Classifying and avoiding singularities in the alternative gravity}.
\emph{Phys.\ Rev. D} \textbf{2009}, {\it 79}, 124007--124022. 

\bibitem{singularity3}
 Bamba, K.;  Odintsov, S.D.;  Sebastiani, L.;  Zerbini, S.
{Finite-time future singularities in modified Gauss-Bonnet and F(R,G) gravity and singularity avoidance}.
  \emph{Eur.\ Phys.\ J.\ C} \textbf{2010}, {\it 67}, 295--310. 

\bibitem{singularity4}
Abdalla, M.C.B.; Nojiri, S.; Odintsov, S.D.
{Consistent modified gravity: Dark energy, acceleration and the cosmic doomsday}.
\emph{Class.\ Quant.\ Grav.} \textbf{2005}, {\it 22}, L35--L42.
    \bibitem{Elizalde:2004mq}
Elizalde, E.; Nojiri, S.; Odintsov, S.D. {Late-time cosmology in (phantom) scalar-tensor theory: Dark energy cosmic speed-up}.
\emph{Phys.\ Rev.\ D} \textbf{2004}, {\it 70}, 043539--043558. 
\bibitem{phantom1}
Caldwell, R.R.
{A phantom menace? Cosmological consequences of a dark energy component with super-negative equation of state}
\emph{Phys.\ Lett.\ B} \textbf{2002}, {\it 545}, 23--29. 
\bibitem{phantom2}Caldwell, R.R.; Kamionkowski, M.; Weinberg, N.N.
{Phantom energy and cosmic doomsday}.
\emph{Phys.\ Rev.\ Lett.} \textbf{2003}, {\it 91}, 071301--071304. 
\bibitem{phantom3}~McInnes, B.
{The dS/CFT correspondence and the big smash}.
\emph{J. High. Energy Phys.} \textbf{2002}, {\it 208}, doi:10.1088/1126-6708/2002/08/029.
\bibitem{phantom4} Nojiri, S.; Odintsov, S.D. {Quantum deSitter cosmology and phantom matter}.
\emph{Phys.\ Lett.\ B} \textbf{2003}, {\it 562}, 147--152. 
\bibitem{phantom5}Nojiri, S.; Odintsov, S.D.
{Effective equation of state and energy conditions in phantom
 inflationary cosmology perturbed by quantum effects}.
\emph{Phys.\ Lett.\ B} \textbf{2003}, {\it 571}, 1--10. 
\bibitem{phantom6}Gonzalez-Diaz, P.F.
{K-essential phantom energy: Doomsday around the corner?}
\emph{Phys.\ Lett.\ B} \textbf{2004}, {\it 586}, 1--4. 
\bibitem{phantom7}Gonzalez-Diaz, P.F.
{On tachyon and sub-quantum phantom cosmologies}.
\textbf{2004}, arXiv:hep-th/0408225. 
\bibitem{phantom8}Sami, M.; Toporensky, A.
{Phantom field and the fate of universe}.
\emph{Mod.\ Phys.\ Lett.\ A} \textbf{2004}, {\it 19}, doi:10.1142/S0217732304013921. 
\bibitem{phantom9}Stefancic, H.
{Generalized phantom energy}.
\emph{Phys.\ Lett.\ B} \textbf{2004}, {\it 586}, 5--10.
\bibitem{phantom10}Chimento, L.P.; Lazkoz, R.
{Constructing Phantom Cosmologies from Standard Scalar Field Universes}.
\emph{Phys.\ Rev.\ Lett.} \textbf{2003}, {\it 91}, 211301--211303. 
\bibitem{phantom11}Chimento, L.P.; Lazkoz, R. {On big rip singularities}.
\emph{Mod.\ Phys.\ Lett.\ A} \textbf{2004}, {\it 19}, doi:10.1142/S0217732304015646. 
\bibitem{phantom12}Hao, J.G.; Li, X.Z.
{Generalized quartessence cosmic dynamics: Phantom or quintessence
Sitter attractor}. \emph{Phys.\ Lett.\ B} \textbf{2005}, {\it 606}, 7--11. 
\bibitem{phantom13}Babichev, E.; Dokuchaev, V.; Eroshenko, Yu.
{Dark energy cosmology with generalized linear equation of state}.
\emph{Class.\ Quant.\ Grav.} \textbf{2005}, {\it 22}, doi:10.1088/0264-9381/22/1/010. 
\bibitem{phantom14}Zhang, X.F.; Li, H.; Piao, Y.S.; Zhang, X.M.
{Two-field models of dark energy with equation of state across}.
\emph{Mod.\ Phys.\ Lett.\ A} \textbf{2006}, {\it 21}, doi:10.1142/S0217732306018469. 
\bibitem{phantom15}Elizalde, E.; Nojiri, S.; Odintsov, S.D.; Wang, P.
{Dark energy: Vacuum fluctuations, the effective phantom phase,and
holography}.
\emph{Phys.\ Rev.\ D} \textbf{2005}, {\it 71}, 103504--103511. 
\bibitem{phantom16}Dabrowski, M.P.; Stachowiak, T.
{Phantom Friedmann cosmologies and higher-order characteristics of
expansion}.
\emph{Ann. Phys.} \textbf{2006}, {\it 321}, 771--812. 
\bibitem{phantom17}Lobo, F.S.N.
{Phantom energy traversable wormholes}.
\emph{Phys.\ Rev.\ D} \textbf{2005}, {\it 71}, 084011--084018. 
\bibitem{phantom18}Cai, R.G.; Zhang, H.S.; Wang, A.
{Crossing w = -1 in Gauss-Bonnet brane world with induced}.
\emph{Commun.\ Theor.\ Phys.} \textbf{2005}, {\it 44}, doi:10.1088/6102/44/5/948. 
\bibitem{phantom19}Arefeva, I.Y.; Koshelev, A.S.; Vernov, S.Y.
{Exactly solvable SFT inspired phantom model}.
\emph{Theor.\ Math.\ Phys.} \textbf{2006}, {\it 148}, arXiv:astro-ph/0412619. 
\bibitem{phantom20}Elizalde, E.; Nojiri, S.; Odintsov, S.D.; S\'aez-G\'omez, D.;
Faraoni, V. {Reconstructing the universe history, from inflation to
phantom and canonical scalar fields}
\emph{Phys.\ Rev.\ D} \textbf{2008}, {\it 77}, 106005--106020. 
\bibitem{ConformalAnomaly2}
Nojiri, S.; Odintsov, S.D.
{AdS/CFT correspondence, conformal anomaly and quantum corrected
bounds}.
\emph{Int.\ J.\ Mod.\ Phys.\ A} \textbf{2001}, {\it 16}, doi:10.1142/S0217751X0100412.
\bibitem{Sahni:2002dx}
  Sahni, V.; Shtanov, Y.
  {Brane world models of dark energy}.
  \emph{J. Cosmol. Astropart. Phys.} \textbf{2003}, {\it 311}, doi:10.1088/1475-7516/2003/11/014.

\bibitem{Frampton:2011sp}
  Frampton, P.H.; Ludwick, K.J.; Scherrer, R.J.
  {The little rip}.
  \emph{Phys.\ Rev.\ D} \textbf{2011}, {\it 84}, 063003--063007.
\bibitem{LopezRevelles:2012cg}
  Lopez-Revelles, A.J.; Myrzakulov, R.; S\'aez-G\'omez, D.
  {Ekpyrotic universes in $F(R)$ Ho\v{r}ava-Lifshitz gravity}.
  \emph{Phys.\ Rev.\ D} \textbf{2012}, {\it 85}, 103521--103530.
  \bibitem{12071646}
  Houndjo, M.J.S.; Alvarenga, F.G.; Rodrigues, M.E.; Jardim, D.F.
  {Thermodynamics in Little Rip cosmology in the framework of a type of f(R; T) gravity}. \textbf{2012}, arXiv:1207.1646 [gr-qc].

  \bibitem{Bamba:2011dk}
  Bamba, K.; Geng, C.Q.; Lee, C.C.
  {Phantom crossing in viable $f(R)$ theories}.
  \emph{Int.\ J.\ Mod.\ Phys.\ D} \textbf{2011}, {\it 20}, doi:10.1142/S0218271811019517. 

\bibitem{Padmanabhan:2009vy}
  Padmanabhan, T. {Thermodynamical aspects of gravity: New insights}.
  \emph{Rept.\ Prog.\ Phys.} \textbf{2010}, {\it 73}, doi:10.1088/0034-4885/73/4/046901.

\bibitem{Verlinde:2010hp}   Verlinde, E.P.
  {On the origin of gravity and the laws of newton}.
  \emph{J. High Energy Phys.} \textbf{2011}, {\it 1104}, doi:10.1007/JHEP04(2011)029.

\bibitem{CMB_Dombriz} Bourhrous, H.; de la Cruz-Dombriz, A.; Dunsby, P. { 	
CMB tensor anisotropies in metric f(R) gravity}. \emph{AIP Conf. Proc.} \textbf{2011}, \emph{1458}, 343--346.

\bibitem{Collapse_Dombriz_1} Cembranos, J.A.R.; de la Cruz-Dombriz, A.; Nunez, B.M. {Gravitational collapse in f(R) theories}. \emph{J. Cosmol. Astropart. Phys.} \textbf{2012}, {\it 1204}, doi:10.1088/1475-7516/2012/04/021.

\bibitem{Collapse_Dombriz_2} Cembranos, J.A.R.; de la Cruz-Dombriz, A.; Nunez, B.M. {On the collapse in fourth order gravities}.
 \emph{AIP Conf. Proc.} \textbf{2011}, \emph{1458}, 491--494.

\bibitem{Collapse_Dombriz_3} Albareti F. D.; Cembranos, J.A.R.; de la Cruz-Dombriz, A. {Focusing of geodesic congruences in an accelerated expanding Universe}. \textbf{2012}, arXiv:1208.4201.

\bibitem{Oyaizu:2008tb}
  Oyaizu, H.; Lima, M.; Hu, W.
  {Nonlinear evolution of f(R) cosmologies. 2. Power spectrum}.
  \emph{Phys.\ Rev.\ D} \textbf{2008}, {\it 78}, 123524--123532.
\bibitem{Schmidt:2008tn}
  Schmidt, F.; Lima, M.V.; Oyaizu, H.; Hu, W.
  {Non-linear evolution of f(R) cosmologies III: Halo statistics}.
  \emph{Phys.\ Rev.\ D} \textbf{2009}, {\it 79}, 083518--083530.

  \bibitem{Schmidt:2008hc}
  Schmidt, F. {Weak lensing probes of modified gravity}.
  \emph{Phys.\ Rev.\ D} \textbf{2008}, {\it 78}, 043002--043015.
 \bibitem{Capozziello:2011nr}
  Capozziello, S.; de Laurentis, M.; Odintsov, S.D.; Stabile, A.
  {Hydrostatic equilibrium and stellar structure in f(R)-gravity}.
  \emph{Phys.\ Rev.\ D} \textbf{2011}, {\it 83}, arXiv:1101.0219 [gr-qc].
\bibitem{microBH}  Dimopoulos, S.;  Landsberg, G.L. {Black Holes at the Large Hadron Collider}. \emph{Phys.\ Rev.\ Lett.} \textbf{2001}, {\it 87}, 161602--161605.

\bibitem{microBH1} Alberghi, G.L.;  Casadio, R.;  Tronconi, A. {Quantum gravity effects in black holes at the LHC}. \emph{J.\ Phys.\ G} \textbf{2007}, {\it 34}, 767--778.




































%
%
%
%
%
\end{thebibliography}

\end{document}